\documentclass[pra, aps, twocolumn, nolongbibliography, superscriptaddress, reprint, nofootinbib]{revtex4-2} 

\newcommand{\tr}{\text{tr}}

\newcommand{\ii}{\text{i}}
\newcommand{\dd}{\text{d}}

\renewcommand{\Im}{\text{Im}}

\newcommand{\curlE}{\mathcal{E}}

\newcommand{\ph}{\varphi}
\newcommand{\omegaeg}{\omega_{\text{eg},0}}

\newcommand{\ms}{\text{MS}}

\newcommand{\hc}{\text{h.c.}}

\newcommand\brpm{\mathbin{\vcenter{\hbox{\oalign{$\scriptstyle({+})$\cr
					\noalign{\kern-.3ex}
					\hfil$\scriptscriptstyle-$\hfil\cr}}}}}

\let\originalleft\left
\let\originalright\right
\renewcommand{\left}{\mathopen{}\mathclose\bgroup\originalleft}
\renewcommand{\right}{\aftergroup\egroup\originalright}

\usepackage[pdftex]{graphicx}
\usepackage{tikz}
\definecolor{myblue}{rgb}{0.12156862745098039, 0.4666666666666667, 0.7058823529411765}
\usepackage{hyperref}
\hypersetup{
     colorlinks=true,
     linkcolor=myblue,
     filecolor=magenta,      
     urlcolor=myblue,
     pdftitle={Strategies for practical advantage of fault-tolerant circuit design in noisy trapped-ion quantum computers},
     citecolor = myblue
}
\usepackage{amsmath}
\usepackage{mathtools}
\usepackage{amssymb}
\usepackage{xcolor}
\usepackage{nicefrac}
\usepackage{siunitx}
\usepackage{bbold}
\usepackage{braket}
\usepackage{hhline}

\usepackage[caption=false]{subfig}
\usepackage{ragged2e} \DeclareCaptionJustification{justified}{\justifying}

\begin{document}

\title{Strategies for practical advantage of fault-tolerant circuit design\\ in noisy trapped-ion quantum computers}

\author{Sascha Heu\ss en}
\email{sascha.heussen@rwth-aachen.de}
\affiliation{Institute for Quantum Information, RWTH Aachen University, Aachen, Germany}
\affiliation{Institute for Theoretical Nanoelectronics (PGI-2), Forschungszentrum J\"{u}lich, J\"{u}lich, Germany}

\author{Lukas Postler}
\affiliation{Institut f\"{u}r Experimentalphysik, Universit\"{a}t Innsbruck, Innsbruck, Austria}

\author{Manuel Rispler}
\affiliation{Institute for Quantum Information, RWTH Aachen University, Aachen, Germany}
\affiliation{Institute for Theoretical Nanoelectronics (PGI-2), Forschungszentrum J\"{u}lich, J\"{u}lich, Germany}

\author{Ivan Pogorelov}
\affiliation{Institut f\"{u}r Experimentalphysik, Universit\"{a}t Innsbruck, Innsbruck, Austria}

\author{\mbox{Christian D. Marciniak}}
\affiliation{Institut f\"{u}r Experimentalphysik, Universit\"{a}t Innsbruck, Innsbruck, Austria}

\author{Thomas Monz}
\affiliation{Institut f\"{u}r Experimentalphysik, Universit\"{a}t Innsbruck, Innsbruck, Austria}
\affiliation{Alpine Quantum Technologies GmbH, Innsbruck, Austria}

\author{Philipp Schindler}
\affiliation{Institut f\"{u}r Experimentalphysik, Universit\"{a}t Innsbruck, Innsbruck, Austria}

\author{Markus M\"{u}ller}
\affiliation{Institute for Quantum Information, RWTH Aachen University, Aachen, Germany}
\affiliation{Institute for Theoretical Nanoelectronics (PGI-2), Forschungszentrum J\"{u}lich, J\"{u}lich, Germany}

\begin{abstract}
Fault-tolerant quantum error correction provides a strategy to protect information processed by a quantum computer against noise which would otherwise corrupt the data. 
A fault-tolerant universal quantum computer must implement a universal gate set on the logical level in order to perform arbitrary calculations to in principle unlimited precision. In this manuscript, we characterize the recent demonstration of a fault-tolerant universal gate set in a trapped-ion quantum computer [Postler et al.~Nature 605.7911 (2022)] and identify aspects to improve the design of experimental setups to reach an advantage of logical over physical qubit operation.
We show that various criteria to assess the break-even point for fault-tolerant quantum operations are within reach for the ion trap quantum computing architecture under consideration. Furthermore, we analyze the influence of crosstalk in entangling gates for logical state preparation circuits. 
These circuits can be designed to respect fault tolerance for specific microscopic noise models. 
We find that an experimentally-informed depolarizing noise model captures the essential noise dynamics of the fault-tolerant experiment that we consider, and crosstalk is negligible in the currently accessible regime of physical error rates. For deterministic Pauli state preparation, we provide a fault-tolerant unitary logical qubit initialization circuit, which can be realized without in-sequence measurement and feed-forward of classical information. 
Additionally, we show that non-deterministic state preparation schemes, i.e.~repeat until success, for logical Pauli and magic states perform with higher logical fidelity over their deterministic counterparts for the current and anticipated future regime of physical error rates. 
Our results offer guidance on improvements of physical qubit operations and validate the experimentally-informed noise model as a tool to predict logical failure rates in quantum computing architectures based on trapped ions. %
 \end{abstract}

\maketitle

\section{Introduction}

The toolbox of quantum fault tolerance provides a key on the way towards universal quantum computation~\cite{Campbell2017}. By careful circuit design this allows one to contain the effect of faults stemming from the fundamentally noisy hardware of real physical quantum systems. Here, the ideal computation takes place in a subspace (dubbed the \emph{logical} subspace) of the (much larger) \emph{physical} Hilbert space, where the logical information is typically encoded in non-local degrees of freedom of a quantum error correcting (QEC) code and protected against local noise~\cite{Terhal2015}. Avenues to experimental investigation of fault-tolerant (FT) design principles have been opened up by recent leaps in quantum computing experiments and the development of the theory of flag fault tolerance, where dedicated auxiliary qubits flag the presence or absence of dangerous error patterns~\cite{Chao2018, Chamberland2018,chamberland2019fault,reichardt2020fault,Chao2020}. In trapped-ion systems, code state preparation~\cite{Nigg2014}, FT error detection~\cite{Linke2017}, FT stabilizer readout~\cite{hilder2022fault}, FT operation of one logical qubit~\cite{Egan2021} as well as logical entangling gates~\cite{Erhard2021} and repetitive QEC cycles~\cite{ryan2021realization} were achieved. The state of the art now lies in FT universal gate sets~\cite{postler2022demonstration} conjoined with repetitive QEC cycles~\cite{Ryan-Anderson2022}. In superconducting qubits, this evolution is paralleled, where code state preparation~\cite{Takita2017,Satzinger2021}, error detecting QEC cycles~\cite{Andersen2020}, logical gates in an error detecting code~\cite{Marques2022} and the operation of a surface code with QEC cycles~\cite{Krinner2022} and higher-distance surface codes~\cite{Acharya2022} were demonstrated. Other qubit platforms are showing greatly increasing capabilities recently along similar directions~\cite{Abobeih2022,Bluvstein2022}.

Central to the task of FT operation of a quantum processor are a) the ability to initialize logical states, i.e.~QEC code states, b) to measure their error syndrome, c) to perform logical gates using a universal set of gates, and d) to determine logical measurement outcomes. All these tasks have to be implemented fault-tolerantly, i.e.~in such a way that they do not introduce errors beyond what can be tolerated by the QEC code. Furthermore the noise level of all operations needs to be below a (model-dependent) threshold~\cite{Aharonov2008, Aliferis2006}. A major concern is the proliferation of errors due to the application of entangling operations when implementing a logical gate. A landmark result that emerged from fault tolerance theory is that FT logical gate operations typically fall into two categories. On the one hand, some gate operations can be relatively straightforward to compose by transversal implementations, where the logical gate operation can be synthesized by independent bit-wise action on the qubit register, thus avoiding any need for entangling operations within the logical qubit block. On the other hand, there are always gates that defy this realization and require special treatment, as dictated by the no-go theorem of Eastin and Knill~\cite{Eastin2009, Bravyi2005}. For the platform of trapped-ion qubits, the ability to perform a universal gate set on the logical level has recently become experimental reality as part of a demonstration by Ref.~\cite{postler2022demonstration}. In the present work, we provide an extensive analysis to put this experiment into a broader context of current and projected experimental capabilities. 

\section{Outline and summary of main results} 
This paper is structured as follows: In Section~\ref{sec:experimental}, we discuss the trapped-ion setup and give an overview over the physics that provide the basis for defining qubit states as well as single-qubit and entangling operations. We lay out how this leads us to an experimentally-motivated noise model, building on and extending the model used in Ref.~\cite{postler2022demonstration}. Also, we introduce the circuit sampling technique of subset sampling and discuss how it fares compared to conventional Monte Carlo sampling. In Section~\ref{sec:prot_sims}, we discuss one of the key aspects of FT circuit design, namely in what scenario and parameter regime they become useful by outperforming their non-FT or bare physical counterparts. We discuss which parameters the logical qubit performance can and should be compared to and present how FT circuits for logical Pauli eigenstate preparation as well as logical magic state preparation perform on those scales. We find in Section~\ref{sec:nondet}, that under our current noise levels, the Pauli state preparation is already on the edge to the break-even point of outperforming the physical initialization operation. The logical magic state preparation is already below one of the relevant break-even points, namely the physical entangling gate error rate with current noise parameters. By a scaling analysis of the physical error rates, we find that both will be brought to the sub-threshold regime with moderate hardware improvements. We extend the discussion in Section~\ref{sec:determ} by comparing non-deterministic circuits, where runs with flag events are discarded, to deterministic circuits, where runs with flag events are instead treated with further circuitry to maintain fault tolerance. We find that the added circuitry reduces the logical fidelity substantially and discuss the scenarios where either might be preferential. In Section~\ref{sec:ctr}, we discuss the relevance of crosstalk, where we explain the notion of entangling crosstalk and the corresponding error channel. We study its potentially detrimental effect on QEC under current and projected experimental noise. We find that it does not constitute a major noise source at current noise levels but might become relevant at lower error rates. Nevertheless, we demonstrate how a specific type of entangling crosstalk can be mitigated by carefully designing the circuit. In Section~\ref{sec:fidelity}, we calculate the quantum state fidelity of a single logical qubit under different noise models. From the comparision of different performance metrics we conclude that the logical fidelity is the appropriate measure and the central figure of merit used for quantifying the logical qubit performance.

 \section{Trapped-ion based quantum processors}
\label{sec:experimental}
One of the most promising system architectures for fault-tolerant quantum information processors is trapped-ion based devices~\cite{Leibfried2003Quantum, Haffner2008Quantum, Ozeri2011Trapped, Bruzewicz2019Trapped}. These devices offer mature hardware, high-fidelity operations and all-to-all qubit connectivity. For register sizes of up to around 20 qubits~\cite{pogorelov2021compact, Ryan-Anderson2022} any arbitrary pair of qubits in the register can be natively entangled with a single quantum operation, facilitating certain quantum algorithms or rather reducing the overhead of their implementation drastically~\cite{linke2017experimental}. This is achieved by exploiting a long-range interaction between the ions mediated by a collective motional mode of the ion Coulomb crystal. Even larger registers can be realized by subdividing the register into smaller segments, each providing all-to-all connectivity~\cite{kielpinski2002architecture}. Interactions between such subsections can be realized by spatially rearranging the segments and single ions within the segments. Individual ions can be moved within the device for reconfiguration, an operation referred to as shuttling, via the application of time-dependent voltages to electrodes of the ion trap. Fault-tolerant gadgets have already been demonstrated in setups following this ion-shuttling based approach~\cite{ryan2021realization, hilder2022fault, Ryan-Anderson2022}. For the remainder of this section we will focus on a system hosting a static ion string that provides all-to-all connectivity in a register of 16 qubits~\cite{pogorelov2021compact}. In the following we will discuss the native gateset and noise processes of the device.

\subsection{Static ion chain quantum processor} \label{sec:expops}

\begin{figure}
    \centering
    \includegraphics[width=\linewidth]{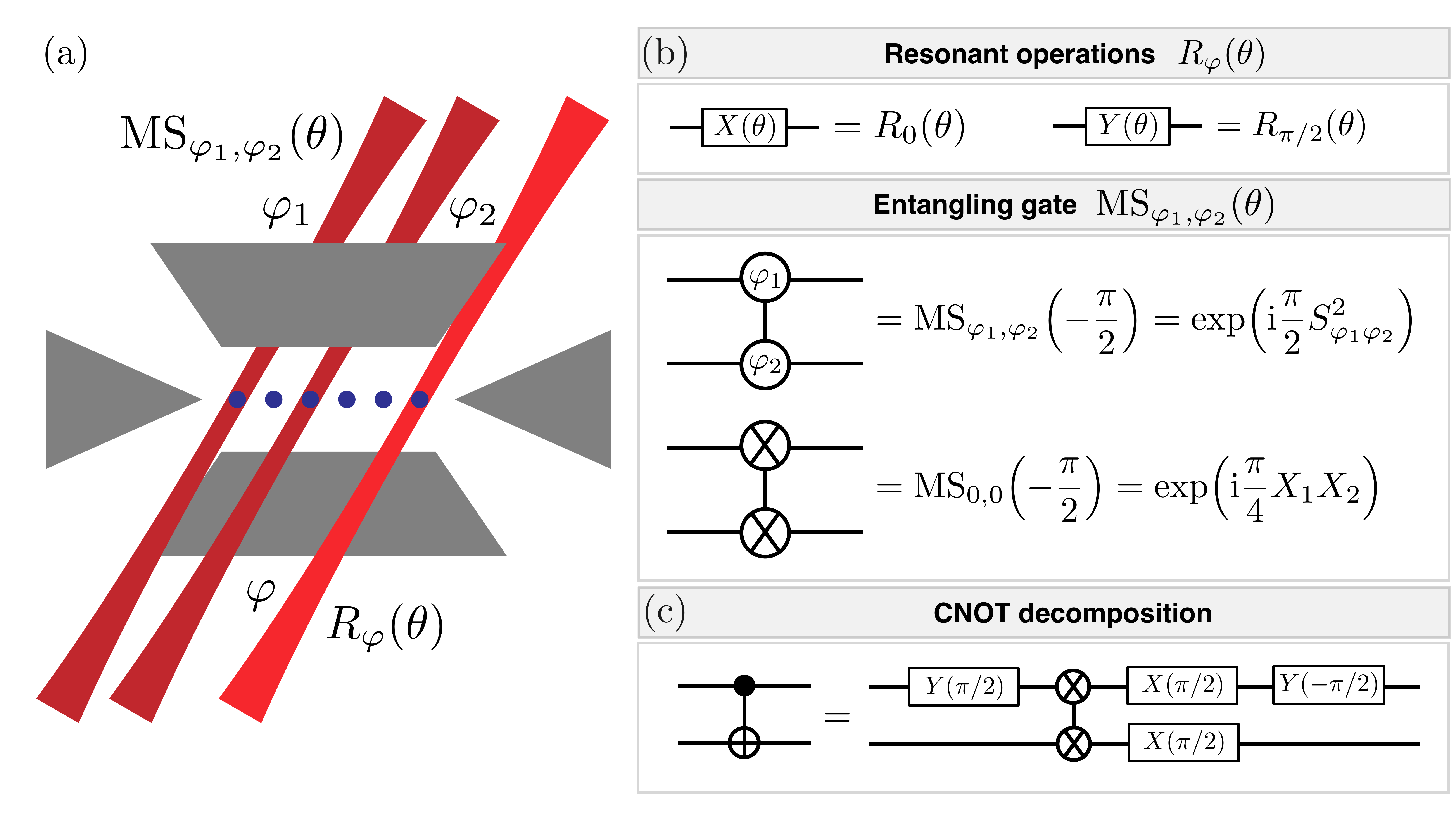}
    \caption{\textbf{Trapped-ion device architecture and native gate set.}
    (a) Trapped ions (blue dots) are suspended in a macroscopic linear Paul trap. Tightly focused laser beams allow for the implementation of single-qubit rotations (lighter-shaded red laser beam) and entangling operations on arbitrary pairs (darker-shaded beams). (b) The native gateset consists of resonant single-qubit operations implementing rotations around an axis in the equatorial plane of the Bloch sphere, where the rotation axis in controlled via the phase $\varphi$ of the laser pulse. The rotation angle $\theta$ is controlled via the pulse area of the laser pulse. The native entangling operation is realized via a M{\o}lmer-S{\o}rensen-type interaction. $Z$-operations can be implemented in software by updating the phase of subsequent light pulses on the respective ion through individual control of the light phase. This allows for the implementation of generalized M{\o}lmer-S{\o}rensen-type gates $\ms_{\varphi_1, \varphi_2}(\theta)$. (c) Decomposition of a CNOT gate into a M{\o}lmer-S{\o}rensen gate and local operations.}
    \label{fig:architecture}
\end{figure}

The system under consideration uses a macroscopic Paul trap~\cite{pogorelov2021compact}. A suitable set of radio frequency and static voltages applied to the trap electrodes ensures that trapped ions form a one-dimensional crystal, where their equilibrium positions are determined by the interplay of the trapping forces and the Coulomb interaction between the ions~\cite{james1998quantum}. Each trapped $^{40}$Ca$^+$ ion hosts a qubit in the Zeeman sublevels 4$S_{\nicefrac{1}{2}, m_j = -\nicefrac{1}{2}} = \ket{0}$ and 3$D_{\nicefrac{5}{2}, m_j = -\nicefrac{1}{2}} = \ket{1}$ of the ground state and a metastable excited state with a lifetime of $T_1 \approx \SI{1.2}{\s}$~\cite{Barton2000Measurement}. As can be seen in Fig.~\ref{fig:architecture}, a tightly focused laser beam addressing this quadrupole transition allows for individual control of the qubits in the register. The native gateset of the apparatus consists of the following three types of operations:

\begin{itemize}
\item \textbf{Resonant operations:} A laser pulse resonant to the qubit transition with variable phase and pulse area implements rotations \mbox{$R_{\varphi}^{(i)}(\theta) = \mathrm{exp}(-\ii\frac{\theta}{2} (X_{i} \cos{\varphi} + Y_{i} \sin{\varphi}))$}
around an axis in the equatorial plane of the Bloch sphere, where $X_{i}$ and $Y_{i}$ are single-qubit Pauli matrices acting on qubit $i$. The rotation angle $\theta$ is controlled via the duration and intensity of the laser pulse, and the angle of the rotation axis with respect to the $X$-axis $\varphi$ is controlled via the pulse phase.
A pulse length of about \SI{15}{\micro\s} is required to implement a rotation angle of $\pi/2$.
\item \textbf{Entangling operations:} Entangling operations acting on an arbitrary pair of ions are realized by illuminating the respective ions with a bichromatic light field slightly detuned from a center-of-mass radial mode, effectively applying a M{\o}lmer-S{\o}rensen (MS) interaction to the respective ions~\cite{sorensen2000entanglement}. The phase of the light illuminating the ion pair can be controlled individually, which allows for the implementation of the unitary operations $\ms_{\varphi_1, \varphi_2}(\theta) = \exp\left(-\ii \theta S_{\varphi_1, \varphi_2}^2 \right)$ with $S_{\varphi_1, \varphi_2} = \frac{1}{2} \left( X_1 \cos \varphi_1  + X_2 \cos \varphi_2 + Y_1 \sin \varphi_1 + Y_2 \sin \varphi_2 \right)$. A rotation angle of $\theta = -\pi/2$ renders the operation maximally entangling and makes the MS gate operation equivalent to a CNOT up to local operations~\cite{maslov2017basic}. The native implementation of this entangling gate we use provides only negative values of $\theta$ due to the spectral structure of the collective motional modes. We want to note that the symbol introduced in the second panel of Fig.~\ref{fig:architecture}b widely used throughout this work refers to an $XX$-rotation with $\theta = -\pi/2$.
\item \textbf{Virtual $Z$-operations:} $Z$-rotations are implemented in software by manipulating a phase register in the classical control hardware~\cite{negnevitsky2018feedback} that keeps track of $Z$-operations for each ion. The phases of all subsequent single-qubit and entangling operations are shifted according to the state of the phase register~\cite{mckay2017efficient}.
\end{itemize}

Currently, the setup under consideration does not allow for parallel execution of gate operations, as a simultaneous illumination of only up to two ions is possible. This restriction is mainly due to a limited number of RF sources controlling the beam steering optics available in the control hardware and laser power limitations, as the light intensity illuminating an ion decreases quadratically with the number of addressed ions~\cite{ringbauer2022universal}. Modifications to the addressing setup would eliminate this technical limitation and facilitate parallel execution of gate operations~\cite{figgatt2019parallel, Ryan-Anderson2022}.

\subsection{Noise modeling and simulation}\label{sec:noise_main}

In this section, we discuss noise processes affecting the performance of the quantum processor under consideration and introduce theoretical models describing these processes. We analyze their influence on the performance of fault-tolerant circuits and estimate necessary improvements to achieve a break-even of fault-tolerant encoded qubits with respect to bare physical qubits.

\textbf{Idling noise.} A fundamental noise process affecting all implementations of physical qubits is idling noise altering the quantum state of a qubit, which is not target of an operation at the respective time. Thereby the effect on idling qubits is not dependent on the target qubits of the respective operation, in contrast to crosstalk discussed later in this section. For trapped-ion architectures utilizing metastable electronic states, three processes are affecting the state of idling qubits: As the qubit state $|1\rangle$ is encoded in a metastable excited state its population decays exponentially. First, it either decays to $|0\rangle$, referred to as amplitude damping, or, second, it leaks out of the computational subspace while decaying to the Zeeman sublevel 4$S_{\nicefrac{1}{2}, m_j = +\nicefrac{1}{2}}$. The rates of these processes are governed by the lifetime of the metastable state $T_1$. Third, fluctuations in the laser frequency or magnetic field during idle time lead to dephasing on a timescale of $T_2 \approx \SI{100}{\milli\s}$\footnote{Typical values for the experimental dephasing time vary between \SI{30}{\milli\s} and \SI{200}{\milli\s} from day to day depending on the electromagnetic environment being present.}.

Due to the predominance of dephasing over amplitude damping and leakage, the incoherent noise channel for idling qubits can be modeled by Pauli-$Z$ faults and reads
\begin{align}
    \curlE_\text{idle,deph}(\rho) &= (1-p_\text{idle})\rho + p_\text{idle} Z \rho Z. \label{eq:noise_deph}
\end{align}
A more accurate model could also include effects of correlated dephasing which were reported in previous investigations~\cite{rivas2015quantifying, postler2018experimental, pal2022relaxation}. However, as idling is only a weak source of failure in our setup we do not expect a difference between correlated and uncorrelated idling noise. Thus, we choose to model the dephasing noise as uncorrelated on the individual physical qubits. The physical error rate for idling faults $p_\text{idle}$ depends on the execution time $t$ of the gate performed on a subset of ions and the coherence time $T_2$. The incoherent probabilities for the dephasing process on idling qubits is given by 
\begin{align}
	p_\text{idle} &= \frac{1}{2} \left( 1 - \exp \left( - \frac{t}{T_2} \right) \right). \label{eq:pidle}
\end{align}

The table below shows typical execution times of various operations in the setup considered and the associated physical error rates for idling qubits during these operations.

\begin{center}
\begin{tabular}{ c | c | c}
operation & time $t$ & $p_\text{idle}$ \\ \hline 
single-qubit rotation & \SI{15}{\micro\s} & $7.5 \times 10^{-5}$ \\ 
MS gate & \SI{200}{\micro\s} & $1.0 \times10^{-3}$  \\
measurement & \SI{300}{\micro\s} & $1.5 \times 10^{-3}$  
\end{tabular} \label{tab:idle_times}
\end{center}
In the current ion trap architecture, all experiments for FT state preparation are performed with both auxiliary and data qubit measurements deferred to the end of the circuit, as described in Ref.~\cite{postler2022demonstration}, so that no idling faults occur during the measurements. Simulations presented in Sec.~\ref{sec:prot_sims} partly contain in-sequence measurements which are modelled with the respective idling error rate. All in-sequence measurements are modeled to have the same idling error rate, although measurements showing at least one bright ion are usually followed by a recooling sequence with a duration on the order of milliseconds~\cite{ringbauer2022universal}. As idling is not the dominant error source we neglect this dependence of the idling error rate on the outcome of a measurement.

\textbf{Single-qubit operations.}
\begin{figure}
    \centering
    \includegraphics[width=\linewidth]{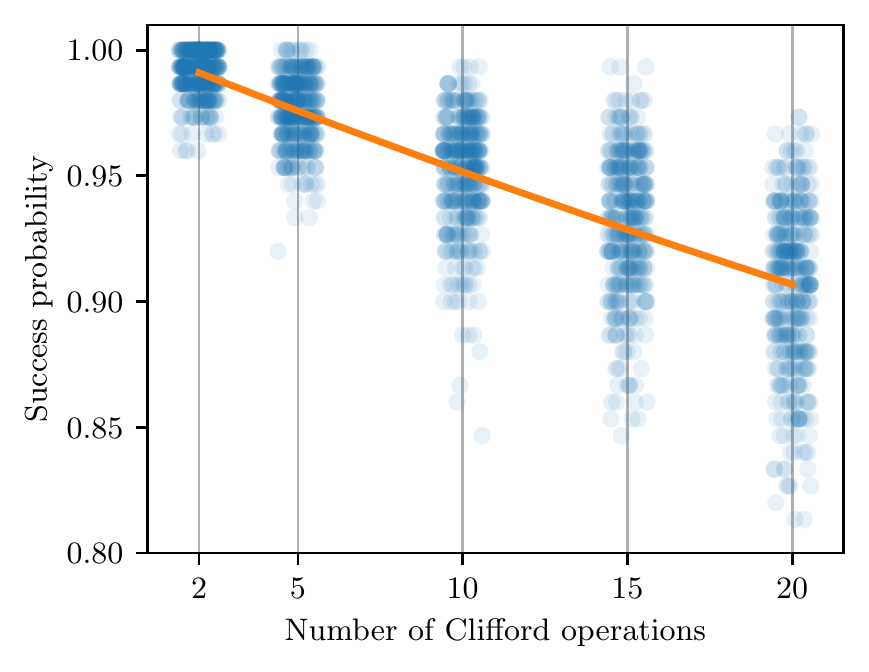}
    \caption{\textbf{Single-qubit gate benchmarking.} Experimental success probabilities of randomized benchmarking sequences containing up to 20 Clifford operations in a 16-qubit register. The scatter on the horizontal axis around the sequence lengths 2, 5, 10, 15 and 20 is introduced for better visibility of the success probability of the individual random sequences. The discretization on the vertical axis is given by averaging over 150 executions per random sequence. For brevity data from 16 qubits is combined to a single dataset. The underlying data for individual qubits can be found in Appendix~\ref{sec:rb}. The decay fitted to the combined data suggests a single-qubit gate fidelity of $0.99760(8)$, where the given error is the 95\% confidence interval.}
    \label{fig:rb}
\end{figure}
As virtual $Z$-operations are noiseless~\cite{mckay2017efficient}, the only erroneous single-qubit operations are resonant operations. We characterize resonant single-qubit operations experimentally via randomized benchmarking~\cite{emerson2005scalable}. For different ions in a 16-qubit register, the fidelity of a single-qubit rotation ranges from $0.9969(4)$ to $0.9980(3)$ with a mean of $0.9976$ and a standard deviation of $2.4 \times10^{-4}$. Combined randomized benchmarking data for sequences of up to 20 Clifford operations per qubit for all 16 qubits are shown in Fig.~\ref{fig:rb}, data for individual qubits can be found in Appendix~\ref{sec:rb}. Faults affecting single-qubit operations acting on a state $\rho$ are modeled as depolarizing noise, hence the modeled noise channel reads

\begin{equation}
    \curlE^{(1)}_\text{dpl}(\rho) = (1-p_1)\rho + \frac{p_1}{3}(X\rho X + Y\rho Y + Z \rho Z).
    \label{eq:depol_single}
\end{equation}

\noindent
With a probability $1-p_1$ the ideal operation is implemented and with a probability $p_1$ a fault operator, randomly drawn from the set of Pauli operations $\{X,\,Y,\,Z\}$, is applied subsequently to the ideal gate. For the theory model we choose $p_1 = 0.005$ for better comparability with Ref.~\cite{postler2022demonstration} although recent improvements on the experimental setup slightly increased the fidelity of single-qubit operations.

\textbf{Entangling operations.} In the system under consideration, entangling operations are based on the center-of-mass motional mode, which offers equal coupling to all qubits in the register. Nevertheless, unwanted coupling to higher order modes with different coupling strengths along the ion string can potentially lead to a varying fidelity for different qubit pairs in the register. To avoid benchmarking on all possible ion pairs, the mean fidelity of a single entangling gate $\mathcal{F}_\mathrm{tq} = 0.975(3)$ is estimated from the quantum state fidelity of the GHZ state $\ket{\psi_{\mathrm{GHZ}}} = (\ket{0}^{\otimes 16} - i \ket{1}^{\otimes 16})/\sqrt{2}$ prepared across the entire register. A more detailed description of this procedure can be found in Appendix~\ref{sec:ghz}. Although this method does not constitute a rigorous characterization of the underlying individual gates, it can still provide insights about the system performance in terms of entanglement generation~\cite{wei2020verifying}.

Microscopic noise models have been derived in previous works, considering amplitude fluctuations or gate miscalibrations in particular~\cite{bermudez2019fault} as well as thermal errors or motional heating~\cite{ballance2017high} and incoherent overrotations~\cite{li2017fault}. However, for simplicity, we apply depolarizing noise to two-qubit gates, as we do for single-qubit operations since our arguments of advantageous FT quantum computation primarily regard the appropriate FT design of quantum circuits. Depolarizing noise is considered the most general and architecture-agnostic incoherent noise channel because the fault operators of the depolarizing noise channels form a basis in the space of single- and two-qubit unitaries respectively. The modeled error channel for depolarizing noise on entangling gates reads

\begin{align}
    \curlE^{(2)}_\text{dpl}(\rho) &= (1-p_2)\rho + \frac{p_2}{15} \sum_{i=1}^{15} E_2^{(i)} \rho E_2^{(i)} \label{eq:depol_two} \\
    E_2 &= \{\sigma_k \otimes \sigma_l, \forall k,l \in \{0,1,2,3\}\}~\backslash~\{I \otimes I\} \notag.
\end{align}

\noindent
With a probability $p_2$ one of fifteen non-trivial weight-2 Pauli faults is added to the ideal entangling gate. We choose $p_2 = 0.025$ as estimated from the GHZ state preparation. Although overrotations have been identified as a dominant source of error in ion trap quantum processors before~\cite{debroy2020logical, zhang2021improving}, we find in Sec.~\ref{sec:prot_sims} that depolarizing noise does not perform worse at estimating logical failure rates than an incoherent overrotation noise model. The latter takes into account the physical nature of optical qubit operations, i.e.~laser driven rotations around a given Pauli axis (see App.~\ref{sec:noise_app}).  We provide further comparison between overrotations and depolarizing noise through quantum state fidelity calculations in Sec.~\ref{sec:fidelity}. 

\begin{figure}[ht!]
    \centering
    \includegraphics[width=\linewidth]{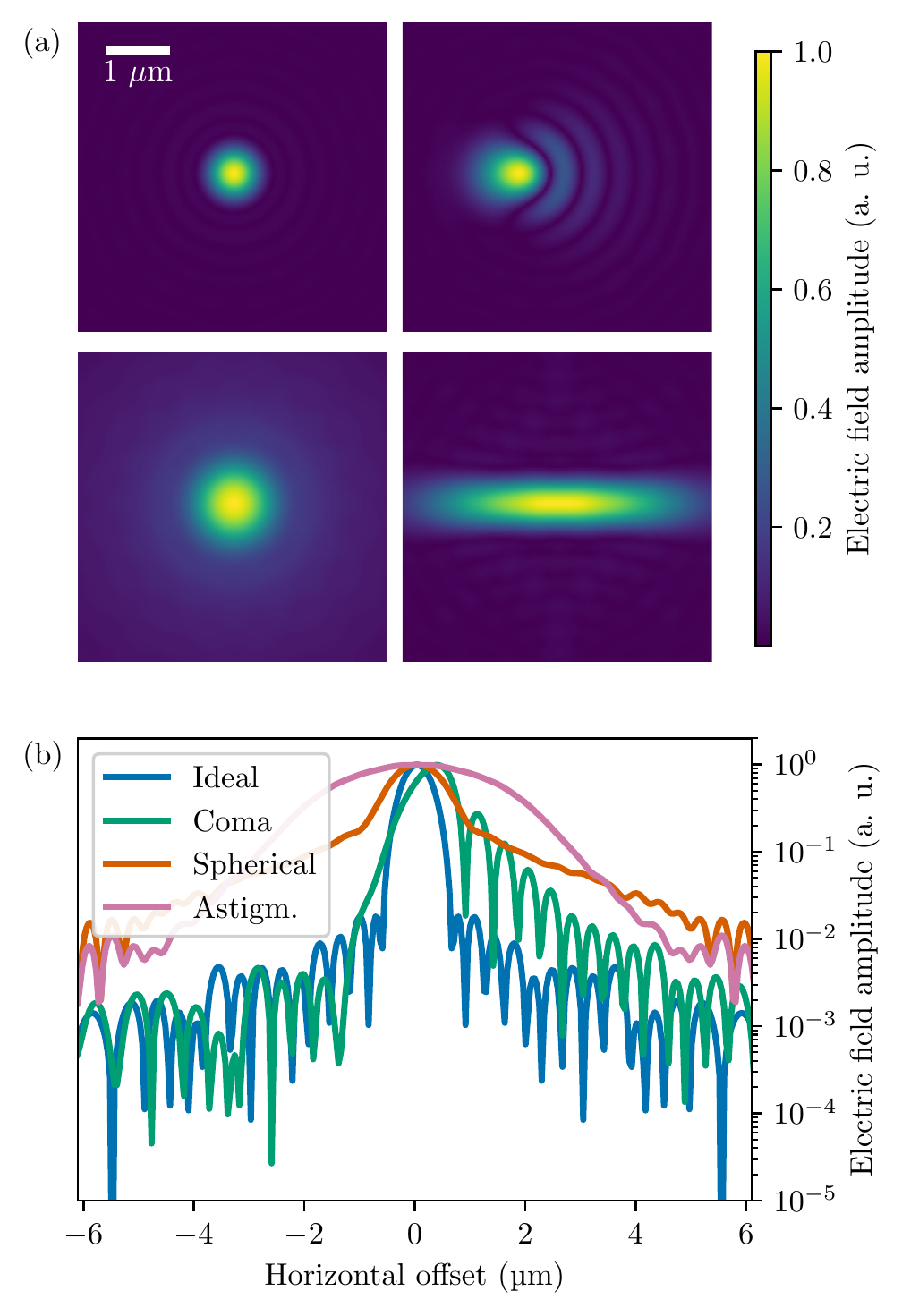}
    \caption{\textbf{Electric field crosstalk.} Fourier optics calculation of aberrations affecting a tightly focused laser beam. We simulate light with a wavelength of $\lambda = \SI{729}{\nano\meter}$ illuminating an objective with an aperture diameter of \SI{40}{\milli\meter} and a focal length of \SI{20}{\milli\meter}. Aberrations are introduced by distorting the input wavefront, where the peak-to-valley phase deviation compared to a plane wave amounts to $2\lambda$. (a) In the upper left image the ideal electric field amplitude $E(\mathbf{x})$ of an aberration-freely focused Gaussian beam is depicted. The other color plots show the effect of different types of aberrations, namely coma (upper right), spherical aberration (lower left) and astigmatism (lower right). The increased diameter of the field distribution leads to increased leakage light at neighboring ions. (b) Cut along the horizontal axis of the field distributions shown in a) through the maximum intensity point for the ideal and the three aberrated spots.
}
    \label{fig:aberrations}
\end{figure}

\begin{figure}
    \centering
    \includegraphics[width=\linewidth]{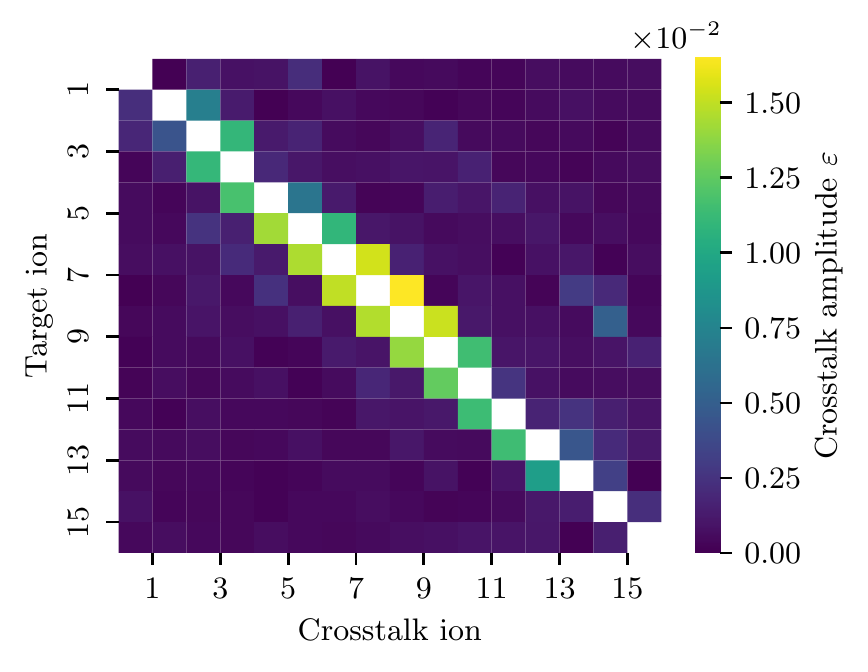}
    \caption{\textbf{Ion-string crosstalk amplitude.} Measured ratio $\varepsilon$ of crosstalk to target Rabi frequency for resonant operations acting on all 16 qubits, with a maximum and mean next-neighbor crosstalk ratio of $1.6 \times 10^{-2}$ and $0.9 \times 10^{-2}$ respectively.}
    \label{fig:crosstalk}
\end{figure}

\begin{figure}
    \centering
    \includegraphics[width=\linewidth]{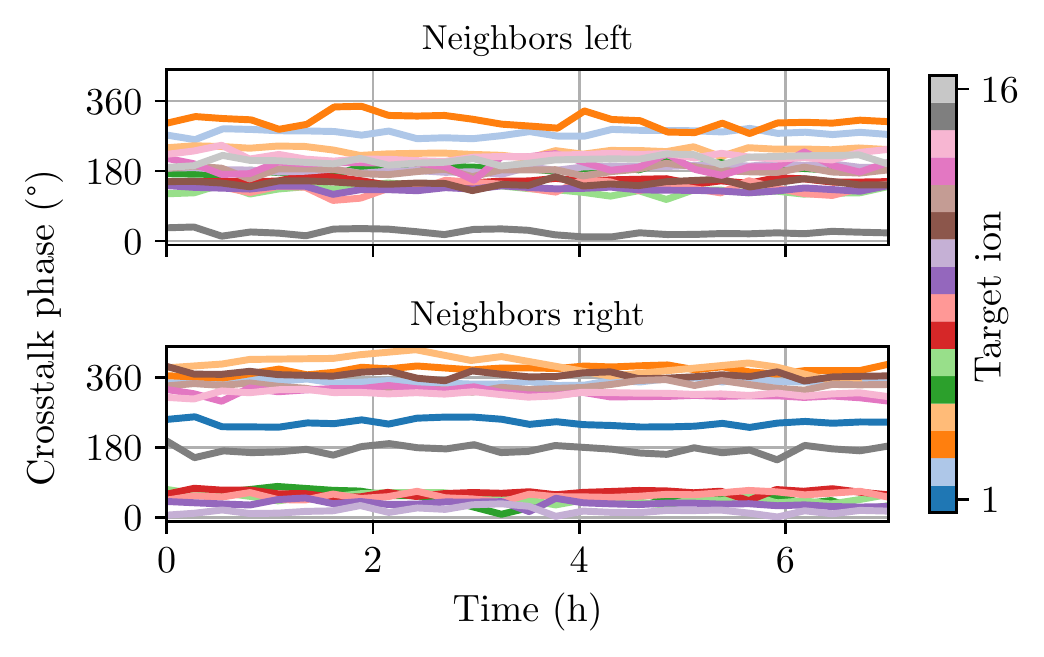}
    \caption{\textbf{Ion-string crosstalk phase.} Measurement of the phase difference between target and crosstalk light field in a 16-qubit ion crystal. The crosstalk phase covers the whole interval $[0, 2\pi]$ for different target-neighbor-pairs, but is stable up to tens of degrees over hours.}
    \label{fig:crosstalk_phase}
\end{figure}

\textbf{Crosstalk.} Another noise process is the unintended manipulation of qubits in spatial proximity to a target qubit, which we refer to in this work as crosstalk. The physical process causing this is leakage light from the tightly focused laser beam, where the main contributions are aberrations caused by imperfect optical systems. In Fig.~\ref{fig:aberrations}a we depict a Fourier optics calculation~\cite{hecht2016optics} of the profiles of the electric field amplitude being proportional to the Rabi frequency of a resonant operation. We show the electric field around the target ion position for an ideally focused Gaussian beam, but also for beams affected by coma, spherical aberration and astigmatism~\cite{wyant1992basic}. The parametrization of the electric field amplitude is $E_\varphi(\mathbf{x}) = E(\mathbf{x})\exp{(i\varphi)}$, where $\mathbf{x}$ is the position in a plane orthogonal to the beam propagation at the ion location and $E(\mathbf{x})$ is a positive, real number. The magnitudes of the aberrations in this example are chosen to give peak-to-valley wavefront distortions of $2\lambda$ and do not necessarily reflect the situation in the experiment. Figure~\ref{fig:aberrations}b shows the calculated electric field amplitude along the ion string, where an offset of zero corresponds to the position of the target ion. In a 16-ion crystal the distances to neighboring ions in the discussed setup are typically around $\SI{4}{\micro\m}$. As a figure of merit for the magnitude of the effect of this crosstalk we use the ratio $\varepsilon = \Omega_n/\Omega$ of the Rabi frequencies of the unintended manipulation at a neighboring ion $\Omega_{n}$ and the target operation $\Omega$. In the experimental setup under consideration the maximum nearest-neighbor crosstalk ratio is $\varepsilon_{\mathrm{max}} = 1.6 \times 10^{-2}$ while the mean over the register is $\varepsilon_{\mathrm{mean}} = 0.9 \times 10^{-2}$ in a 16-qubit register with an axial trap frequency of \SI{400}{\kilo\hertz}. The inter-ion distances range from \SI{3.6}{\micro\meter} in the center to \SI{5.7}{\micro\meter} at the edge of the ion chain. Crosstalk ratios $\varepsilon$ for the 16-qubit register are shown in Fig.~\ref{fig:crosstalk}. We neglect crosstalk to non-nearest neighbors in our model as the measured mean Rabi frequency ratio is more than an order of magnitude lower than between direct neighbors.

It is crucial to note that the phase of the leakage light can significantly differ from the phase of the light at the target ion position. Aberrations that distort the wavefronts at the input of the focusing optics propagate to the ion string in the focal plane and lead to an electric field distribution around the target ion with spatially variable phase. This phase is experimentally accessible via a Ramsey-type experiment, where a superposition state is prepared with leakage light by illuminating a neighboring ion and subsequently its phase is analyzed by applying a resonant single-qubit operation with varied phase to the qubit affected by crosstalk. As can be seen in Fig.~\ref{fig:crosstalk_phase}, the measured phase difference between target and neighboring ion varies across the whole interval of all possible values $[0, 2\pi]$ for different target ions. The wavefront distortions likely stem from non-ideal alignment of the optical setup and surface imperfections in the beam path, and therefore the phase difference of neighboring ions is stable on the timescale of hours.

Based on the aforementioned experimental observations we model crosstalk noise as follows in simulations: 
When a resonant single-qubit gate is applied to a target ion with a rotation angle $\theta = \Omega t$, where $\Omega$ is the Rabi frequency and $t$ is the gate duration, nearest neighbor ions see a resonant operation with a rotation angle \mbox{$\theta_n = \varepsilon \theta$}. After Pauli twirling (see App.~\ref{sec:noise_app}), this leads to the incoherent error process
\begin{align}
	\curlE(\rho) &= \cos^2 \frac{\varepsilon \theta}{2} \rho + \sin^2 \frac{\varepsilon \theta}{2} \left( \cos^2 \varphi X \rho X + \sin^2 \varphi Y \rho Y \right)
\end{align}
for the neighboring ions, where $\varphi$ is the light phase at the respective neighbor ion position. As the phase relation between the light at the target ion and neighbor ion position varies along the ion chain (see Fig.~\ref{fig:crosstalk_phase}), we average over all possible crosstalk phases to obtain the incoherent noise channel
\begin{align}
	\curlE_{\text{c}_1}(\rho) &= (1-p_{\text{c}_1}) \rho + \frac{p_{\text{c}_1}}{2} \left( X \rho X + Y \rho Y \right)
	\label{eq:ct1av}
\end{align}
for each single-qubit crosstalk location. Here $p_{\text{c}_1} = \sin^2 \frac{\varepsilon \theta}{2}$ with $\varepsilon = 1 \times 10^{-2}$ is the probability that crosstalk induces an error on a neighboring qubit. Applying the same reasoning to model crosstalk errors for two-qubit gates gives the channel
\begin{align}
    \curlE_{\text{c}_2}(\rho) &= (1-p_{\text{c}_2}) \rho + \frac{p_{\text{c}_2}}{4} \left( X_tX_n\rho X_tX_n + X_tY_n\rho X_tY_n \right. \notag  \\
    &~~~~~~~~~~~~~~~~~~~~+ \left. Y_tX_n\rho Y_tX_n + Y_tY_n\rho Y_tY_n\right) \label{eq:ct2av}
\end{align}
for any pair of target and neighbor ions denoted by subscripts $t$ and $n$ respectively with $p_{\text{c}_2} = \sin^2 \frac{\varepsilon \pi}{4}$. An illustration of all target-neighbor locations can be found in Fig.~\ref{fig:circ_ctfaults}.

\textbf{State preparation and measurement.}
Measurements in the $Z$-basis are performed by illuminating the ion chain with light resonant to the 4$S_{\nicefrac{1}{2}}$ to 4$P_{\nicefrac{1}{2}}$ transition, leading to fluorescence light emitted by ions projected to $\ket{0}$ and no emitted photons from ions projected to $\ket{1}$~\cite{schindler2013quantum}. Measurement errors are caused by the overlap between bright and dark count distributions originating from the intrinsic overlap of the Poissonian distributions of dark and bright state fluorescence counts and by the probability that an ion decays from the meta-stable excited state during the detection time~\cite{roos2000controlling}. State initialization of the qubit to 4$S_{\nicefrac{1}{2}, m_j = -\nicefrac{1}{2}}$ is achieved by frequency-resolved optical pumping on the quadrupole transition. The ions are illuminated with light resonant to the transition from 4$S_{\nicefrac{1}{2}, m_j = +\nicefrac{1}{2}}$ to 3$D_{\nicefrac{1}{2}, m_j = -\nicefrac{3}{2}}$, while a repumping laser is broadening the transition~\cite{roos2006designer}. Typical probabilities for initialization and measurement faults in the setup considered are around $3 \times 10^{-3}$~\cite{schindler2013quantum}. Both initialization and measurement errors are again modeled as depolarizing noise. Therefore, the model is the same as in Eq.~(\ref{eq:depol_single}) with error probabilities $p_i = p_m = 4.5 \times 10^{-3}$, corresponding to a flip error probability of $3 \times 10^{-3}$ for initialization and measurement, respectively.

All of the above-mentioned noise models are discussed in more detail in Appendix~\ref{sec:noise_app} alongside coherent overrotations and coherent crosstalk on MS gates. Since it is known that QEC decoheres noise through encoding and stabilizer measurement, although coherent by nature~\cite{beale2018quantum, iverson2020coherence}, we mainly focus on incoherent noise in this manuscript. 

\textbf{Numerical methods.} In Sec.~\ref{sec:prot_sims}, we estimate logical failure rates of logical state preparation protocols by performing numerical simulations of both stochastic incoherent Pauli noise models and coherent noise as decribed above. We provide results of numerical simulations for logical failure rates under both \emph{depolarizing noise} on single-qubit gates, two-qubit gates, physical qubit initialization and measurement as well as an \emph{extended noise} model. It includes dephasing noise on idling qubits and crosstalk on both single- and two-qubit gates on top of said depolarizing noise. We use stabilizer simulations~\cite{aaronson2004improved} for Pauli state preparation with incoherent noise and statevector simulations otherwise, i.e.~either for magic state preparation or when applying coherent noise to either type of state preparation. If applicable, stabilizer simulations are advantageous since they allow for simulation of Clifford circuits in polynomial time according to the Gottesman-Knill-theorem \cite{gottesman1998heisenberg}. The exponentially large $n$-qubit Hilbert space of dimension $2^n$ poses a numerical challenge for statevector simulations which run slowly and consume an exponential amount of memory with growing number of qubits $n$. All simulations in this work are performed using a modified version of the python package ``PECOS'' \cite{pecos, ryan2018quantum}. The effect of incoherent noise is treated by means of direct Monte Carlo sampling (MC) and subset sampling (SS) which is an importance sampling technique. Both methods have a preferential range of applicability: MC is used for larger physical error rates, SS achieves accurate estimates with well-defined confidence intervals for lower physical error rates and is especially useful for extracting scaling behavior (see App.~\ref{sec:sim_methods} for details on both methods). \section{Protocols for FT advantage over physical qubits}\label{sec:prot_sims}
The paradigm of FT circuit design holds the promise to maintain coherence within a quantum computation where many physical qubits are involved and suffer the influence of noise \cite{preskill1998reliable}: Faults on individual components of a quantum circuit must not cause errors, which cannot be corrected by the QEC code, on the qubits holding the logical information. There exist errors $E$ \emph{at the end} of the circuit, resulting from faults which happen at locations \emph{within} the circuit, that have weight $\text{wt}(E)$ larger than $t = \lfloor\frac{d-1}{2}\rfloor$. They are thus uncorrectable and will lead to failure of the QEC procedure. Here $d = 2t+1$ is the distance of the QEC code and the weight is the number of qubits on which the error $E$ acts. There will always exist configurations of $t+1$ faults that cause logical failure, i.e.~lead to application of an unintended logical operator when performing QEC because $\text{wt}(E) > t$. 

Up to $t$ faults can be in principle prevented from propagating to cause more than the correctable amount of $t$ errors by advantageous circuit design. By unfortunate circuit design though, large distance logical states could also be corrupted by propagation of lower order faults. In this case, one could encode into lower distance logical states directly instead of using such circuits. We denote fault tolerance towards up to $t$ faults as ``level-$t$ FT'' or ``FT$t$''.  Assume that faults at any circuit location happen independently with probability $p$. Then the logical failure rate $p_L$ of FT implementations of a distance $d$ QEC code scales as $p_L \propto p^{t+1}$ in the limit of low physical error rate $p \rightarrow 0$. For level-$t$ FT all fault configurations up to order $p^t$ must only cause correctable errors\footnote{While there might exist particular higher order fault configurations where faults annihilate each other and do not cause uncorrectable errors, it cannot be guaranteed that \emph{all} such faults only cause correctable errors.}. Note that the weight of the \emph{error} determines whether or not it is correctable and the probability of occurrence for the microscopic \emph{fault} configuration that propagates to an uncorrectable error determines its order in the polynomial for $p_L$.

In this work, we discuss FT schemes of level $t=1$ which thus display a quadratic dependence of the logical failure rate $p_L \propto p^{2}$ as $p \rightarrow 0$. This scaling of FT implementations is contrasting non-FT circuits or operation of physical qubits where single faults can cause uncorrectable errors, thus leading to a linear scaling of the logical failure rate $p_L \propto p$ at low physical error \mbox{rates $p$}. Although FT circuits may involve more (noisy) qubits and gates than their non-FT counterparts, fault tolerance ensures that there exists a regime of physical error rates where the polynomial dependence leads to lower logical failure rates than non-FT and physical qubit implementation \cite{Aliferis2006, Aharonov2008}.

\begin{figure}\centering
    \includegraphics[width=\linewidth]{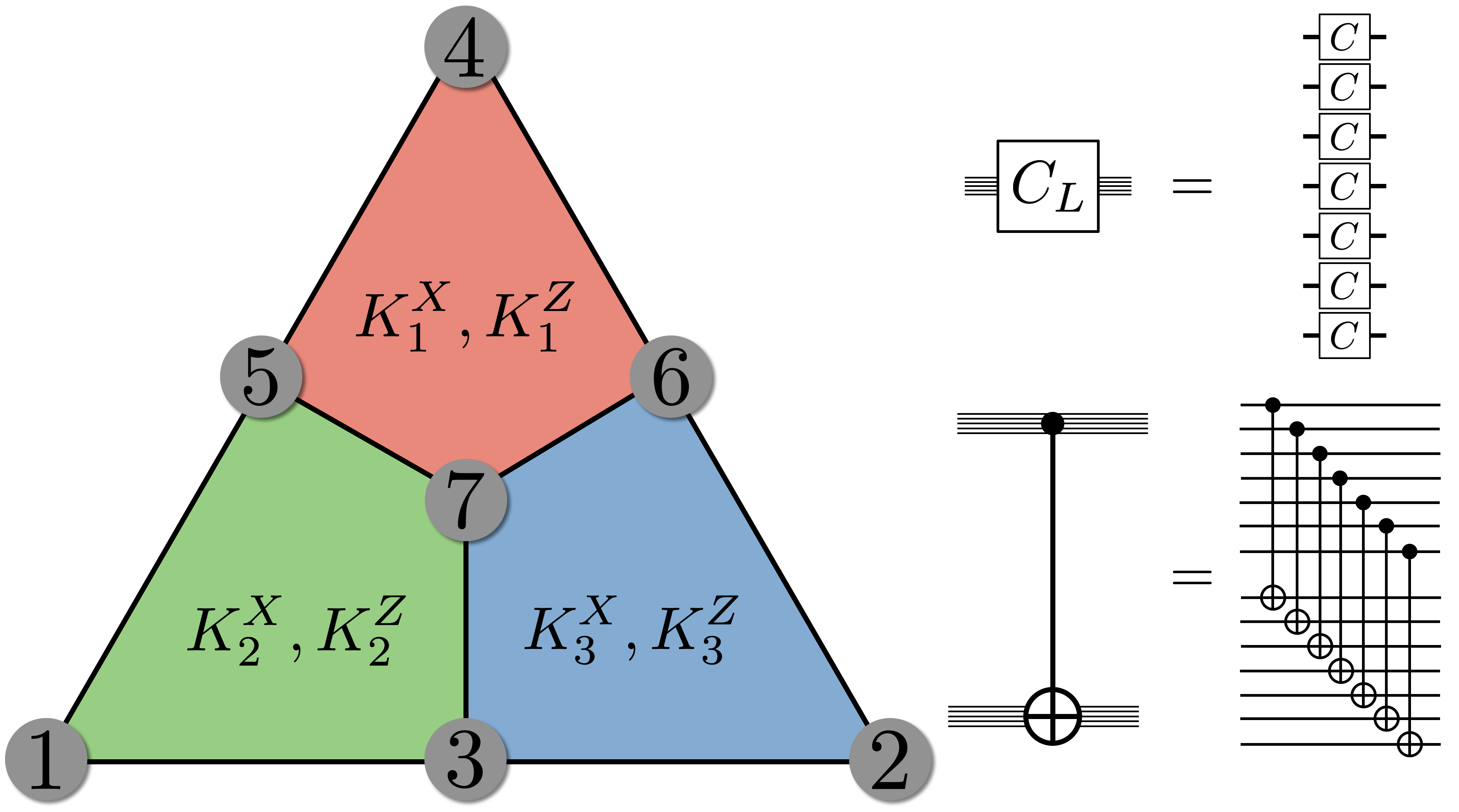}
    \caption{\textbf{Steane code.} \textit{Left:} The Steane code is the smallest representative of the family of topological color codes. As a [[7,1,3]] code, it uses seven physical qubits to encode a single logical qubit with distance $d=3$ allowing for correction of $t = \lfloor\frac{d-1}{2}\rfloor$ = 1 arbitrary Pauli error. Physical qubits sit on the vertices of the graph. Stabilizer generators $K_i$ are plaquettes spanning four physical qubits with mutual overlap of two qubits and are given by Eq.~(\ref{eq:stabs}). The X- and Z-type stabilizers are symmetric. \textit{Right}: All gates of the Clifford group can be implemented transversally and thus fault-tolerantly (FT) in the Steane code and larger distance 2D color codes.}
    \label{fig:steane}
\end{figure}

The Steane code \cite{Steane1996} shown in Fig.~\ref{fig:steane} is the smallest representative of the family of topological color codes \cite{bombin2006topological, bombin2006distillation}. As a [[7,1,3]] code, it encodes $n = 7$ physical qubits into a single logical qubit with distance $d=3$ allowing for correction of $t = 1$ arbitrary Pauli errors while $t + 1 = 2$ or more errors lead to logical failure \cite{nielsen2010quantum}. It has low resource overhead needed for FT universal qubit operations: Not only are all Clifford gates transversal and thus inherently FT in the Steane code. Also, the non-Clifford $T$-gate can be added to the logical gate set, for instance, by magic state injection \cite{Bravyi2005}. Pauli rotations with angle $\pi/4$, i.e.~the $T$-gate, can be performed fault-tolerantly in this way as long as an appropriate magic state is available as a resource. The injection circuit then only requires Clifford operations, which are suitable for the Steane code as they respect the FT requirements stated above.
Different strategies for logical qubit initialization in the logical zero state and a logical magic state are addressed in this section.

The logical qubit is encoded in the $[[7,1,3]]$ Steane code defined by the six stabilizer generators
\begin{align}
    K_1^X &= X_4X_5X_6X_7 & K_1^Z &= Z_4Z_5Z_6Z_7 \notag \\
    K_2^X &= X_1X_3X_5X_7 & K_2^Z &= Z_1Z_3Z_5Z_7 \label{eq:stabs} \\
    K_3^X &= X_2X_3X_6X_7 & K_3^Z &= Z_2Z_3Z_6Z_7 \notag
\end{align}
which are symmetric under exchange of $X$ and $Z$. Any code state $\ket{\psi}_L$ is a +1 eigenstate of all stabilizers and thus stays invariant under application of any stabilizer. As a consequence, Pauli operators acting on code states can be multiplied by stabilizers without changing their effect on the code state. Two Pauli operators that only differ by multiplication with stabilizers are thus called stabilizer equivalent. Since the stabilizer generators exclusively consist of $X$- or $Z$-operators each, the Steane code belongs to the class of CSS (Calderbank-Shor-Steane) codes \cite{calderbank1996good, steane1996error}. The transversality of the Hadamard and the CNOT gates follows directly from these two properties, respectively. The logical operators can be chosen as $X_L = X^{\otimes 7}$ and $Z_L = Z^{\otimes 7}$. By multiplication with stabilizers they can be expressed as weight-3 operators reflecting the fact that the Steane code can correct a single Pauli error. Single Pauli errors $X_i$ and $Z_j$ on any two single qubits $i \neq j$ can be corrected independently, or -- as a consequence -- a single Y-type error since $Y_i \simeq X_iZ_i$ (for $i=j$). Each possible syndrome measurement outcome is mapped to a unique recovery operation, which guarantees the correction of all single Pauli errors, with a look up table as shown in Tab.~\ref{tab:steane_lut}.

\begin{table}[!ht]
\centering
    \begin{tabular}{c|c}
        $K_1^Z,K_2^Z,K_3^Z$ & recovery $R$ \\ \hline
        $+++$ & $I$ \\
        $++-$ & $X_2$ \\
        $+-+$ & $X_1$ \\
        $+--$ & $X_3$ \\
        $-++$ & $X_4$ \\
        $-+-$ & $X_6$ \\
        $--+$ & $X_5$ \\
        $---$ & $X_7$
    \end{tabular}
    \caption{Look up table for the seven qubit Steane code as shown in Fig.~\ref{fig:steane}. $+$ and $-$ indicate a positive and negative expectation value of the respective stabilizer operator. All six of them form the error syndrome $(K_1^X,K_2^X,K_3^X,K_1^Z,K_2^Z,K_3^Z)$. Only the Z-type syndromes and corresponding X-type recoveries are shown. Since the Steane code is symmetric under exchange of $X$ and $Z$, the Z-type recoveries from $X$-syndrome measurements can be applied analogously. The two three-bit syndromes $(K_1^X,K_2^X,K_3^X)$ and $(K_1^Z,K_2^Z,K_3^Z)$ are sufficient to correct all single Pauli errors.}
    \label{tab:steane_lut}
\end{table}

However, this mapping becomes non-unique if \mbox{weight-2} errors can also occur. If two different errors map to the same syndrome then the recovery operation may cause erroneous application of a logical operator as a result of the error correction (EC) attempt. As an example of such a logical failure, consider the error $E = X_3X_5$. The $Z$-syndrome will be measured as $-+-$ and by the look up table we would apply $R = X_6$ as a recovery operation. The total operator $RE = X_3X_5X_6$ is a logical operator since it is stabilizer equivalent to $X_L$ given above.\footnote{The chained error and correction operators $RE$ correspond to $X$ applied to all physical qubits multiplied by all three X-type generators $RE = K_1^XK_2^XK_3^X X_L$.} 

Transversal implementation such as for Clifford gates shown in Fig.~\ref{fig:steane} directly ensures that single faults will at most cause a weight-1 error in each encoded logical qubit because transversal gates never couple two qubits from the same block. The weight-1 errors in each block can then independently be corrected in QEC. 

In this work, we use unitary encoding circuits for the initialization of logical qubits. This is in contrast to initialization procedures which rely on in-sequence stabilizer measurements and feed-forward of syndrome information. Unitary encoding circuits typically prepare logical states with fewer entangling gates at the cost of needing large connectivity between the data qubits which is provided natively in our trapped-ion architecture. These circuits allow for deterministic preparation of the code state since they avoid data qubit measurements altogether. Nonetheless, due to the large degree of inter-qubit connectivity, faults that happen on entangling gates might propagate throughout the circuit and cause uncorrectable errors as a result. This is illustrated e.g.~for the encoding circuit in Fig.~\ref{fig:circ_pauli}. For FT state preparation in the Steane code the goal is to avoid such single fault events being able to cause weight-2 errors to occur on the final data qubit state. We can achieve FT by making use of recently introduced flag circuits \cite{Chao2018, Chamberland2018}. Here, additional auxiliary qubits called flag qubits are coupled to the data qubit block. Their measurement outcomes herald the potential presence of uncorrectable errors on the data qubit state. We refer to a flag measurement of $-1$ as a ``triggered flag'' and call the $+1$ measurement outcome a ``clear flag''. 

In the remainder of this section, we analyze both deterministic and non-deterministic protocols for FT state preparation. Deterministic protocols, although they may contain measurement operations, always terminate with the data qubits prepared in the desired logical state in each individual protocol execution. They are designed to tolerate all possible faults of order $\mathcal{O}(p^1)$ (FT1). With non-deterministic protocols instead, a fraction of preparation runs is discarded when measurements of one or more flag qubits indicate that an erroneous state has been prepared. This cannot be foreseen a priori due to the stochastic nature of noise. Depending on whether the chosen protocol is deterministic or not, a flagged state is either corrected using an appropriate recovery operation, or is discarded. This recovery operation is chosen conditioned on triggered flags and is different from the look up table \ref{tab:steane_lut} used when flags are clear. Non-deterministic protocols typically exhibit lower logical failure rates as they contain fewer gates at the cost of repeatedly executing the circuit in case of triggered flags. On the other hand, deterministic protocols perform worse due to their larger qubit overhead or circuit depth. In the remainder of this section we theoretically investigate both types of protocols for Pauli and magic state preparation. For the non-deterministic preparation circuits used in the experiment \cite{postler2022demonstration}, we provide a scaling analysis of their logical failure rates dependent on physical error parameters in order to estimate how much improvement of physical operations is needed to experimentally achieve lower infidelities than physical qubits.

\subsection{Non-deterministic state preparation}\label{sec:nondet}

In the following, we introduce and discuss circuits for non-deterministic FT state preparation for both Pauli and magic states. For both types of states we provide an evaluation of logical failure rate scaling with physical error parameters obtained via numerical simulations of the two different noise models described in Sec.~\ref{sec:noise_main}. We assess the performance of FT protocols compared to physical qubit operations to estimate break-even points of FT advantage, i.e.~identify for which physical error parameters the infidelity of logical states is lower than their respective physical qubit counterparts.

\begin{figure}\centering
    \includegraphics[width=\linewidth]{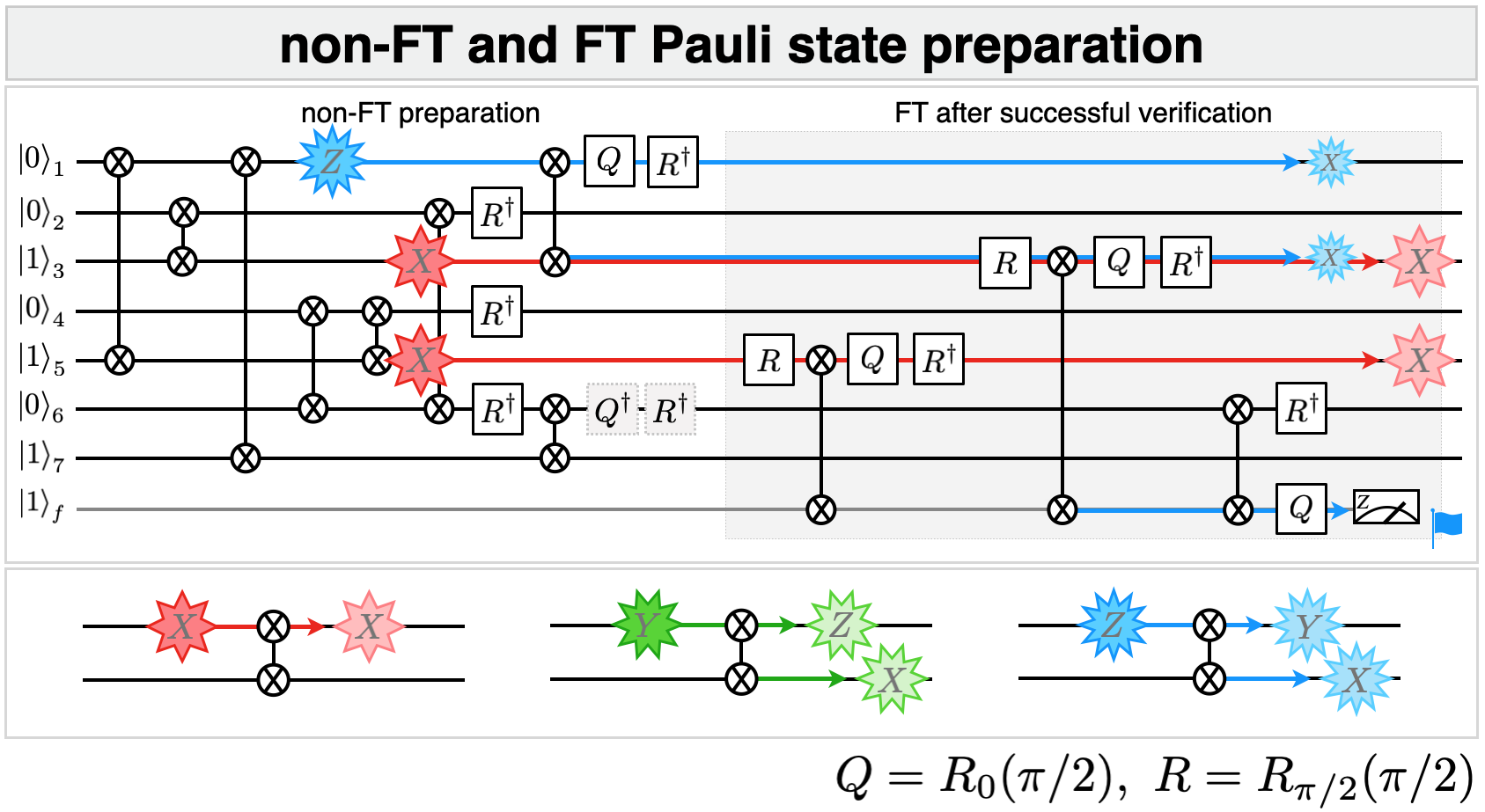}
    \caption{\textbf{Pauli state preparation circuits.} The logical zero state of the Steane code can be initialized using MS gates and single-qubit rotations about the $X$- and $Y$-axes. After the first block the $\ket{0}_L$ state is prepared non-fault-tolerantly (non-FT) on the data qubits 1 to 7. An example of a single $Z_1$ fault which can cause an uncorrectable error is shown as 12-cornered star (blue). The second block, shaded gray, couples to an additional flag qubit which heralds successful FT state preparation. The $Z_1$ will propagate and trigger the flag. When the flag qubit is clear, it is guaranteed that $\ket{0}_L$ is prepared up to a weight-1 error. Crosstalk faults, such as $X_3X_5$ (red 8-cornered stars), can devastate the FT property (cf.~Sec.~\ref{sec:ctr}). Initialization of physical qubits as $\ket{1}$ is done by first initializing them as $\ket{0}$ and then performing an $X$-rotation of angle $\pi$. The last two gates $Q^\dagger$ and $R^\dagger$ (shaded grey) of the first block are only needed for non-FT but not for FT state preparation. General propagation rules for Pauli faults through MS gates are shown in the lower panel.}
    \label{fig:circ_pauli}
\end{figure}

\textbf{Logical Pauli states.} The circuit shown in Fig.~\ref{fig:circ_pauli} is used to prepare the $\ket{0}_L$ state which is the $+1$ eigenstate of the logical $Z$-operator $Z_L$ and also -- as any code state -- the $+1$ eigenstate to all stabilizers including the generating plaquette operators in Fig.~\ref{fig:steane} \cite{goto2016minimizing, bermudez2019fault}. The entangling MS gates prepare the plaquette eigenstates in an interleaved way which minimizes the number of gates. MS gates 1, 3 and 7 prepare $K_2$, MS gates 2, 6 and 8 are needed for preparation of $K_3$ and MS gates 4, 5 and 8 are involved in preparing $K_1$ (counting left to right and top to bottom) \cite{amaro2020scalable}. After executing the first block of the circuit, the state is prepared non-fault-tolerantly, meaning that single faults can still corrupt the $\ket{0}_L$ state, e.g.~the fault $Z_1$ after the third MS gate would propagate to the uncorrectable error $X_1X_3$ at the end of the first block. FT is achieved by running the second block which acts as verification. Here, the flag qubit couples to the data qubits, effectively measuring a weight-3 logical $Z$-operator. This logical $Z$-operator must be chosen such that any weight-2 error resulting from a single fault will trigger the flag. If the flag is triggered the state is discarded and another trial must be run until the flag is clear. The flag qubit measurement heralds uncorrectable errors such as the one caused by the aforementioned $Z_1$ fault. The error will propagate through the second MS gate of the verification block to $X_1X_3X_f$ so that the flag will be triggered.

Crosstalk is known to be a major source of failure in ion trap quantum computers as described in Sec.~\ref{sec:experimental}. The effect of crosstalk in general does not respect the FT circuit design principle \cite{schindler2013quantum, sarovar2020detecting, parrado2021crosstalk}. As an example, consider the FT Pauli preparation circuit in Fig.~\ref{fig:circ_pauli}. Here, a $X_3X_5$ crosstalk fault can occur after the fifth MS gate under the noise channel in Eq.~\eqref{eq:ct2av}. It will propagate through the circuit and cause an uncorrectable weight-2 $X$-error on the data qubits without triggering the flag. This illustrates that even though logical failure rates of FT circuits are expected to scale quadratically, there exists a linear term in the expansion of $p_L$ caused by dangerous crosstalk fault locations which will eventually destroy the advantageous scaling behavior (for more details on the microscopic crosstalk noise model and its fault operators see App.~\ref{sec:noise_app}).

After successfully preparing the logical zero state $\ket{0}_L$, any of the remaining five cardinal states on the Bloch sphere $\ket{1}_L,\,\ket{+}_L,\,\ket{-}_L,\,\ket{+\ii}_L$ and $\ket{-\ii}_L$ can be reached by subsequently applying the appropriate logical single-qubit rotation to $\ket{0}_L$. As all Clifford gates can be realized transversally and are thus FT in the Steane code, so is the full preparation procedure for any of the six Pauli states.

\textbf{Logical magic state.} 
It is known that Clifford gates are not sufficient to implement single-qubit rotations of an arbitrary angle on the Bloch sphere. Therefore, the Clifford gates alone cannot be used for universal quantum computation. In order to reach universality, the Solovay-Kitaev-theorem states that any point on the logical Bloch sphere can be reached with in principle arbitrary precision when a $\pi/4$-rotation about an arbitrarily-chosen axis is available \cite{solovay1995lie, kitaev1997quantum}. We choose to implement a logical $T$-gate as 
\begin{align}
    T_L = \exp \left( -\ii \frac{\pi}{8}Y_L \right)
\end{align}
a rotation about the $Y$-axis because the corresponding magic state
\begin{align}
    \ket{H}_L &= \cos\left(\frac{\pi}{8}\right)\ket{0}_L + \sin\left(\frac{\pi}{8}\right)\ket{1}_L.
\end{align}
is the $+1$ eigenstate to the logical Hadamard operator $H_L$. Thus, the logical magic state can be prepared by FT measurement of $H_L$ which will project the data qubit state onto $\ket{H}_L$ if the measurement outcome is $+1$. When $\ket{H}_L$ is available, $T_L$ can be implemented by magic state injection, which only requires Clifford gates \cite{chamberland2019fault}. Because all Clifford gates are transversal in the Steane code, preparing $\ket{H}_L$ fault-tolerantly with high fidelity is the crucial step for implementing the FT universal gate set. 

\begin{figure}\centering
    \includegraphics[width=\linewidth]{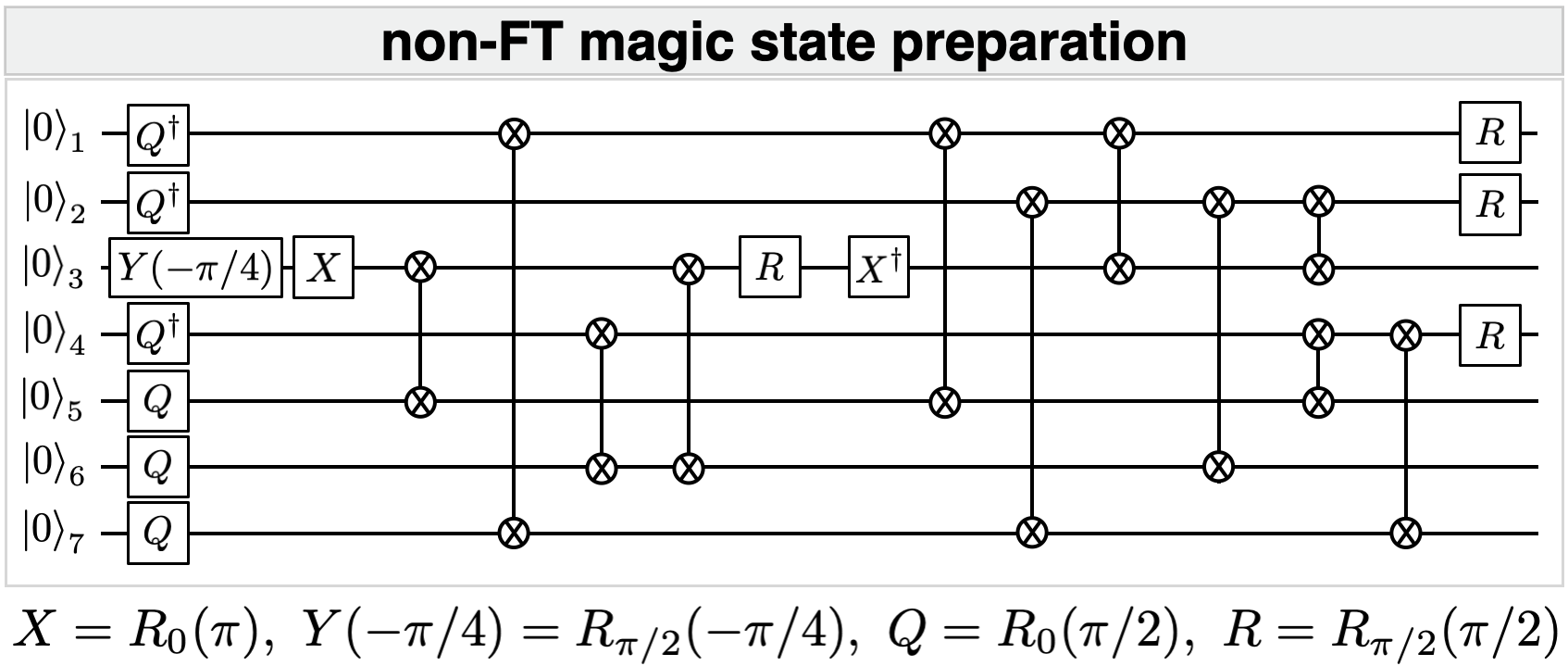}
    \caption{\textbf{Non-FT magic state preparation circuit.} The physical magic state is prepared on qubit 3 and then grown into the encoded $\ket{H}_L$ state of the Steane code. Hermitian-conjugate rotation operators amount to rotations in the respective opposite direction. For coherent rotation noise simulation, the direction of rotation affects the overall logical failure rate.}
    \label{fig:circ_magic_nonft}
\end{figure}

\begin{figure*}\centering
    \begin{minipage}{\linewidth}
    \includegraphics[width=0.7\linewidth]{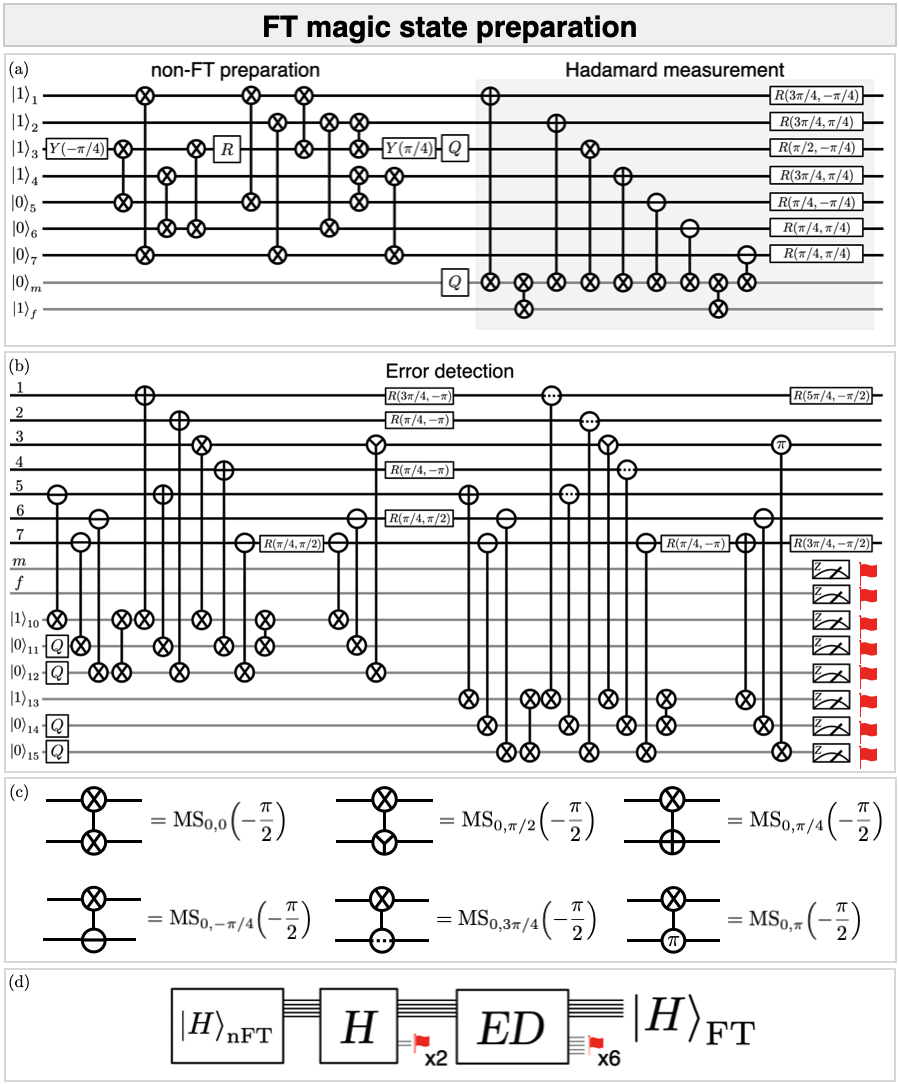}
    \caption{\textbf{FT magic state preparation circuit.} The logical magic state is prepared fault-tolerantly after executing all three circuit blocks. Single-qubit $Z$-rotations are absorbed into phases of MS gates and single-qubit rotations\protect\footnote{For an accepted state the single-qubit $Z$-rotations $R_Z^{(1)}(-3\pi/4)R_Z^{(2)}(-3\pi/4)R_Z^{(3)}(-\pi)R_Z^{(4)}(-3\pi/4)R_Z^{(5)}(-3\pi/4)$ $R_Z^{(6)}(\pi/4)R_Z^{(7)}(-\pi/4)$ need to be applied (in software) to the data qubits 1 to 7.}. \mbox{(a) Non-FT} magic state preparation is followed by a flag-FT measurement of the logical Hadamard operator. The flag qubits herald dangerous faults which may happen during preparation or measurement. Note that the single-qubit rotations in the non-FT preparation block differ from Fig.~\ref{fig:circ_magic_nonft} since they were optimized in conjunction with the subsequent Hadamard measurement block. \mbox{(b) Flag-FT} parallel syndrome readout circuit. Auxiliary qubits act as flags. If any flag is measured as $-1$ the state is discarded. \mbox{(c) Phase-shifted} MS gates with six different phases on their respective data qubit are used in the circuit (cf.~Fig.~\ref{fig:architecture}). \mbox{(d) Sequence} of logical building blocks of the FT magic state preparation protocol acting on data qubits and flag qubits.}
    \label{fig:circ_magic_ft}
    \end{minipage}
\end{figure*}

The principle of repeat until success is also employed for magic state preparation in the non-deterministic protocol given by Ref.~\cite{chamberland2019fault}. The circuit in Fig.~\ref{fig:circ_magic_nonft} prepares the magic state $\ket{H}_L$ non-fault-tolerantly, analogous to the first step of FT Pauli state preparation. Verification of the prepared state consists of two steps. First, the logical Hadamard operator is measured, which projects the data qubit state to the $H_L$-axis. The flag circuit shown as part of the sequence in Fig.~\ref{fig:circ_magic_ft} is used to measure $H_L$ fault-tolerantly. Any dangerous fault which could occur on the measurement qubit in this block will trigger the flag. Transversality of $H_L$ ensures that faults on single data qubits will not spread to higher-weight errors. The measurement qubit itself is also interpreted as a flag in this protocol so that a run that prepares the $-1$ eigenstate of $H_L$ is discarded as well. Second, one round of FT parallel stabilizer readout, given in a CNOT version by Ref.~\cite{reichardt2020fault}, flags all other potentially dangerous faults. In this step, we measure $X$- and $Z$-stabilizers in an interleaved way, which is more resource-efficient because of its reduced number of 28 entangling gates compared to sequential stabilizer measurements (at least 48 entangling gates).
Firstly two $Z$- and one $X$-stabilizer, $K_2^X,\,K_1^Z$ and $K_3^Z$, are measured; then, in the second half, the remaining stabilizers $K_2^Z,\,K_1^X$ and $K_3^X$ are measured via one auxiliary qubit each. The auxiliary qubits are coupled to each other by four additional entangling gates. The interleaved arrangement of entangling gates used for each of the individual stabilizer measurements permits that the auxiliary qubits act simultaneously as both readout and flag qubits. This means that the circuit can be used for error detection: If an error is already present before running the circuit, the auxiliary qubits will indicate a non-trivial syndrome. If a dangerous fault happens during the circuit and it acts on an otherwise ideal input state, the auxiliary qubits act as flags and will be triggered. Thus the circuit can be used to verify that the logical qubit is in the $+1$ eigenstate of all stabilizers, without introducing additional faults if all flags are clear.
All three blocks as shown in Fig.~\ref{fig:circ_magic_ft} need to be run and the state is accepted only if none of the eight flag qubits is triggered. The compiled version of this protocol into MS gate circuits contains single-qubit $Z$-rotations and thus the phases $\varphi_1, \varphi_2$ of the MS gates $\ms_{\varphi_1, \varphi_2}(-\pi/2)$ and the phase $\varphi$ of single-qubit rotations $R(\varphi, \theta)$ in Fig.~\ref{fig:circ_magic_ft} are adjusted as described in Sec.~\ref{sec:expops} (also see Eqs.~\eqref{eq:msz} and \eqref{eq:rz}).

\newpage \textbf{Scaling results}. As illustrated above, the regime of advantageous FT implementation is to be found at low physical error rates due to its quadratic scaling behavior with physical error rate as compared to linear scaling of physical qubits or logical error rates of non-FT protocols. In order to demonstrate the capabilities of FT state preparation protocols to outperform non-FT implementations, we show the scaling of logical failure rates dependent on the set of physical error rates described above. We provide an easily accessible overall idea of scaling behavior, such that we can estimate the necessary improvements of trapped-ion operation fidelities, by introducing a single parameter $\lambda$ to uniformly scale all physical error parameters as
\begin{align}
\lambda \cdot (p_1, p_2, p_i, p_m, \dots). \label{eq:lambda}
\end{align} 

Claiming FT advantage over physical qubits must be specifically justified for a given hardware implementation because in different experimental setups one encounters different physical phenomena, which realize the physical gate operations or even the physical qubit to begin with. One criterion to judge upon FT advantage, suggested in Ref.~\cite{gottesman2016quantum}, is that the logical operation realized within a given hardware architecture should be compared to the corresponding physical operation as it could be realized in exactly that same hardware architecture. For the initialization of the logical qubit, we compare logical zero state preparation to the physical qubit initialization error rate and logical magic state preparation to first initializing the physical qubit to $\ket{0}$ followed by a physical $Y$-rotation by an angle $\pi/4$, which is the most straightforward way to prepare the physical magic state $\ket{H} = T\ket{0}$. Here we opt to provide the said comparison with the same physical error parameters achieved in our ion trap setup for both the logical and corresponding physical operation.\footnote{Another possibility is to compare to the best possible hardware implementation of the corresponding physical operation. In our ion trap even lower physical error rates could be achieved with smaller ion registers.} Additionally to the, more rigorous, comparison of logical to physical operations, another break-even criterion is derived from comparing logical failure rates with the MS gate error rate $p_2$, as done previously e.g.~in Refs.~\cite{trout2018simulating} and \cite{gutierrez2019transversality}, since the overall noise in our experiment is dominated by the error rate $p_2$.\footnote{Rigorous comparison with $p_2$ in the sense of Ref.~\cite{gottesman2016quantum} would require comparing a physical CNOT gate to an error-corrected logical CNOT gate. In our architecture we may realize the logical gate by seven transversal CNOT gates followed by a round of QEC on both logical qubits. We note that the logical error rate of this approach is dominated by the QEC block since it contains most of the procedure's entangling gates. As a consequence, the logical error rate of a QEC block serves as a proxy to the full logical CNOT gate error rate.} 

Our definition of the logical failure rate $p_L(\lambda)$ is the logical infidelity $1-F_L$. It reflects the probability to falsely conclude, by measurement of logical operators, that the desired state has been prepared correctly (up to correctable errors) when in fact the wrong logical information is output on the data qubits. The logical fidelity $F_L$ is determined by the expectation value of the projector $P_{\pm O}$ onto the respective axis $O \in \{Z_L, H_L\}$ of the logical Bloch sphere
\begin{align}
    P_{\pm O} = \frac{I \pm O}{2} \end{align}
for the logical zero or the logical magic state respectively.
For a single preparation of the $\ket{0}_L$ state, the expectation value $\langle P_{Z_L} \rangle$ after one round of ideal EC may only take the values 0 or 1. Dangerous $X$-errors are either correctly recovered from or will result in a logical $X$-operator after ideal EC ($\braket{0|1} = 0,\,\braket{0|0} = 1$). For the logical magic state, logical errors of all three Pauli types $X,\,Y$ and $Z$ can be present on the state after ideal EC. A logical $Y$-flip causes the output state to flip from the correct magic state $\ket{H}_L$ to the orthogonal $-1$ Hadamard eigenstate $\ket{-H}_L$ ($\braket{H|Y|H} = 0$). Logical $X$- and $Z$-flipped states still have finite overlap with $\ket{H}_L$ thus contributing a finite value to the logical infidelity ($\braket{H|X|H} = 1/\sqrt{2},\,\braket{H|Z|H} = 1/\sqrt{2}$). We discuss fidelity measures further in Sec.~\ref{sec:fidelity}. For flag circuits, all preparation rounds, which trigger a flag and are thus discarded, do not contribute to the logical failure rate. 

\begin{figure*}\includegraphics[width=0.95\linewidth]{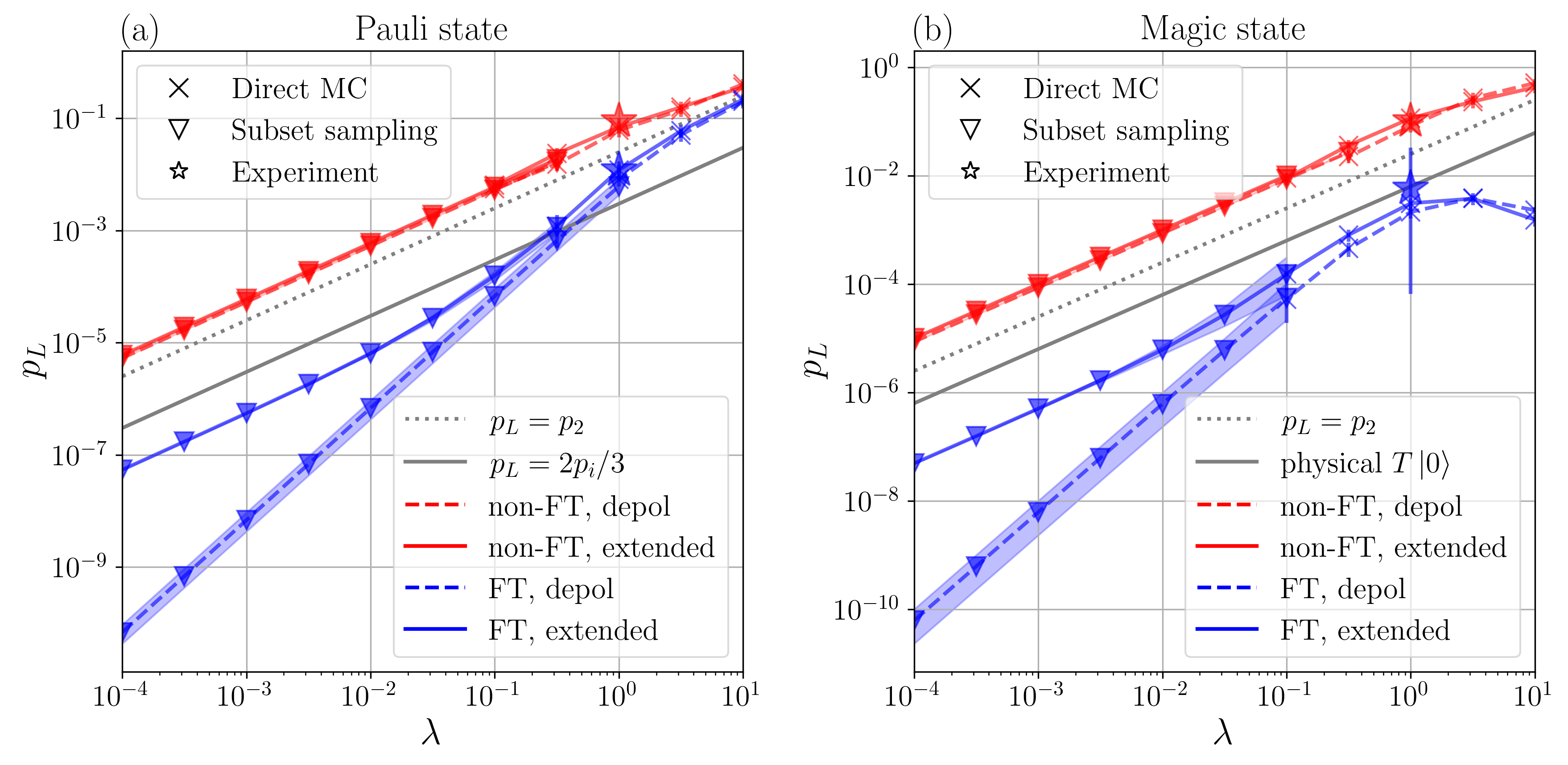}
\caption{\textbf{Logical state scaling.} Uniform scaling with a factor $\lambda$ of all physical error parameters in the non-FT and FT state preparation circuits alongside with parameters of physical qubits (both initialization -- gray, dotted -- and entangling operation -- gray, solid). For numerical simulations, we employ direct Monte Carlo (MC, cross markers) and subset sampling (SS, triangle markers) with subsets up to $w_\text{max} = 3$ in their preferential domain of physical error rates (see App.~\ref{sec:sim_methods} for a more detailed discussion). At the experimentally achieved rates $\lambda = 1$ (star marker) the models coincide in their prediction of logical failure rates within uncertainty intervals. \mbox{(a) \textbf{Pauli state.}} We compare the extended noise model containing idling and crosstalk (solid lines with markers) to depolarizing noise on single and two-qubit gates, initialization and measurements (dashed lines with markers). For each MC data point and subset failure rate we sample at least 100 times and until the uncertainty of the respective logical failure rate estimator is below a relative error of 0.5 but at most $10^4$ times. \mbox{(b) \textbf{Magic state.}} Logical failure rates using the extended noise model and the depolarizing noise model are shown. We sample at least 100 times for each MC data point and subset failure rates of the non-FT circuits and the FT circuit with extended noise. For the FT circuit with depolarizing noise we use at least 1000 samples for each subset failure rate. We sample at most $10^4$ times for the non-FT circuits and up to $10^5$ times for the FT circuits or until a relative error of 0.3 for the FT circuit under depolarizing noise and 0.5 for the other cases is reached. The left-most MC data point of the FT depolarizing line is obtained from $2 \times 10^5$ samples. For FT preparation at $\lambda = 10^1$ the logical failure rate decreases again which is related to the fact that most runs are discarded.} \label{fig:scaling}

\end{figure*}

In Fig.~\ref{fig:scaling}a we show the uniform scaling of all physical error parameters with the scaling parameter $\lambda \in [10^{-4}, 10^1]$ for the non-FT and FT Pauli state preparation compared to physical qubit parameters. The first is the rate $2p_i/3$ at which depolarizing noise of strength $p_i$ causes failure of initializing a physical qubit to $\ket{0}$. The second is the MS gate error rate $p_2$. We observe that the FT preparation achieves lower logical failure rates than both the non-FT preparation and physical MS gate error rate for all values of $\lambda$. It is larger than the physical qubit initialization error rate for $\lambda \gtrsim 0.3$ and lower than the physical qubit initialization error rate for $\lambda \lesssim 0.3$. Within the interval $\lambda \in [10^{-1}, 10^1]$, i.e.~with one order of magnitude stability around the experimentally achieved physical error parameters at $\lambda = 1$, the simulations with the four parameter depolarizing noise model quantitatively agree with the extended noise model. It is only at very low physical error parameters $\lambda \lesssim 10^{-1}$ that the extended noise simulation deviates from the depolarizing noise estimation. This is because crosstalk, which does not respect the FT properties of the circuit, becomes the dominant source of failure in this domain. The scaling becomes linear here with extended noise whereas the quadratic scaling of depolarizing noise continues for all $\lambda \rightarrow 0$. In this regime of low $\lambda$, we cannot rely on predictions made from the depolarizing noise model. In the experimentally accessible regime around $\lambda = 1$ the depolarizing noise prediction is as reliable as the extended noise model.

It is known from previous investigations of incoherent noise in general and crosstalk in particular that incoherent Pauli noise may underestimate logical failure rates \cite{katabarwa2015logical, gutierrez2016errors, darmawan2017tensor, parrado2021crosstalk}. For the experimental error parameters at $\lambda = 1$ coherent overrotation noise on MS gates in the FT Pauli state preparation circuit causes an infidelity of $0.0116(7)$ which is larger than the incoherent depolarizing noise $0.0076(5)$ or an incoherent XX-overrotation channel $0.0082(6)$. When also adding coherent XX-type crosstalk, as given by Eq.~(\ref{eq:cct}), infidelity increases to $0.0141(7)$, while the experimentally measured value is $0.012\substack{+5 \\ -4}$.

The scaling behavior of the magic state preparation protocols, which we show with depolarizing and extended noise in Fig.~\ref{fig:scaling}b, exhibits qualitatively similar features as the Pauli state preparation described above. In our setup, the physical qubit criterion of first initializing the qubit to $\ket{0}$ and then applying a physical $T$-gate is stricter than claiming to beat the MS gate error rate $p_2$ for our specific physical error parameter values. Both physical qubit criteria yield lower $p_L$ than the non-FT circuit for all observed values of the uniform scaling parameter $\lambda$. Remarkably, the simulation data for the FT magic state preparation suggests that its logical failure rate $p_L$ is lower than for both physical operations within the full $\lambda$-interval. In the regime of low physical error rates $\lambda \lesssim 0.03$ we find that the advantage of the FT implementation over both physical qubit criteria, i.e.~the offset between the parallel lines, is of more than one order of magnitude. This implies that we surely beat the physical qubit criteria despite the destructive phase averaged crosstalk noise.

From the preceding analysis we conclude that the depolarizing noise model is well suited to predict experimentally measured logical infidelities. With future improvements of physical ion trap operations, more complex noise models should be taken into account. Only moderate experimental improvements, smaller than one order of magnitude, are needed in order to reach FT advantage over physical qubits judged by comparison with the corresponding physical qubit state preparations. We now move on to discuss deterministic protocols for FT state preparation. 

\subsection{Deterministic state preparation}\label{sec:determ}

The FT state preparation procedures discussed so far can be modified such that state preparation is deterministic, i.e.~states never need to be discarded. If the acceptance rate of a non-deterministic protocol becomes too low, they might become experimentally unfeasible, e.g.~due to cycle time constraints, although the fidelity of accepted states is high. With sufficiently low physical error parameters, the additional qubit and gate overhead that deterministic protocols require may not cause a severe increase of logical failure rates. The deterministic protocols for Pauli and magic state preparation, which we will discuss, make use of the fact that the flag has been triggered which limits the number of errors which can be present on the data qubits. The measurement information of the flag qubit is used to conditionally apply additional operator measurements. As long as all errors that are not stabilizer equivalent can be distinguished by those measurements, the combined flag and syndrome information can then be used to correct all errors that are caused by single faults in the circuit, thus preserving the FT property. 

\begin{figure*}[!htbp]
    \centering
    \includegraphics[width=0.8\linewidth]{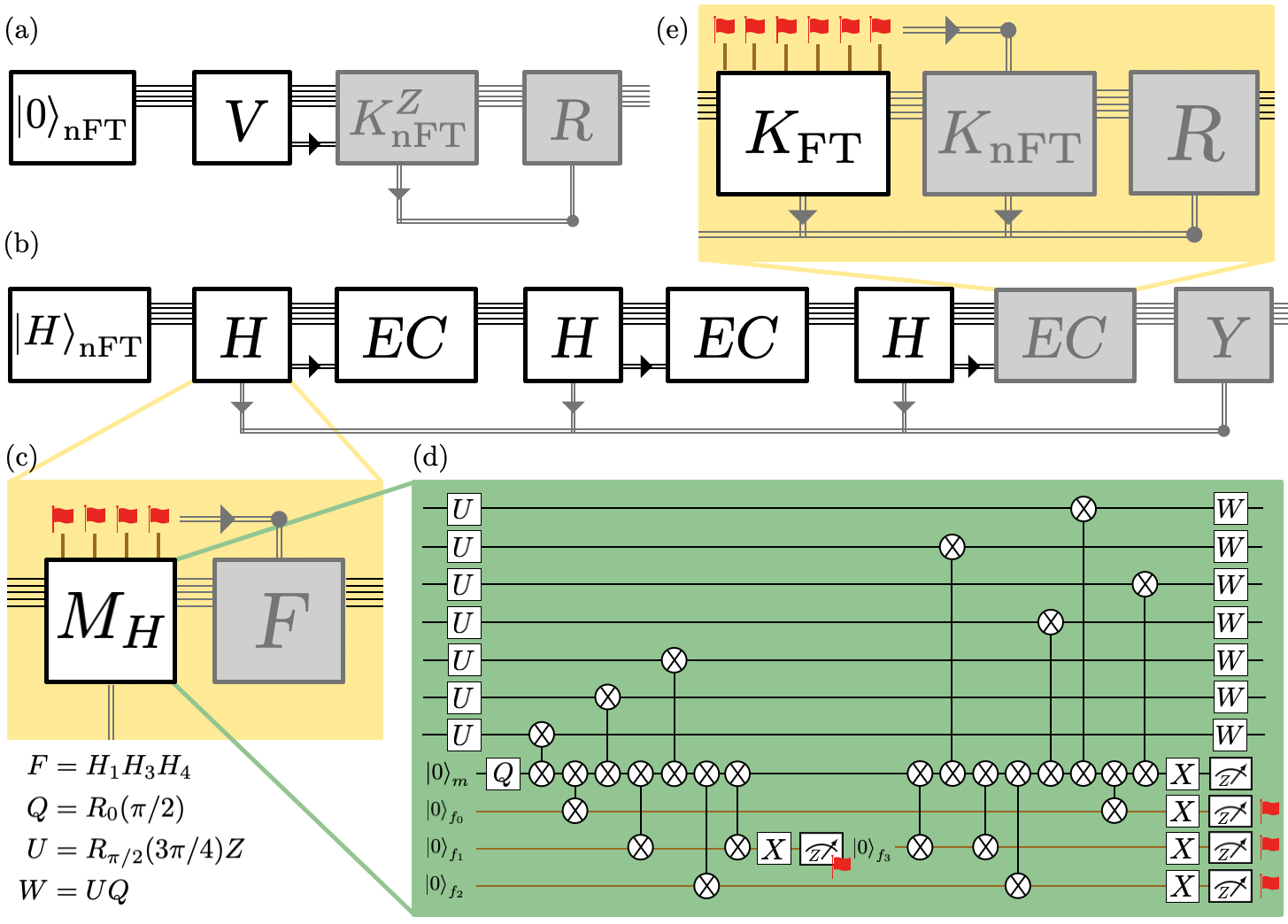}
    \caption{\textbf{Deterministic FT state preparation.} Schemes with logical building blocks acting on registers of data qubits and auxiliary qubits. Shaded blocks are only applied conditioned on classical measurement information. (a) $\ket{0}_L$: The non-FT encoding and verification blocks ($\ket{0}_\text{nFT}$ and $V$, see Fig.~\ref{fig:circ_pauli}) are followed by additional measurements of $Z$-stabilizers if the flag is triggered. Measuring with a single auxiliary qubit is sufficient to preserve the FT property of the scheme. A recovery operation $R$ is applied according to the modified look up table \ref{tab:pauliflaglut} depending on how many stabilizers are measured (block $K^Z_\text{nFT}$). When the flag is clear no additional measurements and recovery are performed. (b) $\ket{H}_L$: Non-FT magic state preparation is followed by three repetitions of Hadamard measurement and FT EC. The last EC block is only executed if the third Hadamard measurement yields a non-trivial result. Finally, a logical $Y$-flip (block $Y$) is applied to the data qubits if the Hadamard expectation value is measured as $-1$ in the second and the third round. (c) Four flag qubits are necessary to correct all dangerous errors that can happen during the Hadamard measurement. Our compiled MS gate circuit used to measure the logical Hadamard operator is shown in (d). If and only if the flag pattern $f_0, f_2, f_3 \in \{-+-, --+, --- \}$ the extra operation $F = H_1H_3H_4$ must be applied immediately after the Hadamard measurement to guarantee error distinguishability (see example in App.~\ref{sec:detFTm}). (e) If and only if the FT parallel syndrome readout (block $K_\text{FT}$, see Fig.~\ref{fig:circ_kft}) flags we proceed by measuring the syndrome with single auxiliary qubits ($K_\text{nFT}$, see Fig.~\ref{fig:circ_knft}). The recovery $R$ is chosen from the Hadamard error set (Tab.~\ref{tab:hlut}) when any flag of $M_H$ is triggered. Otherwise, $R$ is determined by the flag error set $\{X_3X_7, X_4X_6, Z_3Z_7, Z_4Z_6 \}$ if a matching syndrome, $-++$ or $++-$ for $X$- or $Z$-stabilizers respectively, is measured, otherwise the standard Steane code look up table \ref{tab:steane_lut} is applied.}
    \label{fig:det_seq}
\end{figure*}

\textbf{Logical Pauli state.} In the following we lay out a new protocol for deterministic FT Pauli state preparation. The desired $\ket{0}_L$ state can still be recovered even when the flag is triggered instead of discarding the flagged state as in the non-deterministic case. While a single error is still tolerable, a weight-2 error leads to application of an erroneous recovery operation which causes logical failure when using the look up table decoder from Tab.~\ref{tab:steane_lut}. Instead, we may extend the look up table decoder to prioritize two-qubit recovery operations when the flag is triggered. These two-qubit errors make up the so-called flag error set. By exhaustively placing all single faults on the FT encoding circuit, we find that only two dangerous data qubit errors, namely $X_1X_3$ and $X_4X_5$, that are not stabilizer equivalent can propagate to the final data qubit state. For example, they can be caused respectively by faults $Y_1X_3$ on the last MS gate and $Z_4X_5$ on the fifth MS gate of the non-FT block in Fig.~\ref{fig:circ_pauli} which also trigger the flag. All other resulting data qubit errors are, if not stabilizer equivalent to either $X_1X_3$ or $X_4X_5$, equivalent to a weight-1 error or a logical $Z$-operator. The latter acts trivially on the logical zero state that is being prepared. Additionally, given that the flag is triggered, we find that the only weight-1 errors that can result on the data qubits from a single fault are $X_3$, $X_5$ and $X_6$. 

\begin{figure}[!htbp]
    \centering
    \includegraphics[width=\linewidth]{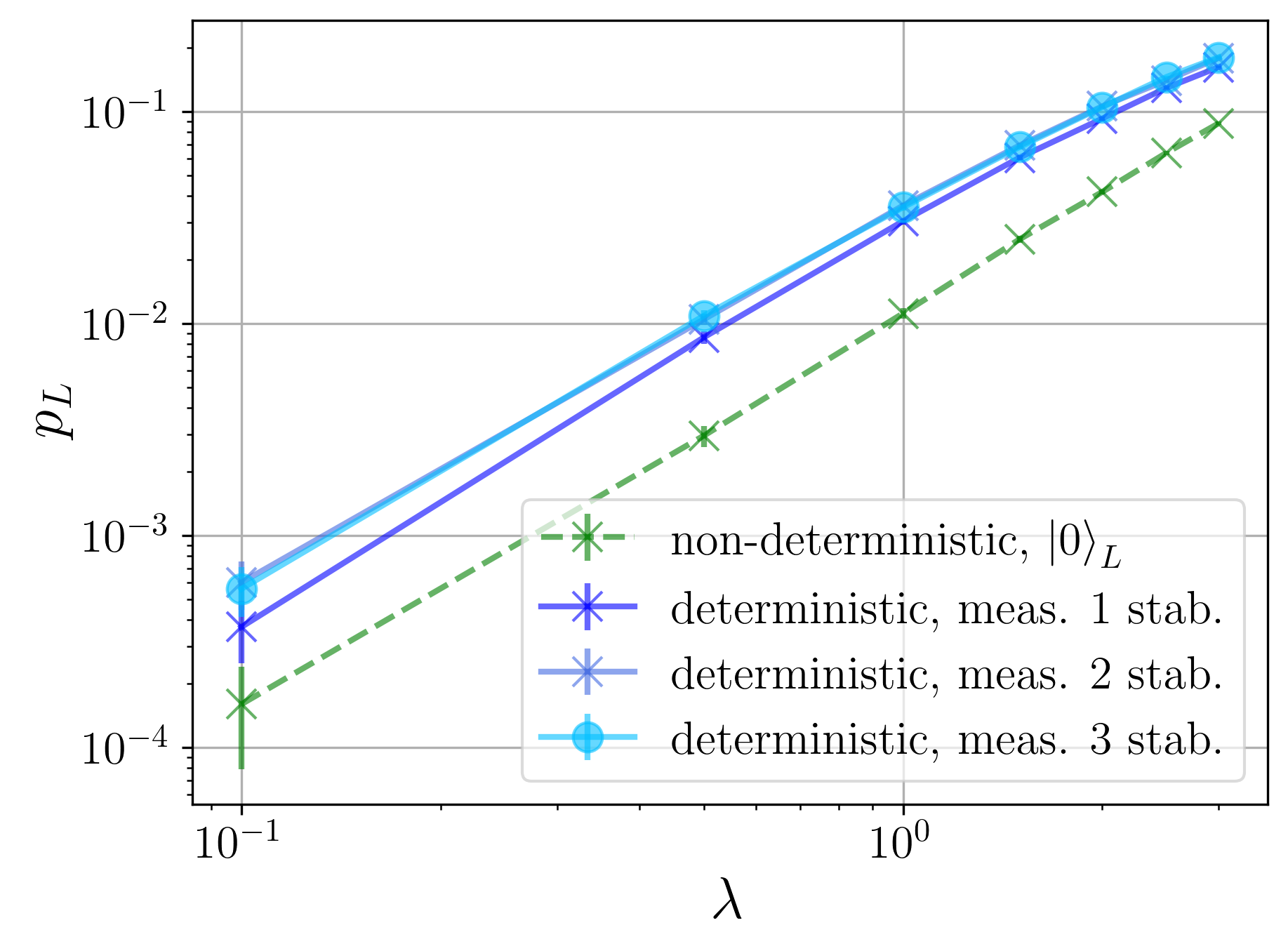}
    \caption{\textbf{Deterministic Pauli state scaling.} Uniform scaling under extended noise with a factor $\lambda$ of all physical error rates for the FT flag preparation circuit of $\ket{0}_L$ and the deterministic extension where stabilizers are measured with single auxiliary qubits. For each of the $10^{5}$ MC samples, the preparation is repeated until the flag qubit is clear. The non-deterministic circuit yields lower logical error rates then the deterministic procedure. Since both scale quadratically, there will not be a crossover point at lower $\lambda$. The lines for measuring two and three stabilizers lie on top of each other.}
    \label{fig:det_pauli}
\end{figure}

The two dangerous errors $X_1X_3$ and $X_4X_5$ can be distinguished by measuring only two additional stabilizers. Their syndrome will not be confused with the syndromes of the single-qubit errors because the triggered flag restricts the number of errors that can occur. A pictorial illustration of the protocol with stabilizer measurement conditioned on the classical flag information is shown in Fig.~\ref{fig:det_seq}a. For the correction procedure, the look up table \ref{tab:pauliflaglut} can be applied. 
\begin{table}[!ht]
\centering
    \begin{tabular}{c||c|c|c||c|c}
         data qubit error & $K_1^Z$ & $K_2^Z,K_3^Z$ & $R_2$ & $K_{123}^Z$ & $R_1$ \\ \hline
         $X_1X_3$ & $+$ & $+-$ & $X_1X_3$ & $-$ & $X_7$\\
         $X_4X_5$ & $+$ & $-+$ & $X_4X_5$ & $-$ & $X_7$\\
         $X_6X_7$ & $+$ & $-+$ & $X_4X_5$ & $-$ & $X_7$\\
         \hline
         $X_6$ & $-$ & $+-$ & $X_1X_3$ & $+$ & $I$\\
         $X_5$ & $-$ & $-+$ & $X_4X_5$ & $+$ & $I$\\
         $X_3$ & $+$ & $--$ & $X_3$ & $+$ & $I$
    \end{tabular}
    \caption{Modified look up table for deterministic Pauli state preparation. It is used instead of Tab.~\ref{tab:steane_lut} if and only if the flag is triggered. All errors can be corrected, allowing for a residual weight-1 error, when measuring either the single stabilizer $K_{123}^Z = Z_1Z_2Z_4Z_7$ or the two stabilizers $K_2^Z = Z_1Z_3Z_5Z_7$ and $K_3^Z = Z_2Z_3Z_6Z_7$ or all three stabilizer generators, including $K_1^Z=Z_4Z_5Z_6Z_7$. The recovery $R_1$ is applied when only $K_{123}^Z$ is measured. $R_2$ is the recovery operation when $K_2^Z$ and $K_3^Z$ are measured.}
    \label{tab:pauliflaglut}
\end{table}

Applying the recovery $R_2 = X_4X_5$ when measuring the reduced syndrome $(K_2^Z,K_3^Z) = -+$ will not cause a logical failure because either the $X_4X_5$ error is corrected or the product of error and recovery will be $X_4X_5X_5$ in case the data qubit error was $X_5$. The result is the weight-1 error $X_4$ so FT is respected. From this example, we see that measuring $K_1^Z$ is not necessary to correct the weight-2 errors. The same holds if the actual error is $X_6X_7$ since it is stabilizer equivalent to $X_4X_5$. 

Moreover, we find that measuring only the stabilizer operator $K_{123}^Z = K_1^Z \times K_2^Z \times K_3^Z = Z_1Z_2Z_4Z_7$ is sufficient to neutralize the dangerous weight-2 errors. As shown in Tab.~\ref{tab:pauliflaglut}, its expectation value is $+1$ for the correctable weight-1 errors and $-1$ for both uncorrectable weight-2 errors. By applying the recovery operation $R_1 = X_7$, both errors $X_1X_3$ and $X_6X_7$ are turned into correctable weight-1 errors $X_5$ and $X_6$ respectively. Note that a single auxiliary qubit is sufficient for syndrome readout since an additional fault happening in this step -- on top of the fault that already happened to trigger the flag -- would render the overall fault configuration to be of order $p^2$ so FT1 is not violated. The result of these additional measurements is a deterministic fault-tolerant way to prepare the logical zero state of the Steane code. Given the flag has been triggered, we are able to correct all weight-2 errors possibly present on the data qubit state by measuring a reduced set of stabilizers\footnote{We note that the measurement of the flag qubit and the auxiliary qubit for $K_{123}^Z$ can be avoided completely so that the state preparation circuit works deterministically without in-sequence measurements or feed-forward of measurement information by applying the $R_1$ recovery through a Toffoli gate controlled by the two auxiliary qubits. The FT property remains intact this way since the Toffoli only couples to a single data qubit.}. Of course, it is also possible to measure all three stabilizer generators and by the full three-bit syndrome uniquely distinguish all weight-1 and weight-2 errors given in Tab.~\ref{tab:pauliflaglut}. 

\begin{figure}[!htbp]
    \centering
    \includegraphics[width=\linewidth]{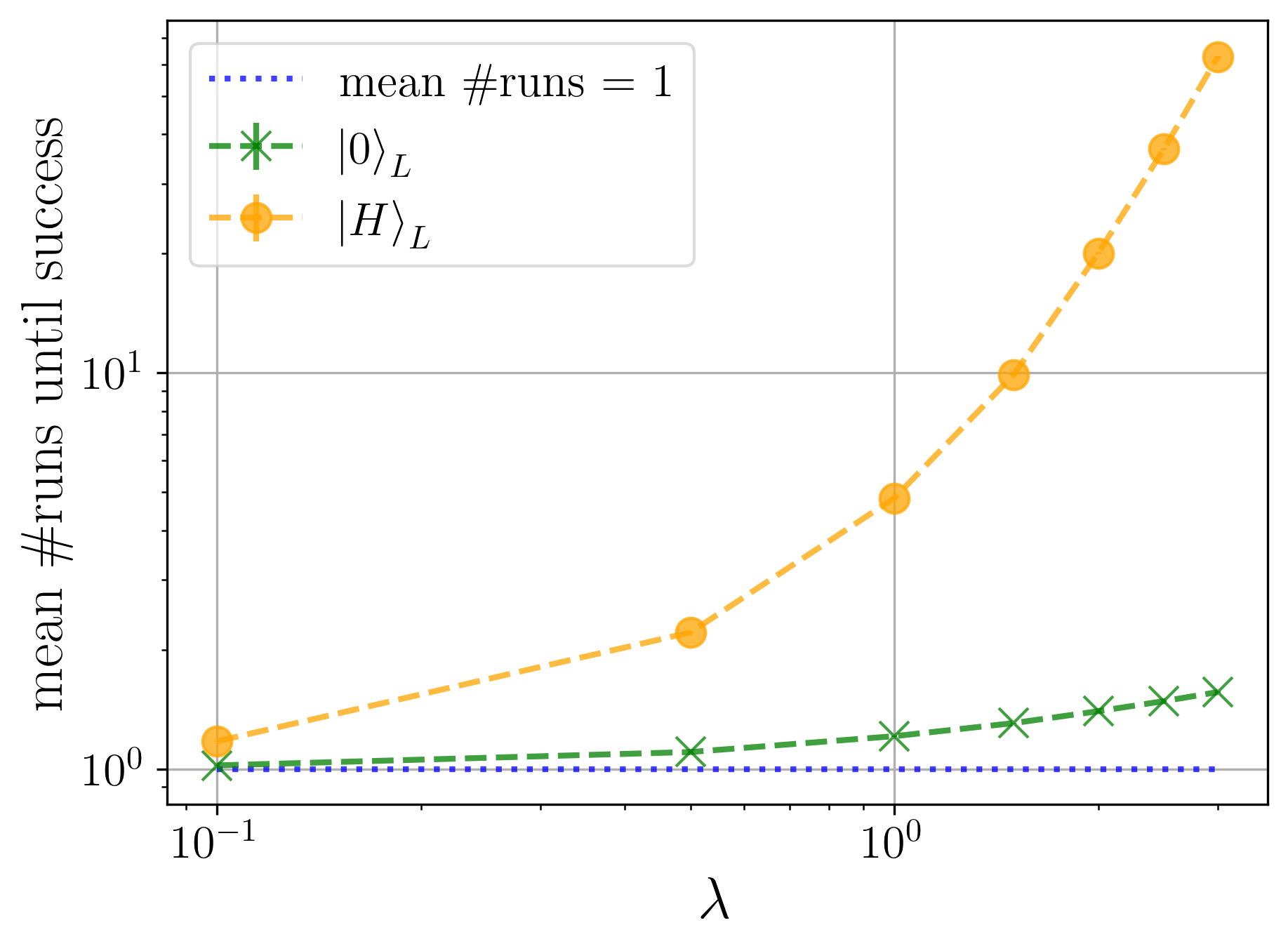}
    \caption{\textbf{Repetition overhead for non-deterministic state preparation.} Averaged number of repetitions until the prepared state is accepted, i.e. all flags measured as $+1$. Preparation of the magic state $\ket{H}_L$ on average needs more trials than the Pauli state $\ket{0}_L$. Deterministic state preparation schemes succeed after a single run by definition. Uncertainties on data points are smaller than the marker sizes.}
    \label{fig:reptimes}
\end{figure}

\begin{figure}[!htbp]
    \centering
    \includegraphics[width=\linewidth]{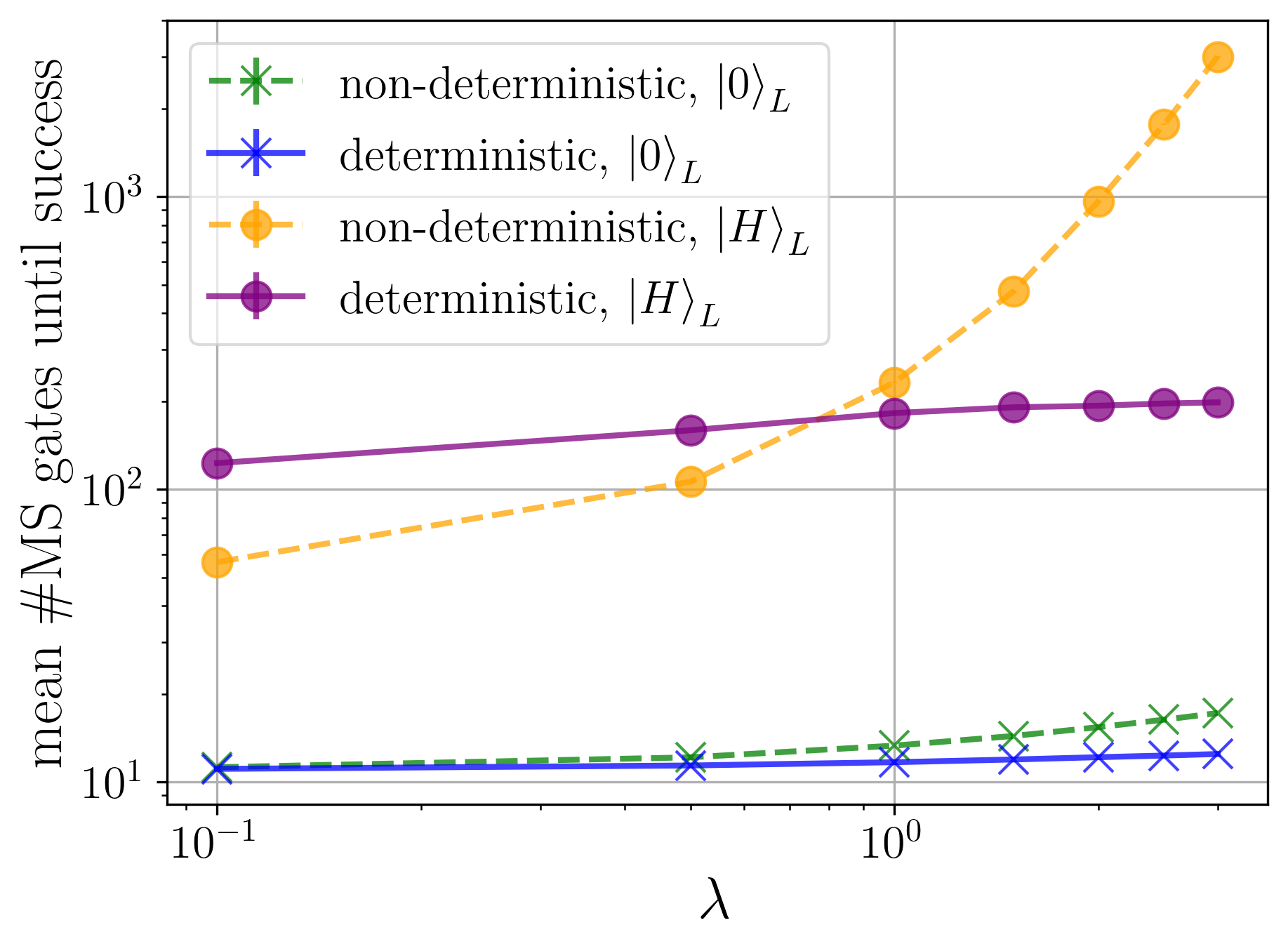}
    \caption{\textbf{Entangling gate overhead for FT logical state preparation.} Averaged number of entangling gates needed until the prepared state is accepted. For the non-deterministic protocols this amounts to all flags being clear. Deterministic protocols realize different circuit sequences depending on in-sequence measurement outcomes. The increase in average MS gate count is moderate for the deterministic protocols. For the non-deterministic FT magic state preparation the increase is over two orders of magnitude. It requires fewer MS gates on average than the deterministic FT magic state preparation protocol for $\lambda \lesssim 0.8$. Uncertainties on data points are smaller than the marker sizes.}
    \label{fig:gatecounts}
\end{figure}

Analogously to the previous scaling analysis, we show the scaling behavior of the deterministic and non-deterministic FT Pauli state preparation in Fig.~\ref{fig:det_pauli}. The uniform scaling parameter $\lambda$ of all physical error parameters, including crosstalk, varies between $0.1$ and $3$. Both schemes scale quadratically in this interval due to their FT property. Nonetheless the non-deterministic scheme, where only non-flagged states are accepted, has a logical failure rate one order of magnitude lower than the deterministic schemes where either two or three stabilizer measurements are performed in case the flag is triggered. It is ensured that there cannot be another crossover point at lower values of $\lambda$ since the vertical offset between the curves is determined by the coefficient of the quadratic term. Figure \ref{fig:reptimes} shows the average number of times the non-deterministic preparation needs to be repeated until the state is accepted. While for the deterministic scheme this value is equal to one by construction, we see that for increasing $\lambda \in [0.1, 3]$ the mean number of necessary repetitions moderately grows from $1.020(1)$ to $1.563(6)$ which is feasible for experimental implementation. The number of repetitions translates to an increase of required entangling gates, shown in Fig.~\ref{fig:gatecounts}, from approximately 11 to 17.2 for the non-deterministic protocol and from approximately 11 to 12.5 for the deterministic protocol on average. In case of a triggered flag, the deterministic protocol will proceed with just the measurement of $K_{123}^Z$, requiring 4 additional MS gates, instead of repeating the preparation circuit with 11 MS gates.

We conclude that using the non-deterministic state preparation protocol is preferable in the examined range of $\lambda$ and below since it yields logical failure rates one order of magnitude lower than the deterministic state preparation at the cost of a moderate number of repetitions, given that the necessary repetition times are permitted by other experimental constraints.

\textbf{Logical magic state.} The protocol for deterministic FT magic state preparation has been pointed out in Ref.~\cite{chamberland2019fault}. We provide a compiled version into MS gates visualized in Fig.~\ref{fig:det_seq}b and discuss the expected performance for current and anticipated future trapped-ion physical error parameters. 

After preparing the logical magic state non-fault-tolerantly, we measure the logical Hadamard operator three times, each involving the use of four flag qubits to distinguish all possible errors resulting from a fault triggering flags. The measurement circuit labeled $M_H$ is shown in Fig.~\ref{fig:det_seq}c with the detailed MS compilation given in Fig.~\ref{fig:det_seq}d. For the flag patterns, i.e.~combinations of flag qubit measurement outcomes, $f_0, f_2, f_3 \in \{-+-, --+, --- \}$ an additional operator $F = H_1H_3H_4$ must be applied to guarantee error distinguishability. Hadamard-type errors on four data qubits, which can arise from a single $X$-fault on the measurement qubit of the circuit in Fig.~\ref{fig:det_seq}d, cannot be corrected using the six-bit syndrome if $F$ were not applied. Triggered by the aforementioned flag patterns, $F$ transforms a dangerous error into a lower weight error which can then be corrected by the subsequent EC block (see an explicit example in App.~\ref{sec:detFTm}). 

After each Hadamard measurement, a full round of FT error correction must be performed before the logical Hadamard can be measured again. The EC block, shown in Fig.~\ref{fig:det_seq}e consists of the flag-FT parallel readout circuit ($K_\text{FT}$) which we previously used to discard erroneous states in the non-deterministic protocol. Now, it is followed by an additional block of syndrome readout with single auxiliary qubits ($K_\text{nFT}$, compiled with the CNOT decomposition of Fig.~\ref{fig:architecture}c) in case any flag is triggered. If the flags of $M_H$ are triggered and the syndrome is not trivial we apply a recovery operation according to the Hadamard look up table \ref{tab:hlut} given in App.~\ref{sec:detFTm}. Here, the full six-bit syndrome is necessary to identify the correct recovery operation despite the CSS property of the Steane code. If all Hadamard flags are clear or the syndrome is not in the Hadamard look up table but the parallel readout circuit $K_\text{FT}$ yields a triggered flag, we make use of the flag error set FES $= \{X_3X_7, X_4X_6, Z_3Z_7, Z_4Z_6 \}$ to correct weight-2 errors of both X- and Z-type informed by the $Z$- and $X$-syndrome measured by $K_\text{nFT}$ respectively. The flag error set is formed by all dangerous errors that can result from single faults in the $K_\text{FT}$-block that trigger a flag. If the syndromes measured by the two blocks $K_\text{FT}$ and $K_\text{nFT}$ agree, we apply the recovery from the standard look up table (Tab.~\ref{tab:steane_lut}). The third EC block can be omitted in case the third Hadamard measurement yields a $+1$ measurement outcome and no flags are triggered. 

In the end, a logical $Y_L$-correction is applied dependent on the three Hadamard measurement outcomes. It is applied if the three consecutive Hadamard measurements are either $---$ or $+--$, otherwise no additional correction is applied. These corrections take into account logical operators that can arise from single faults in the non-FT preparation circuit (Fig.~\ref{fig:circ_magic_nonft}) already. A detailed derivation is given in the Appendix of Ref.~\citep{chamberland2019fault}.

\begin{figure}\centering
    \includegraphics[width=\linewidth]{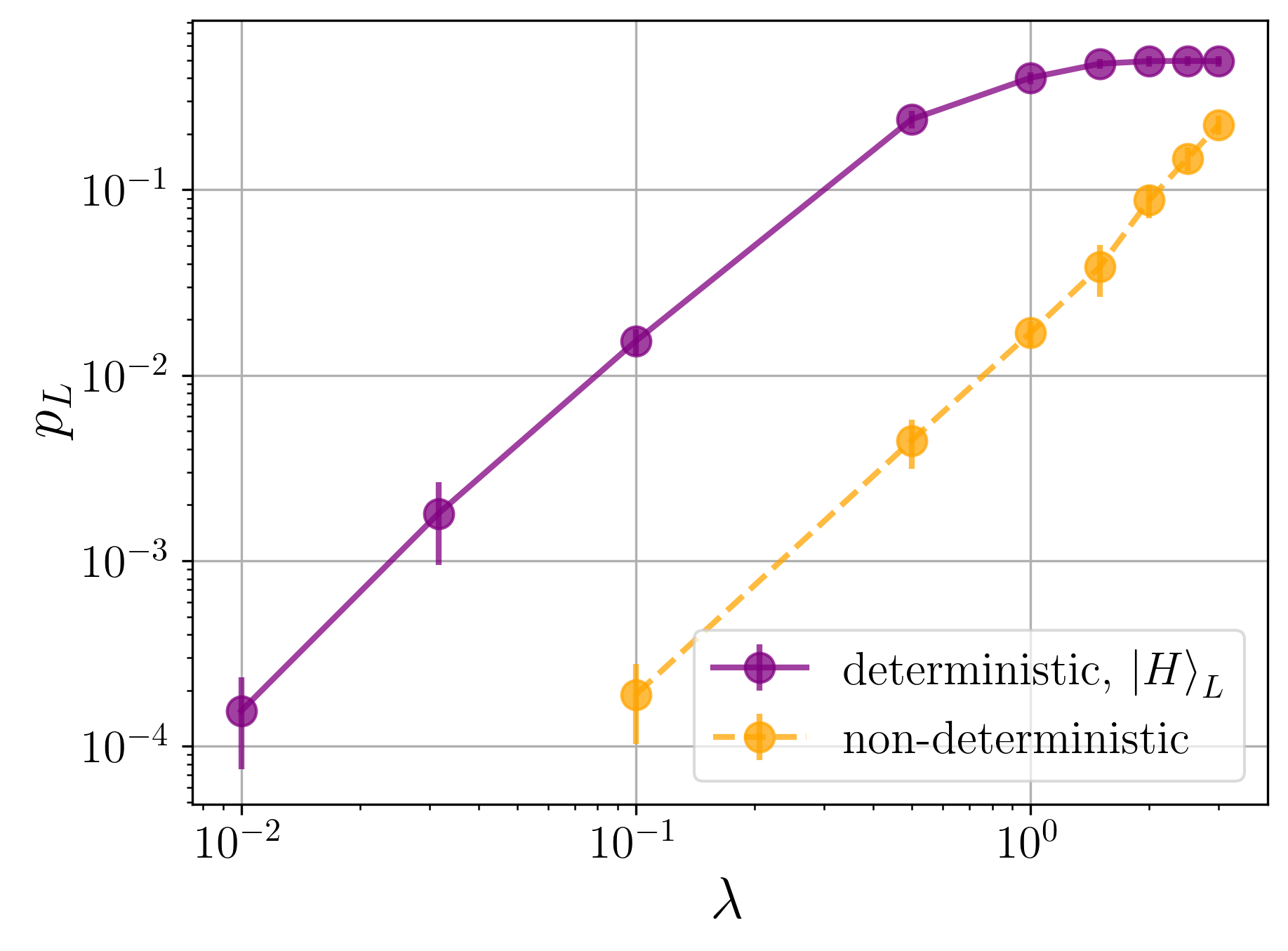}
    \caption{\textbf{Deterministic magic state scaling.} Comparison of deterministic and non-deterministic FT magic state preparation for physical error rates uniformly scaled with parameter $\lambda$. The significant overhead of the deterministic scheme leads to a logical failure rate approximately two orders of magnitude larger than for the non-deterministic scheme at low $\lambda$. For the deterministic scheme we use (1000,\,$10^4$,\,$10^5$) samples for the data points at ($\lambda > 0.1,\,0.01 < \lambda \leq 0.1,\,\lambda = 0.01$). For the non-deterministic scheme we use (1000,\,$10^4$,\,$10^5$) samples for the data points at ($\lambda > 1,\,0.1 < \lambda \leq 1,\,\lambda = 0.1$).}
    \label{fig:det_magic}
\end{figure}

As for the logical zero state, we show the comparison of logical failure rates achieved by the deterministic and non-deterministic protocol over the uniform scaling range $\lambda \in [0.1, 3]$ and subjected to extended noise in Fig.~\ref{fig:det_magic}. While the non-deterministic scheme scales quadratically over the entire range of $\lambda$, the deterministic scheme just transitions towards quadratic scaling at low physical error parameters. For $\lambda \leq 1$ the advantage in logical failure rates of the non-deterministic over the deterministic scheme is as large as approximately two orders of magnitude. This is due to the gate overhead that the deterministic scheme requires. On the other hand, employing the non-deterministic scheme demands a repetition overhead which we show in Fig.~\ref{fig:reptimes}. Although at low $\lambda$ the mean number of repetitions until the FT magic state is accepted approaches 1, at scaling factors $\lambda = 1$ and $\lambda = 3$ we need approximately 5 and 63 repetitions on average respectively. From Fig.~\ref{fig:gatecounts}, it is clear that the number of necessary MS gates also increases drastically through the repetition procedure. While the non-deterministic protocol requires only 56 MS gates at $\lambda = 0.1$ on average, the deterministic protocol uses on average approximately 113 MS gates at $\lambda = 0.1$ and 198 MS gates at $\lambda = 3$ due to more frequent flag events and thus more realizations of the full EC sequence. Due to the larger number of repetitions at $\lambda = 3$ the mean number of MS gates increases to a large value of approximately 3010. At $\lambda = 1$ the deterministic protocol requires approximately 182 MS gates on average; slightly less then the approximately 232 MS gates needed on average for the non-deterministic protocol.

The trade-off between deterministic and non-deterministic protocols includes on the one hand preparing the logical state with high fidelity while on the other hand also keeping acceptance rates high or equivalently keeping the required number of circuit repetitions sufficiently low. For the FT magic state preparation the trade-off between logical fidelity and gate overhead is more pronounced than for the Pauli state. In order to use the non-deterministic protocol in an experimental realization and benefit from its low logical failure rate, one must be able to tolerate the potentially large repetition overhead for the algorithm aimed to be performed. When the deterministic protocol is used, the runtime of a quantum algorithm can be bounded at the expense of the large gate overhead which deteriorates the resulting logical failure rate compared to the non-deterministic protocol. 

\begin{figure*}[!htbp]
  \includegraphics[width=0.95\linewidth]{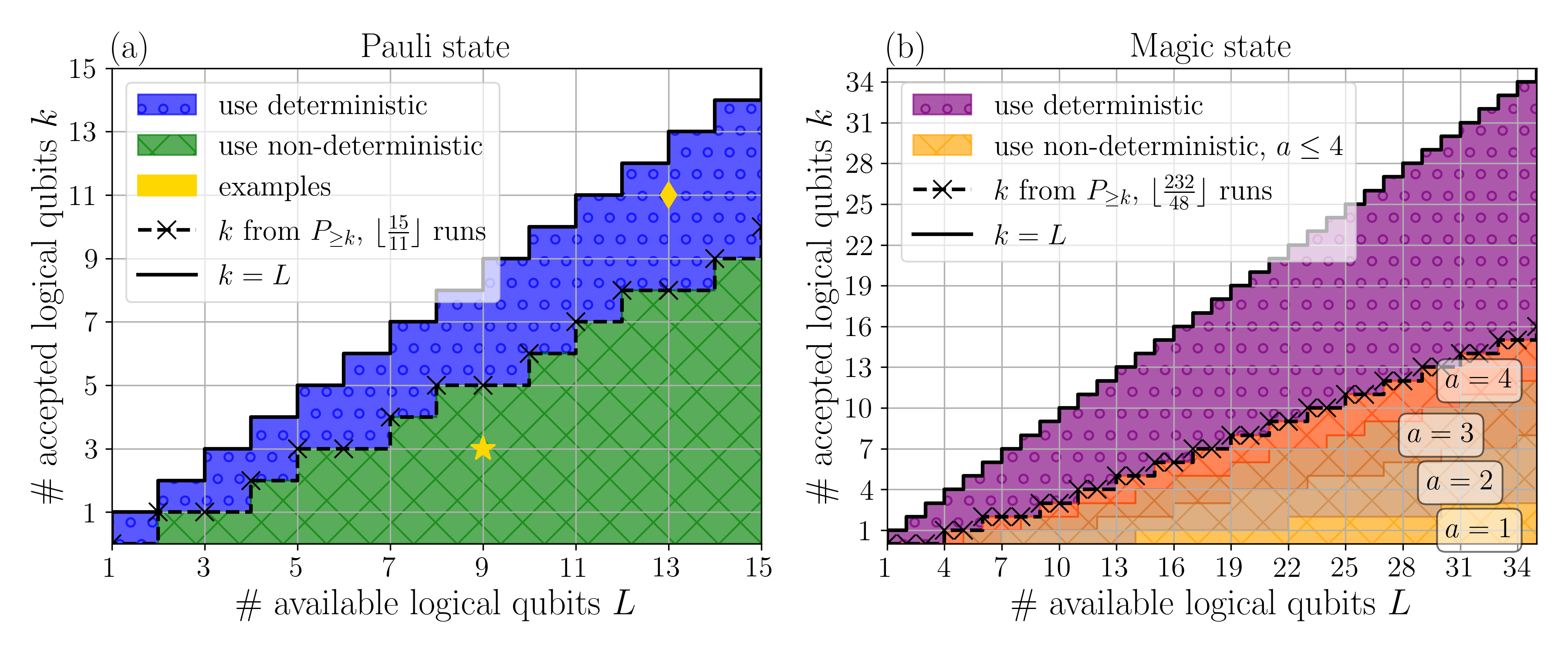}\caption{\textbf{Preparation time advantage}. Non-deterministic logical state preparation schemes (crosses) have a preparation time advantage over the deterministic schemes (small circles) as long as the number of entangling gates needed until $k$ out of $L$ logical qubits are accepted is lower than for the deterministic scheme. The deterministic schemes always prepare $k = L$ logical qubits (solid black lines). The boundaries between regimes of advantage of either scheme (dashed lines with crosses) are calculated at our respective flag rates at $\lambda = 1$ using Eq.~\eqref{eq:betasuccess} with $P_{\geq k} \geq 95\%$ and $a = a^* = \lfloor \nicefrac{t_d}{t_n} \rfloor$. In the region above this line using the deterministic scheme is advantageous since the non-deterministic scheme would take more MS gates for the same result of $k$ logical qubits. (a) \textbf{Pauli state.} Running the non-deterministic scheme twice already takes $2 \times 11$ MS gates -- more than the 15 MS gates needed for the deterministic scheme. Thus we compare the expected number of accepted logical qubits $k$ at our flag rate $f = 0.17$ when $L$ logical qubits can be prepared for one circuit run of either scheme. For example, to prepare at least 3 out of 9 logical qubits correctly the non-deterministic scheme is sufficient (star marker) while the deterministic scheme should be used if, e.g., at least 11 out of 13 logical qubits need to be accepted (diamond marker). (b) \textbf{Magic state.} At our flag rate $f = 0.8$, we can run the non-deterministic magic state preparation at most $\lfloor \nicefrac{232}{48} \rfloor = 4$ times and stay below the number of entangling gates used by the deterministic scheme on average. Regions of accepted logical qubit number after 1, 2, 3 and 4 runs are shown in shades of orange.}
    \label{fig:k_vs_L_fixed_fa}
\end{figure*}

For scale-up to multiple logical qubits, scheduling aspects may become relevant for the specific physical architecture at hand. Deterministic logical state preparation can be performed in parallel, if the experiment permits, and all logical states will be prepared after constant time. When $L$ logical qubits are prepared non-deterministically, the waiting time until all logical qubits are verified is limited by the logical qubit which needs the most repetitions until accepted. On average, preparing the qubits non-deterministically is advantageous if the average number of repetitions (see Fig.~\ref{fig:reptimes}) for a given set of physical error parameters $a_\lambda$ leads to a smaller total state preparation time $a_\lambda t_n < t_d$, assuming that a single trial takes time $t_n$, than using the deterministic scheme taking time $t_d$. Even if one is lacking parallel operation capabilities, the waiting time of the other $L-1$ logical qubits -- while one logical qubit is being prepared -- does not need to be detrimental to the overall fidelity: An additional round of QEC can be performed on each logical qubit before feeding it into a subsequent logical building block. Moreover, it is not required with our protocols that successful state preparations coincide in time. 

Suppose that we are capable of preparing $L$ logical qubits, using the non-deterministic Pauli state preparation protocol, when we only need $k$ accepted logical qubits in order to use them to run a quantum algorithm. With flag rate $f$, the number of logical qubits that are rejected due to flag events after $a$ runs of the non-deterministic encoding circuit is $Lf^a$. As a consequence, the number of logical qubits $L$ needed to accept $k$ logical qubits on average at flag rate $f$ after $a$ trials is given by 
\begin{align}
    k &= L(1 - f^a) \label{eq:k_of_Lfa}
\end{align} 
and the number of runs $a$ needed to accept $k$ out of $L$ logical qubits on average at flag rate $f$ reads
\begin{align}
    a &= \frac{\log (1 - k/L)}{\log f}.
\end{align}

Since all logical qubit preparations are independent from another, the probability $P_{\geq k}$ that at least $k$ out of $L$ logical qubits are prepared correctly after $a$ runs is given by the cumulative binomial distribution with success probability $1-f^a$
\begin{align}
	P_{\geq k} &= \sum_{j=k}^{L}\binom{L}{j} (1-f^a)^{j} (f^{a})^{L-j} \label{eq:cumqubitsuccess} \\
	&= 1 - \sum_{j=0}^{k-1}\binom{L}{j} (1-f^a)^{j} (f^{a})^{L-j} \\
	&= 1 - \sum_{j=0}^{k-1} B(L,j,1-f^a) \\
	&= 1 - I_{f^a}(L-k+1,k) \label{eq:betasuccess}
\end{align}
where we use the regularized incomplete Beta function $I$ \cite{wadsworth1961introduction, lang2018strictly}. We can extract the number of necessary qubits $L$ or the number of circuit runs $a$ to obtain $k$ accepted qubits with a desired probability $P_{\geq k}$ by numerical inspection of Eq.~\eqref{eq:betasuccess}. 

It is advantageous to use the non-deterministic preparation procedure as long as after at most $a^* = \lfloor \nicefrac{t_d}{t_n} \rfloor$ preparation attempts the number of accepted qubits $k$, either on average or with probability $P_{\geq k}$, is sufficient to perform the desired quantum algorithm. For our logical Pauli state preparation schemes, we have $\nicefrac{t_d}{t_n} = \nicefrac{15}{11} \approx 1.4$ when using the number of entangling gates as a proxy for the circuit execution time. So if more than $a^*$ trials were needed, there would be no savings in the number of entangling gates over the deterministic scheme anymore. In Fig.~\ref{fig:k_vs_L_fixed_fa} we show the number of accepted logical qubits $k$ given that $L$ logical qubits can be prepared and highlight which of the two schemes is advantageous in terms of preparation time. While the deterministic scheme will always prepare $k = L$ logical qubits, Eqs.~\eqref{eq:k_of_Lfa} and \eqref{eq:betasuccess} provide the expected number on average or -- here -- with a $95\%$ probability $P_{\geq k}$, which we show for the Pauli state at flag rate $f = 0.17$ in Fig.~\ref{fig:k_vs_L_fixed_fa}a. For the logical magic state, the large number of MS gates used by the deterministic scheme on average at $\lambda = 1$ allows one to run up to 4 trials of the non-deterministic scheme since the fraction of entangling gates is $\nicefrac{t_d}{t_n} = \nicefrac{232}{48} \approx 4.8$. The expected number of accepted logical magic states after up to 4 runs is compared to the deterministic scheme in Fig.~\ref{fig:k_vs_L_fixed_fa}b.

\section{Influence of entangling crosstalk on logical states}\label{sec:ctr} 

We have seen in the previous discussion of FT circuits that crosstalk in general does not respect the FT property and thus can lead to linear scaling effects in the logical failure rates detrimental to the FT property of -- otherwise -- FT circuits. Crosstalk on single-qubit gates does not cause correlated faults but mere single-qubit faults on neighboring qubits. After entangling gates however, crosstalk fault operators of Pauli weight-2 can potentially propagate to cause uncorrectable weight-2 errors at the end of the circuit. Preserving the quadratic scaling behavior in the logical failure rate is thus essential when aiming for advantage of FT circuits over their non-FT counterparts or physical qubits, which scale linearly. In this section we demonstrate that, for the FT Pauli preparation circuit, it is actually possible to find circuit implementations which show quadratic scaling of their logical failure rate and thus respect FT despite the presence of crosstalk. Our argument is derived from the CSS property of the Steane code.

\begin{figure}
    \centering
    \includegraphics[width=\linewidth]{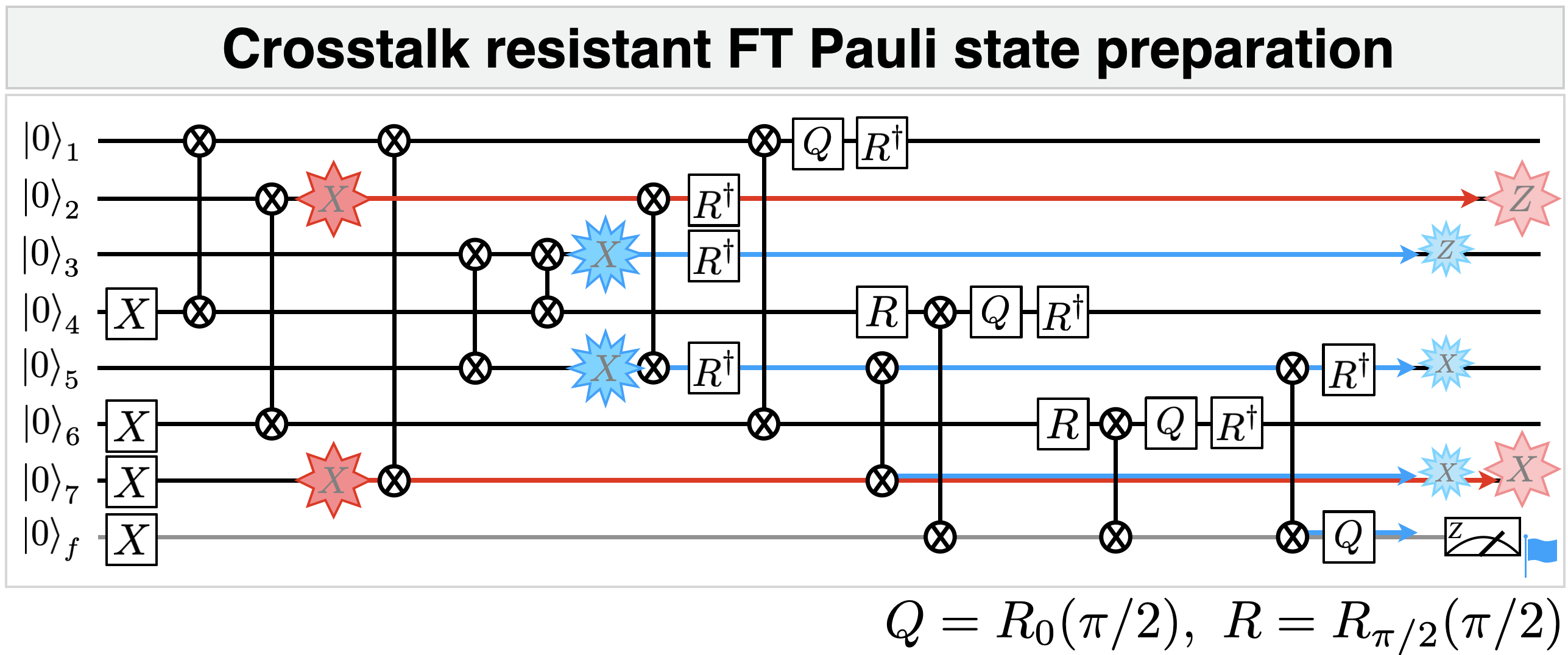}
    \caption{\textbf{Crosstalk-resistant FT Pauli state preparation.} There is no single XX-type fault placed at any crosstalk location which causes an output state with $X$-error of weight greater than one and does not also trigger the flag. The Steane code is capable of correcting a weight-1 $X$- and $Z$-error each. $XX$-faults are prevented by local rotations from resulting in a weight-2 error. An example of such a fault on the second MS gate is depicted by the red 8-cornered stars. The resulting two weight-1 errors are $Z_2$ and $X_7$. The previously (Fig.~\ref{fig:circ_pauli}) dangerous $X_3X_5$ fault after the fifth MS gate (blue 12-cornered stars) now triggers the flag.}
    \label{fig:ms_ctr}
\end{figure}

As long as not more than a single fault occurs, accepted states from the FT Pauli state preparation circuit are guaranteed to be the correct $\ket{0}_L$ state up to a single correctable error. XX-type crosstalk on a target-neighbor location $t,\,n$ of an MS gate acting on target qubits $t_1$ and $t_2$ is described by the channel
\begin{align}
    \curlE_\text{xct}(\rho) &= (1-p_{\text{c}_2}) \rho + p_{\text{c}_2} X_{t}X_{n} \rho X_{t}X_{n}.
\end{align}
In the presence of XX-type crosstalk fault tolerance can be uphold if all XX-faults can be made to propagate to correctable errors at the end of the circuit.\footnote{Note that $\curlE_\text{xct}$ is a special case of Eq.~\eqref{eq:ctincohallphases} with all phases equal to zero. The incoherent error probability is shifted $p_{\text{c}_2} \rightarrow 4p_{\text{c}_2}$ for crosstalk locations which involve common neighbor ions. Also see App.~\ref{sec:noise_app} for a more detailed discussion.} Since the local rotations that stem from the CNOT decomposition into MS-gates rotate $X$-fault operators on the control qubit to $Z$-faults (see Fig.~\ref{fig:circ_pauli}), some of the resulting error operators may be correctable because a single $Z$- and a single $X$-error are correctable distinctly in the Steane code. An example of this effect can be seen for an $X_2X_7$ fault after the second MS gate in Fig.~\ref{fig:ms_ctr} which becomes an $Z_2X_7$ error at the end of the circuit. The Steane code can correct $Z_2$ and $X_7$ independently. Consequentially, it is desirable to choose a qubit mapping of the FT encoding circuit that reduces the number of neighbor locations around the control qubits and allows for detection of dangerous crosstalk faults by the flag verification qubit. Robustness against crosstalk faults via optimal qubit mapping has been shown before by searching for Hamiltonian paths in a qubit mapping graph for a comparative code study with realistic ion trap noise \citep{debroy2020logical}.

We distinguish these types of circuits by calling them ``crosstalk-resistant'' (CTR) and ``non-crosstalk-resistant'' (non-CTR). Qubit indices can be relabeled to obtain a CTR circuit for FT preparation of the $\ket{0}_L$ state using MS gates as given in Fig.~\ref{fig:ms_ctr}. After relabeling, the new stabilizers have support on qubits $(1,4,6,7),\,(2,5,6,7)$ and $(3,4,5,7)$.
The $X_3X_5$ crosstalk fault after the fifth MS gate, discussed as an example in Sec.~\ref{sec:prot_sims}, will now trigger the flag as opposed to the non-CTR circuit in Fig.~\ref{fig:circ_pauli} so that the output state with the dangerous error $Z_3X_5X_7$ will be discarded. A CTR circuit for FT magic state preparation was not found. 

\begin{figure}
    \centering
    \includegraphics[width=\linewidth]{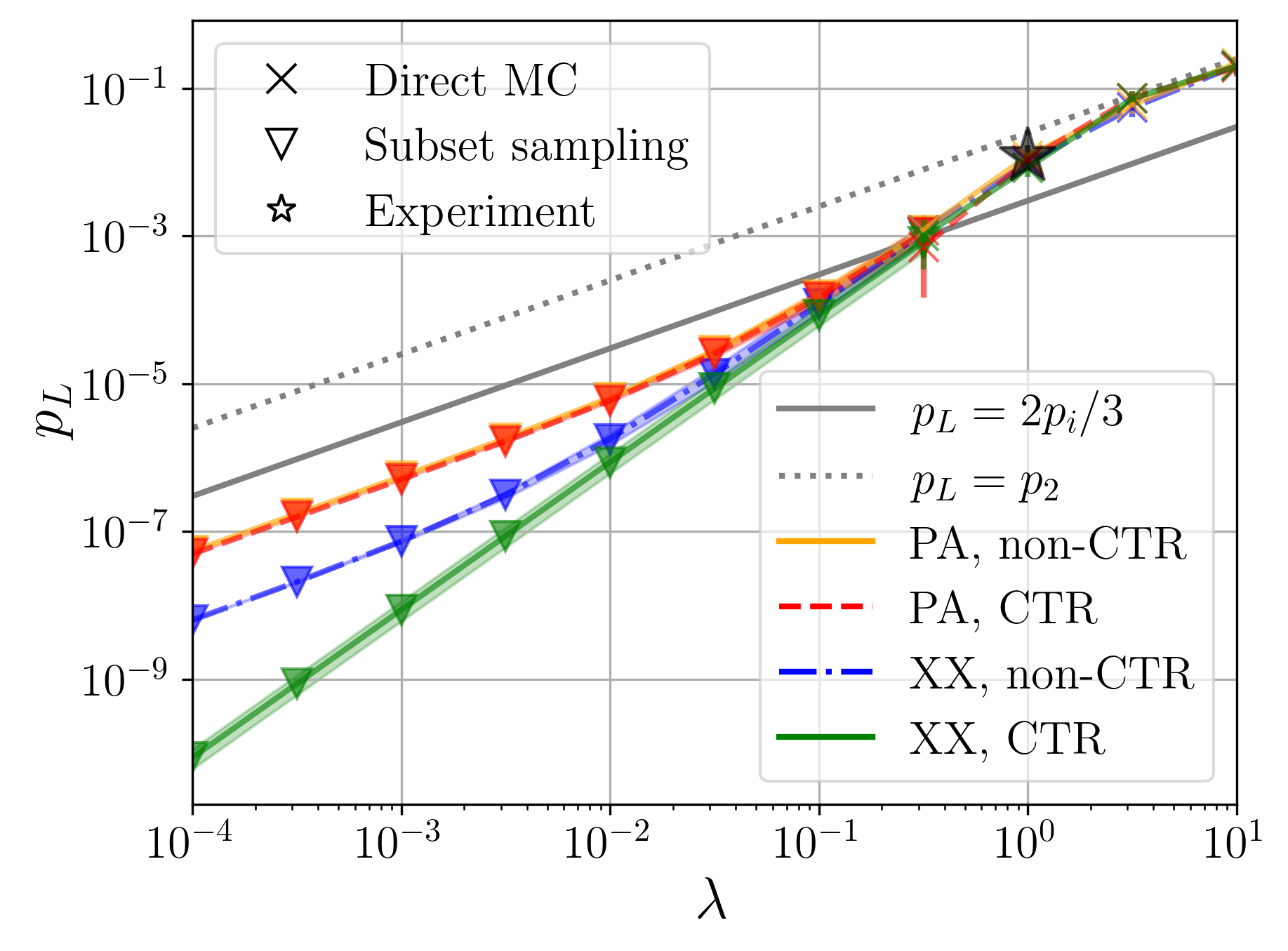}
    \caption{\textbf{Crosstalk-resistant Pauli state scaling.} Uniform scaling with a factor $\lambda$ of all physical error parameters under XX-type crosstalk (XX) and phase averaged crosstalk (PA) in the FT Pauli state preparation circuit with an XX-crosstalk-resistant (CTR) and non-crosstalk-resistant (non-CTR) qubit mapping. Lines for the two state preparations with PA overlap. For numerical simulations, we employ direct Monte Carlo (MC, cross markers) and subset sampling (SS, triangle markers) in their preferential domain of physical error rates. The experimentally measured value (star marker) lies at $\lambda = 1$. In this regime of physical error rates, all four curves coincide within their confidence intervals. At lower values of $\lambda$ crosstalk becomes a dominant source of failure causing linear scaling if CTR does not hold. Error rates of physical operations are shown for comparison (gray lines without markers) as in Fig.~\ref{fig:scaling}. For each MC data point and subset failure rate we sample at least 100 times or until the uncertainty of the respective logical failure rate estimator is below a relative error of 0.5 but at most $10^4$ times. All subsets up to $w_\text{max} = 3$ are taken into account for SS.}
    \label{fig:ctr_uniform}
\end{figure}

In Fig.~\ref{fig:ctr_uniform} we present the CTR property of the Pauli circuit and compare its logical failure rate to the non-deterministic, non-CTR FT Pauli state preparation from Fig.~\ref{fig:scaling}a. Extended noise is applied to both circuits. While, as before, there is no visible distinction between logical failure rates in an interval of approximately \mbox{$\lambda \in [10^{-1}, 10^{1}]$}, the non-CTR circuit transits from quadratic scaling to a linear scaling for $\lambda \lesssim 10^{-2}$ because crosstalk destroys the FT property. The CTR circuit continues to scale quadratically for all $\lambda \rightarrow 0$ under the influence of XX-type crosstalk on MS gates. XX-type crosstalk is only a valid description of the actual physical processes if crosstalk phases are zero on all ions. 

We have shown in Fig.~\ref{fig:crosstalk_phase} that in reality the crosstalk phases, although constant over time, vary over a large range of angles. To take this fact into account, we replace the XX-type crosstalk $\curlE_\text{xct}$ by the phase averaged crosstalk channel (Eq.~(\ref{eq:ct2av}), derived in App.~\ref{sec:noise_app}) which applies fault operators $XX,\,XY,\,YX$ and $YY$ with equal probability to each crosstalk location. In Fig.~\ref{fig:ctr_uniform} we show the scaling behavior for the same two circuits under the influence of the phase averaged (PA) crosstalk channel. Not only is the logical failure rate larger than for XX-type crosstalk alone. Now both the non-CTR and CTR circuit scale linearly at low $\lambda$ because the XY- and YY-type crosstalk faults can cause logical failures. The distinction between the two circuits is barely visible anymore. However, at $\lambda = 1$ all four circuit models agree with the experimentally measured value of logical infidelity.

To conclude this section, we note that the existence of the CTR Pauli encoding circuit is a special case which does not generalize to arbitrary quantum circuits. While fundamentally valid, the CTR characteristic cannot be upheld in our experimental setting since crosstalk phases will always mix the different X- and Y-type contributions even if they are constant over long times. As a consequence, it can not be guaranteed that the quadratic scaling behavior of FT circuits in the presence of crosstalk does actually lead to an advantage over physical qubits; minimization of crosstalk in physical operations is imperative. 

\section{Quantum state fidelity of logical qubits} \label{sec:fidelity}

While the logical fidelity is a good quantity to assess the degree of successful state preparation as a measure of operational performance in QEC, in this section we assess the quantum state of the logical qubit in a more general way by calculating its \emph{quantum state fidelity} \cite{nielsen2010quantum}.

The quantum state fidelity of a stabilizer state is defined as the mean of expectation values of all operators that form the stabilizer group $W_k$
\begin{align}
    \mathcal{F}(\rho_t,\rho) = \frac{1}{128} \sum_{k=1}^{128} \langle W_k \rangle 
\end{align}
with a target state $\rho_t = \ket{t}\bra{t}$ and $\rho = \ket{\psi}\bra{\psi}$ a stabilizer state such that $W_k\ket{\psi} = \pm \ket{\psi}$. The stabilizer group of the Steane code contains $128 = 2^7$ stabilizer operators and is generated by the stabilizer generators in Eq.~\eqref{eq:stabs} that define the logical qubit. The code space population $p_\text{CS}$ is defined analogously but only involves averaging over the 64 code space stabilizer expectation values
\begin{align}
    p_\text{CS} = \frac{1}{64} \sum_{k=1}^{64} \langle W_k \rangle
\end{align}
and contains no logical operators which would fix the logical state within the code space \cite{nigg2014quantum}. More detail on the derivation of the quantum state fidelity of stabilizer states is given in App.~\ref{sec:qsf}.

Since the largest physical error rate in our model is the infidelity of the MS gate, we expect the MS gate dynamics to dominantly influence the quantum state fidelity and, as a consequence, the logical failure rate. Thus, in the following, we compare the quantum state fidelity for noisy logical qubit preparation using depolarizing noise versus incoherent overrotation noise on MS gates. The MS gate is a rotation about the two-qubit XX-axis and it would thus be consistent to model MS gate noise by the overrotation channel given by Eq.~(\ref{eq:incoh_ovr}). The depolarizing noise channel is often used instead due to its general, hardware-agnostic structure but by introducing faults of all Pauli types it might overestimate the effect of MS gate errors compared to overrotations. It was previously expected that overrotation is the more accurate noise model \cite{li2017fault}. 

\begin{figure}
    \centering
    \includegraphics[width=\linewidth]{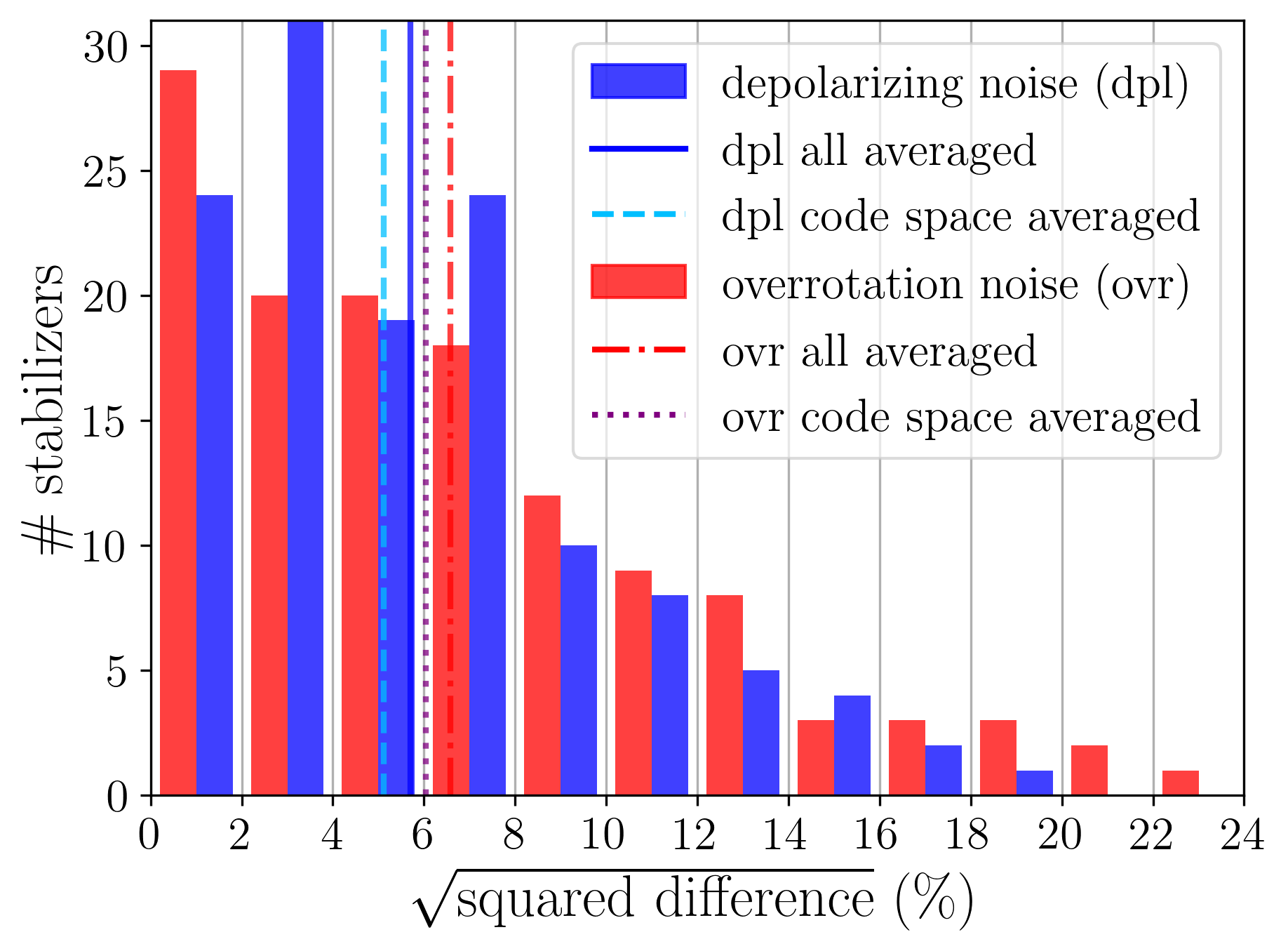}
    \caption{\textbf{Stabilizer estimation under different MS noise models.} Distributions for depolarizing (blue) and overrotation (red) noise models of the deviation $\sqrt{(\langle S_i \rangle_\text{exp}-\langle S_i \rangle_\text{sim})^2}$ of all 128 stabilizer expectation values of the logical qubit in simulation to experiment. Each bin has a width of $2\%$. Mean values which correspond to quantum state fidelities and code space population for both noise models are indicated as vertical lines and deviate from experimental values by approximately~$6\%$. Individual stabilizer expectation value estimates differ to up to $24\%$. All simulation data is generated by direct MC sampling until $10^5$ states are accepted. Each stabilizer has been measured 100 times in the experiment.}
    \label{fig:stab_sqdiff}
\end{figure}

The table below shows values for the quantum state fidelity $\mathcal{F}$ and code space population $p_\text{CS}$ with 95\% confidence intervals of a single logical qubit in the $\ket{0}_L$ state prepared by the FT circuit in Fig.~\ref{fig:circ_pauli}.
\begin{center}
\begin{tabular}{ c | c | c}
noise & $\mathcal{F}$ & $p_\text{CS}$ \\ \hline 
depolarizing & $82.63(3)\%$ & $82.62(4)\%$ \\ 
overrotation & $86.18(3)\%$ & $86.20(4)\%$ \\ \hline 
experiment & $82.7(11)\%$ & $83.1(15)\%$
\end{tabular} \label{tab:qsfpcs}
\end{center}

In Fig.~\ref{fig:stab_sqdiff} we compare experimental data to numerical simulations with the depolarizing noise model as described before for a single logical qubit with MS gate errors modeled as either depolarizing or overrotation noise. For each of the stabilizer operators we determine the deviation $\sqrt{(\langle S_i \rangle_\text{exp}-\langle S_i \rangle_\text{sim})^2}$ of the simulated expectation value $\langle S_i \rangle_\text{sim}$ with both noise models to the experimentally measured expectation value $\langle S_i \rangle_\text{exp}$. We observe that the distribution of deviations is very similar for both noise channels. While most stabilizer expectation values deviate little from the experimental values, individual stabilizer expectation value deviations can be as high as approximately $20\%$ for depolarizing noise and $24\%$ for overrotation noise. The averaged deviations (RMS) for all 64 or 128 expectation values, i.e.~for the code space population and quantum state fidelity respectively, are $6.0(15)\%$ and $6.6(11)\%$ with overrotation noise but for depolarizing noise yield the lower values of $5.1(15)\%$ and $5.7(11)\%$.

It is evident that incoherent overrotation noise does not provide a more accurate description of MS gate errors than depolarizing noise for the circuits used in our experiment. Respecting the FT property of the state preparation circuit on the logical level appears to be the more relevant characteristic of noise than its microscopic structure. This is in stark contrast to the detrimental effect that crosstalk can exert when it does not respect fault tolerance. The effect of crosstalk strongly depends on the microscopic structure which differs between the XX-type and phase average model discussed in the previous section. 
We stress that the logical fidelity is an appropriate quantity to compare the agreement of experimental data to noise simulations and that computing the full quantum state fidelity does not provide additional information about the QEC procedure.

\section{Conclusions \& Outlook}

We provided a detailed numerical study and analysis of future potential for FT universal gate set implementations. Incoherent Pauli noise simulations suggest that reaching thresholds of FT advantage over physical qubits need improvements on physical error rates of less than an order of magnitude. Currently the logical error rate is limited predominantly by entangling gate errors in the experimental setup under consideration in this work. Crosstalk on M{\o}lmer-S{\o}rensen gates is not a substantial source of error for the advantage of FT over non-FT circuit implementation in our ion trap architecture at current noise levels. However, we give a crosstalk resistant qubit mapping for FT Pauli state preparation which keeps scaling quadratically under XX-type crosstalk as physical error rates are scaled to zero opposed to the usual circuits where crosstalk typically breaks the FT property. We showed that the microscopic structure of crosstalk affects the scaling of logical error rates. 

Therefore choosing a different set of physical gates could also make available crosstalk resistant circuits for the realization of other logical building blocks. Furthermore, crosstalk errors could be suppressed by utilizing inherently crosstalk insensitive gate operations like composite pulses~\cite{fang2022crosstalk, torosov2022narrowband}, or active suppression schemes, where additional laser fields are applied to the qubit register that destructively interfere with unwanted leakage light at neighboring ion positions.
Exploiting the fact that for each ion a global phase can be freely chosen might allow for crosstalk resistant qubit mappings even in the case of random but constant crosstalk phases. However, this method does not provide enough degrees of freedom to directly control the effective crosstalk phase for both neighbors of all qubits in the register. Further investigations are needed to clarify if crosstalk resistant mappings for various logical building blocks can be found using this technique.

Also, we have found that deterministic state preparation schemes for Pauli and magic state preparation do not outperform non-deterministic ones at current physical error rates and are not expected to do so even with improvements on physical error rates due to their larger gate count. The repetition overhead needed for non-deterministic state preparation is moderate for both the Pauli and magic state at current noise levels. 

Our analysis validates depolarizing noise as an appropriate effective model for FT logical state preparation in the ion trap system from Ref.~\cite{pogorelov2021compact}. Flag circuits are recognized as a promising paradigm to reach the break-even point where FT circuits will outperform physical qubits \cite{Ryan-Anderson2022}. Not only is the depolarizing noise model sufficient to predict logical failure rates but also the average over stabilizer expectation values for a single logical qubit initialized to its logical zero state. Individual stabilizer expectation values can be estimated to about $24\%$ relative uncertainty. The detailed crosstalk investigation provided in this work illustrates the value of considering aspects specific to the physical architecture realizing the quantum computer. We point out that for long protocols with deep circuits such as the deterministic FT magic state preparation scheme, coherent errors might build up and cause an additional source of logical failure. The effect of coherent noise to the logical failure rate of such circuits is a subject for further studies. 

In the future, effective noise models for different quantum computing architectures and logical building blocks will aid in the characterization of fault-tolerant universal quantum computers. Simulating large distance logical qubits can help to better understand relevant error processes and facilitate practical realization of error-corrected logical qubit operations below the pseudothreshold.

 \section*{Data availability}
The data underlying the findings of this work are available at \url{https://doi.org/10.5281/zenodo.7565571}.

\section*{Code availability}
All codes used for data analysis are available from the corresponding author upon reasonable request.

\section*{Author contributions}
S.H. performed the numerical simulations and analyzed the data. S.H., M.R. and M.M. performed characterization and theory modeling. L.P. and I.P. carried out the experiments. L.P., I.P., C.D.M., P.S. and T.M. contributed to the experimental setup. S.H., L.P. and M.R. wrote the manuscript, with contributions from all authors. T.M., P.S. and M.M. supervised the project.

\section*{Acknowledgements}
We gratefully acknowledge support by the EU Quantum Technology Flagship grant under Grant Agreement No. 820495 (AQTION), the US Army Research Office through grant number W911NF-21-1-0007, the European Union’s Horizon Europe research and innovation program under Grant Agreement No. 101046968 (BRISQ), the ERC Starting Grant QCosmo under Grant No. 948893, the ERC Starting Grant QNets through
Grant Number 804247, the Austrian Science Fund under Project No. F7109 (SFB BeyondC), the Austrian Research Promotion Agency under Contracts No. 896213 (ITAQC) and by the Office of the Director of National Intelligence (ODNI), Intelligence Advanced Research Projects Activity (IARPA), via the U.S. Army Research Office through Grant No. W911NF-16-1-0070. This research is also part of the Munich Quantum Valley (K-8), which is supported by the Bavarian state government with funds from the Hightech Agenda Bayern Plus. We further receive support from the IQI GMBH. The views and conclusions contained herein are those of the authors and should not be interpreted as necessarily representing the official policies or endorsements, either expressed or implied, of the ODNI, IARPA, or the US Government. The US Government is authorized to reproduce and distribute reprints for Governmental purposes notwithstanding any copyright annotation thereon. Any opinions, findings, and conclusions or recommendations expressed in this material are those of the author(s) and do not necessarily reflect the view of the US Army Research Office. S.H. would like to thank Josias Old for valuable discussions on deterministic preparation circuits. S.H. acknowledges funding by the Deutsche Forschungsgemeinschaft (DFG, German Research Foundation) under Germany’s Excellence Strategy ‘Cluster of Excellence Matter and Light for Quantum Computing (ML4Q) EXC 2004/1’ 390534769. The numerical simulations were performed with the aid of computing resources at Forschungszentrum Jülich.
 
\appendix

\section{Noise model details} \label{sec:noise_app}
In the following, we provide more details for the noise model used to perform the simulations of faulty quantum circuits presented in Sec.~\ref{sec:noise_main} of the main text. The four independent physical error rates on single-qubit gates, two-qubit gates, qubit initialization and measurement are the sources of error in the simulations accompanying the experimental fault-tolerant universal gate set realization \cite{postler2022demonstration}. For the extended noise model, we also include dephasing noise on idling qubits as well as crosstalk on single- and two-qubit gates. For the latter, we provide two different descriptions, namely as a coherent noise channel and as an incoherent Pauli channel. Overrotations on Mølmer-Sørensen (MS) gates are also considered in both a coherent and incoherent model. The derivation of generalized crosstalk noise on gates with arbitrary laser phase, Eqs.~(\ref{eq:ct1av}) and (\ref{eq:ct2av}), is the main focus of this Appendix.

The noise channels we state below are examples of quantum operations $\curlE$ which map an initial qubit state $\rho$ to a final state $\rho' = \curlE(\rho)$ and thus allow to formalize evolution of a state under noise. We may express $\curlE$ as a Kraus map
\begin{align}
    \curlE(\rho) &= \sum_i K_i \rho K_i^\dagger
\end{align}
where the Kraus operators $K_i$ describe the noise on $\rho$. 

As discussed in Sec.~\ref{sec:noise_main}, all rotation axes of physical gate operations are parametrized by the phase(s) of the respective qubit laser(s). In the following, we elaborate on the realization of single- and two-qubit gate rotations about axes parametrized by the laser phase(s) which we put to use for the FT magic state preparation circuit in Fig.~\ref{fig:circ_magic_ft}. It is compiled from a circuit built from CNOT gates to a circuit containing only MS gates. The compiled circuit then contains single-qubit $Z$-rotations which need not be performed physically in the ion trap system, e.g.~by AC Stark shifts. All rotation axes, and therefore laser phases, for subsequent gates are changed in order to propagate a $Z$-rotation until the end of the circuit~\cite{mckay2017efficient}. Here they can be accounted for in software (and when measuring in the $Z$-basis they can be omitted entirely). In order to take advantage of this, we need to allow for different phases $\varphi_1$ and $\varphi_2$ on the MS target ions and vary the phase $\varphi$ for single-qubit rotations. We now give a generalization of the standard Pauli-type single- and two-qubit rotations, also including the case of crosstalk. The standard Pauli $X$- and $Y$-gates and the XX-type MS gate shown in Fig.~\ref{fig:architecture}b will be recovered as special cases from this general discussion.

\subsection{Single-qubit gates}

Single-qubit rotations are parametrized as a unitary evolution with the operator
\begin{align}
	R_\varphi(\theta) &= \exp\left(-\ii \frac{\theta}{2} \sigma_\ph \right) \\
	\sigma_\ph &= X \cos \varphi + Y \sin \varphi
\end{align}
where $\sigma_\ph$ describes the rotation axis in the equatorial plane of the Bloch sphere. For example, one recovers the $X(Y)$-gate for $\varphi = 0 (\pi/2)$ and $\theta = \pi$. With $\varphi = \pi/4$ the resulting spin operator is $\sigma_{\pi/4} = \frac{X+Y}{\sqrt{2}}$, implementing a non-Clifford rotation.

Crosstalk occurs on gates when the laser light intended to only shine on ions in order to perform a qubit rotation cannot be focused tightly enough so that a finite electric field is at the position of a non-targeted ion. Then, neighboring ions also receive a fraction of residual laser light and the rotation intended to the gate ions is partly performed as well on the neighbor ions. The coupling of the laser field $\vec{E}$ to the electric quadrupole of the ion state is measured by the Rabi frequency $\Omega$. The Rabi frequency $\Omega$ is proportional to the gradient of the electric field at the location of the neighbor ion. We assume that the main contribution to the gradient of the electric field is given by the longitudinal change in electric field of the electromagnetic wave. Therefore the gradient is proportional to the amplitude of the electric field amplitude. Consequently also the Rabi frequency on a neighbor ion $\Omega_n$ is proportional to the electric field amplitude at the location of the neighbor ion. Since the rotation angle $\theta$ of the single-qubit gate is given by $\theta = \Omega t$, where $t$ is the time the laser light is on, the rotation angle on the neighbor qubit $\theta_n$ is determined by the crosstalk ratio $\varepsilon = \Omega_n/\Omega$ via 
\begin{align}
    \theta_n &= \varepsilon \theta.
\end{align}
In our simulations, we assume an average crosstalk ratio of $\varepsilon = 1 \times 10^{-2}$.

For single-qubit crosstalk, the neighboring ions to the target ion, where a rotation about $\theta$ shall be performed, see residual laser light which causes the crosstalk rotation of angle $\varepsilon \theta$. The rotation on each neighbor location is
\begin{align}
	R_\varphi(\varepsilon \theta) &= \exp \left( -\ii \frac{\varepsilon \theta}{2} \sigma_\ph \right).
\end{align}
The rotation operator $R_\varphi(\varepsilon \theta)$ acts on a single-qubit density matrix $\rho$ like
\begin{alignat}{3}
	\curlE(\rho) &= R_\varphi(\varepsilon \theta)\,\rho\,&R^\dagger_\varphi(\varepsilon \theta) & \notag \\
    &= \cos^2 \frac{\varepsilon \theta}{2} \rho &+ \sin^2 &\frac{\varepsilon \theta}{2} \left( \sigma_\ph \rho \sigma_\ph \right) + \frac{\ii}{2} \sin \varepsilon\theta\, \left[ \rho, \sigma_\ph \right] \notag \\
    &= \cos^2 \frac{\varepsilon \theta}{2} \rho &+ \sin^2 &\frac{\varepsilon \theta}{2} \left( \cos^2 \varphi X \rho X + \sin^2 \varphi Y \rho Y \right. \notag \\ 
    & & &+ \left. \cos \varphi \sin \varphi (X\rho Y + Y\rho X) \right) \notag \\ 
	& &+ \frac{\ii}{2} \sin\, &\varepsilon\theta\,\left( \rho (X \cos \varphi + Y \sin \varphi) \right. \notag  \\ 
    & & &- \left. (X \cos \varphi + Y \sin \varphi) \rho \right). \label{eq:ct1channel}
\end{alignat}
In order to efficiently simulate the above coherent noise channel $\curlE$ in a stabilizer simulation, we now perform the Pauli twirling approximation (PTA) \cite{bennett1996mixed, emerson2007symmetrized, silva2008scalable, dankert2009exact, geller2013efficient} to obtain the (approximate) incoherent channel of the form
\begin{align}
	\tilde{\curlE}(\rho) &= \frac{1}{4} \sum_{P \in \mathcal{P}} P \curlE(P \rho P) P \label{eq:twappr_}
\end{align}
with $\mathcal{P} = \{I, X, Y, Z\}$. Each term in the sum of Eq.~(\ref{eq:twappr_}) of the channel $\tilde{\curlE}(\rho)$ reads
\begin{align}
	P \curlE(P \rho P) P &= \cos^2 \frac{\varepsilon \theta}{2} \rho + \sin^2 \frac{\varepsilon \theta}{2} P \sigma_\ph P \rho P \sigma_\ph P \notag \\ 
    &+ \frac{\ii}{2} \sin \varepsilon\theta \left[ \rho,  P \sigma_\ph P \right] \label{eq:twirlcontrib}
\end{align}
for any Pauli matrix $P \in \mathcal{P}$. With the identities
\begin{align}
	I \sigma_\ph I &= X \cos \varphi + Y \sin \varphi \\
	X \sigma_\ph X &= X \cos \varphi - Y \sin \varphi \\
	Y \sigma_\ph Y &= -X \cos \varphi + Y \sin \varphi \\
	Z \sigma_\ph Z &= -(X \cos \varphi + Y \sin \varphi)
\end{align}
we can calculate the twirled channel. We calculate the sum over the Paulis for each of the three terms in Eq.~(\ref{eq:twirlcontrib}) separately to find the Pauli twirled channel
\begin{align}
	\tilde{\curlE}(\rho) &= \cos^2 \frac{\varepsilon \theta}{2} \rho + \sin^2 \frac{\varepsilon \theta}{2} \left( \cos^2 \varphi X \rho X + \sin^2 \varphi Y \rho Y \right) \label{eq:ct1channeltwirled} \\
	&\equiv (1-p_{\text{c}_1}) \rho + p_{\text{c}_1} \left( r_x X \rho X + (1-r_x) Y \rho Y \right) \end{align}
where we define the physical error rate as before but also introduce the \emph{phase ratios} $r_x = \cos^2\varphi$ and \mbox{$r_y = 1 - r_x = \sin^2\varphi$}. All terms in the commutator and the off-diagonal terms in the $\sin^2$-term cancel. This directly corresponds to taking only the diagonal terms of the process matrix $\chi$ parametrizing the coherent channel of Eq.~(\ref{eq:ct1channel}) in the Pauli basis
\begin{widetext}
\begin{align}
    \chi =
	\begin{pmatrix}
		\cos^2 \nicefrac{\varepsilon \theta}{2} & \nicefrac{\ii}{2}\sin\varepsilon\theta\cos\varphi & \nicefrac{\ii}{2}\sin\varepsilon\theta\sin\varphi  & 0 \\
		-\nicefrac{\ii}{2}\sin\varepsilon\theta\cos\varphi  & \sin^2 \nicefrac{\varepsilon \theta}{2} \cos^2 \varphi & \sin^2 \nicefrac{\varepsilon \theta}{2} \cos \varphi \sin \varphi & 0 \\
		-\nicefrac{\ii}{2}\sin\varepsilon\theta\sin\varphi  & \sin^2 \nicefrac{\varepsilon \theta}{2} \cos\varphi \sin\varphi & \sin^2 \nicefrac{\varepsilon \theta}{2} \sin^2 \varphi & 0 \\
		0 & 0 & 0 & 0
	\end{pmatrix}.
\end{align}
\end{widetext}

As an example for crosstalk on a single-qubit Pauli gate, consider the coherent rotation about a Pauli axis $\sigma \in \{X, Y\}$ (realized via $\varphi \in \{0, \pi/2\}$) as described by the operator
\begin{align}
	R_{\sigma}(\theta) &= \cos \frac{\theta}{2} - \ii \sin \frac{\theta}{2} \sigma. \label{eq:rotop}
\end{align}
With a laser beam that affects three qubits, the target ion $t$ and its two neighbor ions $n_1(t)$ and $n_2(t)$ that are subjected to a fraction $\varepsilon$ of the laser electric field, the total rotation operator is the product of three single-qubit rotations 
\begin{align}
	&R^{(n,t)}_{\sigma}(\theta) \notag \\
    &= \exp\left(-\ii\frac{\theta}{2} \sigma_t\right) \exp\left(-\ii\frac{\varepsilon\theta}{2} \sigma_{\varphi_{n_1(t)}}\right) \exp\left(-\ii\frac{\varepsilon\theta}{2} \sigma_{\varphi_{n_2(t)}}\right)
\end{align}
where the rotation axes of the neighbor ions are determined by the Pauli operators $\sigma_{\varphi_{n_1(t)}}$ and $\sigma_{\varphi_{n_2(t)}}$ independently from the Pauli operator on the target ion.
Let us assume that the phase on neighbor $n_1(t)$ is $\varphi_{n_1(t)} = \pi/2$ so that a $Y$-rotation will be performed. The corresponding rotation operator transforms the state $\rho$ like
\begin{align}
	&R^{(n_1)}_{Y}(\theta)\,\rho\,R^{(n_1)}_{Y}(\theta)^\dagger = \exp\left(-\ii\frac{\varepsilon\theta}{2} Y_{n_1}\right)\,\rho\,\exp\left(+\ii\frac{\varepsilon\theta}{2} Y_{n_1}\right) \notag \\
	&= \cos^2  \frac{\varepsilon\theta}{2} \rho + \sin^2  \frac{\varepsilon\theta}{2} Y_{n_1} \rho Y_{n_1} + \frac{\ii}{2} \sin \varepsilon \theta\, \left[ \rho, Y_{n_1} \right].
\end{align}
Performing the PTA to this transformation amounts to neglecting the third term containing the commutator. The Pauli-twirled channel is then an incoherent error channel of the form
\begin{align}
	\curlE(\rho) &= (1-p_{\text{c}_1}) \rho + p_{\text{c}_1} Y \rho Y
\end{align}
for the respective neighbor ion location and the probability
\begin{align}
    p_{\text{c}_1} &= \sin^2 \frac{\varepsilon\theta}{2} \label{eq:psin_ct1}
\end{align}
of applying the crosstalk fault operator $Y$. 

Since we observe that the phases in Fig.~\ref{fig:crosstalk_phase} are distributed across the whole interval of all possible values $\varphi \in [0, 2\pi]$, we use 
\begin{align}
    \int_0^{2\pi} \dd\varphi \cos^2\varphi = \int_0^{2\pi} \dd\varphi \sin^2\varphi = \pi \label{eq:intphi}
\end{align}
to average over the crosstalk phase $\varphi$ in Eq.~(\ref{eq:ct1channeltwirled}):
\begin{align}
	\langle \tilde{\curlE} \rangle_\varphi (\rho) &= (1-p_{\text{c}_1}) \rho \notag \\ 
    &+ \frac{p_{\text{c}_1}}{2\pi} \int_0^{2\pi} \dd\varphi\, \left( \cos^2 \varphi X \rho X + \sin^2 \varphi Y \rho Y \right).\label{eq:ct1channeltwirledaveraged}
\end{align}
From this we obtain the incoherent noise channel
\begin{align}
	\curlE_{\text{c}_1}(\rho) &= (1-p_{\text{c}_1}) \rho + \frac{p_{\text{c}_1}}{2} \left( X \rho X + Y \rho Y \right)
\end{align}
which we use in our numerical simulations.

Note that for this channel the physical crosstalk error rate $p_{\text{c}_1} = p_{\text{c}_1}(\theta)$ depends on the rotation angle of the gate as opposed to the depolarizing or our dephasing channel. The quantum circuits in this work contain rotation angles $\theta \in \{\pi, \pi/2, \pi/4\}$ for which we list the approximate probabilities according to Eq.~\eqref{eq:psin_ct1} in the table below. 
\begin{center}
\begin{tabular}{ c | c}
rotation angle $\theta$ & physical error rate $p_{\text{c}_1}$ \\ \hline 
$\pi$ & $2.5 \times 10^{-4}$ \\ 
$\pi/2$ & $6.2 \times 10^{-5}$ \\
$\pi/4$ & $1.5 \times 10^{-5}$  
\end{tabular} \label{tab:p_ct1}
\end{center}
For the incoherent channel, both neighbor ions $n_1$ and $n_2$ have their own independent single-qubit crosstalk error channel. 

\subsection{MS gates}

The two-qubit entangling gate in our trapped-ion architecture is the MS gate. We now provide a derivation of our noise model for crosstalk on MS gates based on the gate Hamiltonian. The Hamiltonian of the MS gate reads
\begin{align}
	H(t) &= H_0 + H_\text{int}(t) \\
	H_0 &= \sum_{j=1}^{Q} \frac{\omegaeg}{2} \sigma_{z,j} + \nu \left( a^\dagger a + \frac{1}{2} \right) \\
	H_\text{int}(t) &= \sum_{j=1}^{Q} \frac{\Omega_j(t)}{2} \left( e^{\ii \left( \vec{k}_1\vec{x}_j - \left( \omegaeg + \delta \right) t - \ph_j \right)}  \right. \notag \\
    &+ \left. e^{\ii \left( \vec{k}_2\vec{x}_j - \left( \omegaeg - \delta \right) t - \ph_j \right)} + \hc \right) \left(\sigma_j^+ + \sigma_j^-\right).
\end{align}
with $\sigma_j^\pm = (X_j \pm \ii Y_j)/2$. Here $Q$ is the number of all ions that laser light shines on and the $\Omega_j$ are their respective Rabi frequencies. Using $ \vec{k}_i\vec{x} = \eta_i(a^\dagger + a)$, we operate in a regime where the detuning $\delta \ll \omegaeg$ is much smaller than the qubit frequency so that the Lamb-Dicke parameters $\eta_1,\,\eta_2 \approx \eta$ are assumed to be the same for both target ions 1 and 2. With the rotated spin operator
\begin{align}
	\sigma_{\ph_j} &= X_j \cos \ph_j + Y_j \sin \ph_j
\end{align}
we can write the sum over the ions explicitly as MS gate target ions $t' \in \{1, 2\}$ and neighbor ions $n \in \{n_1(1), n_2(1), n_1(2), n_2(2)\}$ with their Rabi frequencies $\Omega_{t'}(t) = \Omega$ and $\Omega_n = \varepsilon\Omega$:
\begin{align}
	H_\text{int}(t) &\approx -\eta \Omega \left( a e^{-\ii \epsilon t} + a^\dagger e^{\ii \epsilon t} \right) \left( \sum_{t'} \frac{1}{2} \sigma_{\ph_{t'}} + \sum_n \frac{\varepsilon}{2} \sigma_{\ph_n} \right)
\end{align}
where $\epsilon = \nu - \delta$. The final form of the Hamiltonian can now be expressed as
\begin{align}
	H_\text{int}(t) &= -\eta \Omega \left( a e^{-\ii \epsilon t} + a^\dagger e^{\ii \epsilon t} \right) S_{\vec{\ph}}
\end{align}
with the collective spin operator $S_{\vec{\ph}} = \frac{1}{2} \sigma_{\vec{\ph}}$ where $\vec{\ph} = (\ph_{t_1},\,\ph_{t_2},\,\ph_{n_1(1)},\,\ph_{n_2(1)},\,\ph_{n_1(2)},\,\ph_{n_2(2)})$ contains all target and neighbor ion phases. 

From this Hamiltonian follows the time evolution 
\begin{align}
	U(t) &= D(\Gamma(t) \sigma_{\vec{\ph}}) \exp \left( \ii \theta(t) S_{\vec{\ph}}^2 \right)
\end{align}
with $\Gamma(t) = \int_0^t \gamma(t') \dd t'$ and $\theta(t) = \Im \int_0^t \gamma(t') \dd t' \int_0^{t'} \gamma^*(t'') \dd t''$ where $\gamma(t) = \ii \eta \Omega e^{\ii \epsilon t}$ and the displacement operator $D(\alpha) = \exp \left( \alpha a^\dagger - \alpha^* a \right) \sim 1 + \left( \alpha a^\dagger - \alpha^* a \right)$ for which $D(\alpha)D(\beta) = D(\alpha + \beta) \exp \left( \ii\,\Im (\alpha\beta^*) \right)$ holds. The parameters of the gate $\Gamma(t)$ and $\theta(t)$ can be adjusted experimentally to realize the MS gate \cite{martinez2022analytical}.

The collective spin operator contains both the target and their nearest neighbor ions 
\begin{align}
	S_{\vec{\varphi}} &= S_{\vec{\varphi}}\Bigr|_\text{targets} + S_{\vec{\varphi}}\Bigr|_\text{neighbors} \label{eq:spop_twoterms} \\
    &= \frac{1}{2} \left(  \sigma_{\ph_1} + \sigma_{\ph_2} \right. \notag \\ 
    &+ \left. \varepsilon\left( \sigma_{\ph_{n_1(1)}} + \sigma_{\ph_{n_2(1)}} + \sigma_{\ph_{n_1(2)}} + \sigma_{\ph_{n_2(2)}}\right) \right) \label{eq:spop_twotermsexpanded}
\end{align}
where the latter have their Rabi frequencies suppressed by the crosstalk ratio $\varepsilon$. Squaring $S_{\vec{\varphi}}$ will create all combinations of target and neighbor ions in first order of $\varepsilon$ which we can as well express as
\begin{align}
	S_{\vec{\varphi}}^2 &= S^2_{\vec{\varphi}}\Bigr|_\text{intended} + S^2_{\vec{\varphi}}\Bigr|_\text{crosstalk}. \label{eq:spsq_twoterms}
\end{align}
The MS gate 
\begin{align}
    \ms_{\vec{\ph}}(\theta) &= \exp\left(-\ii \theta S_{\vec{\ph}}^2 \right) \label{eq:msgateunitary}
\end{align}
transforms the state $\rho$ as
\begin{align}
    \curlE(\rho) &= \exp\left(-\ii \theta S_{\vec{\ph}}^2 \right)\,\rho\,\exp\left(\ii \theta S_{\vec{\ph}}^2 \right). \label{eq:msgatechannel}
\end{align}

The intended part realizes the MS gate rotation on the target ions. The unitary evolution, which describes the intended MS gate, then reads
\begin{align}
	\ms_{\varphi_1, \varphi_2}(\theta) &= \exp\left(-\ii \theta S_{\varphi_1, \varphi_2}^2 \right) \label{eq:msgateintended}
\end{align}
with the spin operator
\begin{align}
    S_{\varphi_1, \varphi_2} &= \frac{1}{2} \left( \sigma_{\ph_1} + \sigma_{\ph_2} \right).
\end{align}
The MS interaction originates from the square of the spin operator
\begin{align}
	S_{\varphi_1, \varphi_2}^2 &\stackrel{\sim}{=} \frac{1}{2} \sigma_{\ph_1} \sigma_{\ph_2}
\end{align}
where we have omitted terms which either sum to zero as the Pauli operators anticommute on the same qubit or square to the identity and thus only contribute an irrelevant global phase. For the case $\varphi_1 = \varphi_2 = 0$ we find the usual XX-type MS-gate
\begin{align} 
	\ms_{0,0}(\theta) &= \exp \left( -\ii \frac{\theta}{4} (X_1 + X_2)^2 \right) \\
	&\overset{\sim}{=} \exp \left( -\ii \frac{\theta}{2} X_1X_2 \right) \label{eq:ms0_unitary} \\
	&= \cos \frac{\theta}{2} -\ii \sin \frac{\theta}{2} X_1 X_2
\end{align}
which has the same form as Eq.~(\ref{eq:rotop}) with $\sigma = X_1X_2$. Another gate relevant to our simulations is, for example, the non-Clifford gate
\begin{align}
	\ms_{0, \pi/4}(\theta) &= \exp \left( -\ii \frac{\theta}{2} X_1 \left( \frac{X_2 + Y_2}{\sqrt{2}} \right) \right)
\end{align}
which appears in the circuit for deterministic FT magic state preparation in Fig.~\ref{fig:circ_magic_ft}. Here, the identities used for propagation of $Z$-rotations to the end of the circuit are
\begin{align}
	\ms_{0,0}(-\pi/2) R_Z^{(t_1)}(\alpha) &= R_Z^{(t_1)}(\alpha) \ms_{-\alpha,0}(-\pi/2) \label{eq:msz}\\
	R_\varphi(\theta) R_Z(\alpha) &= R_Z(\alpha) R_{\varphi-\alpha}(\theta). \label{eq:rz}
\end{align}

\begin{figure}
    \centering
    \includegraphics[width=0.8\linewidth]{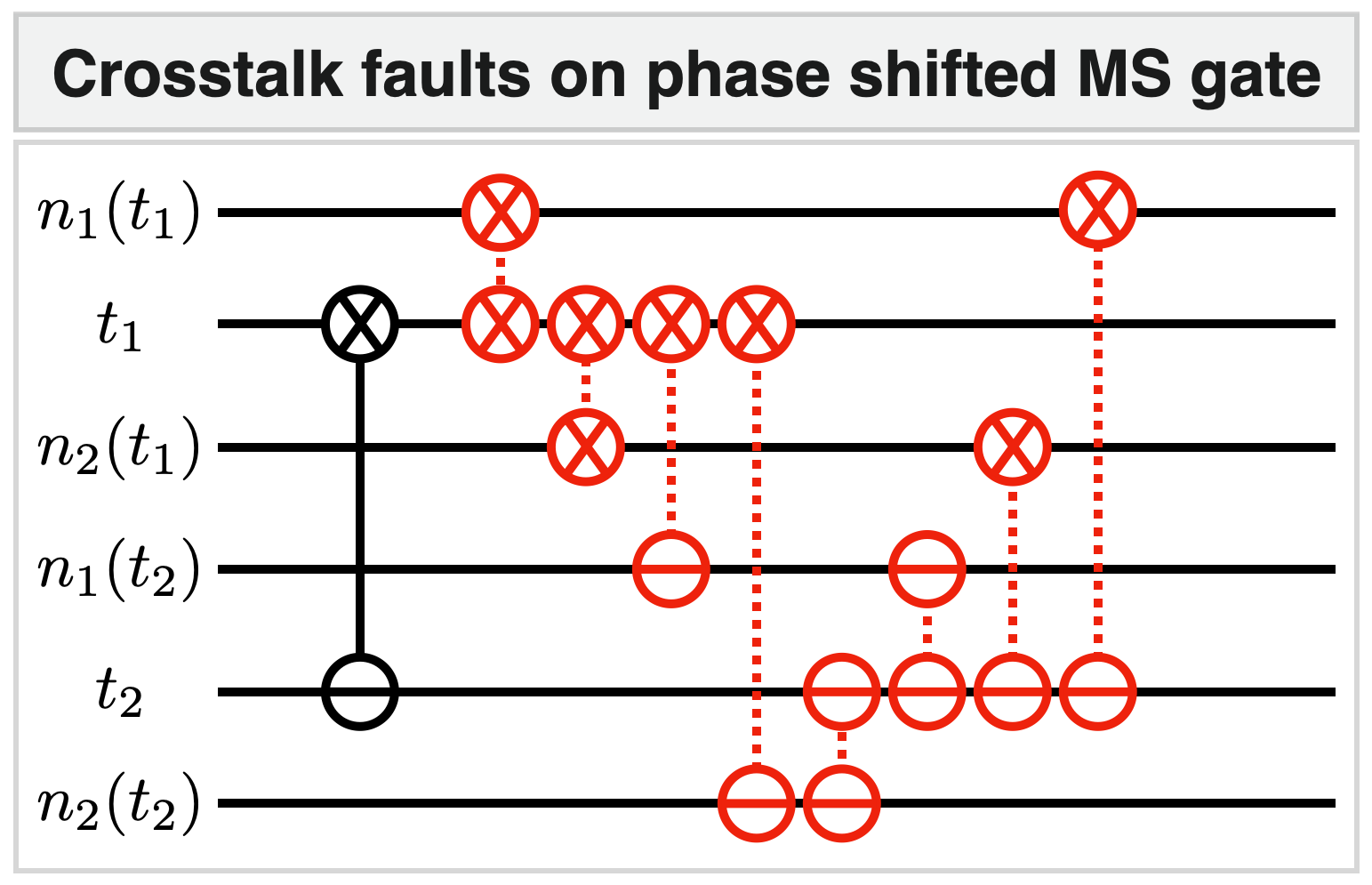}
    \caption{\textbf{Crosstalk faults on phase shifted MS gate.} Fault locations (red, dotted lines) for an $\ms_{0, -\pi/4}\left(-\frac{\pi}{2}\right)$ gate (black, solid vertical line) originating from the square of the spin operator in Eq.~(\ref{eq:spsq_F}). The phase of the crosstalk corresponds to the phase of the associated target ion. }
    \label{fig:circ_ctfaults}
\end{figure}

The crosstalk term in Eq.~\eqref{eq:spsq_twoterms} contains all two-combinations of single-qubit operators in order $\varepsilon$, as depicted as an example in Fig.~\ref{fig:circ_ctfaults}. Neglecting higher orders of $\varepsilon$, each crosstalk location can be treated as an independent coherent two-qubit rotation. For example, the location $t_1,\,n_1(t_1)$ is contained in the squared spin operator as
\begin{align}
	S^2_{\vec{\varphi}}\Bigr|_\text{crosstalk} \supset \frac{\varepsilon}{2} \sigma_{\ph_1} \sigma_{\ph_{n_1(1)}} \label{eq:spsq_onelocation}
\end{align}
and generates the rotation
\begin{align}
	R_{1, n_1(1)} &= \exp\left(-\ii\frac{\varepsilon}{2}\theta \sigma_{\ph_1} \sigma_{\ph_{n_1(1)}} \right). \label{eq:twoqubitrotation}
\end{align} 
For any crosstalk location $t,\,n$ we describe its independent unitary evolution by the coherent channel
\begin{align}
	\curlE(\rho) &= R_{t, n}(\varepsilon \theta)\,\rho\,R^\dagger_{t, n}(\varepsilon \theta) \\
    &= \exp\left(-\ii\frac{\varepsilon}{2}\theta \sigma_{\ph_t} \sigma_{\ph_{n}} \right) \,\rho\, \exp\left(\ii\frac{\varepsilon}{2}\theta \sigma_{\ph_t} \sigma_{\ph_{n}} \right)
\end{align}
(analogously to Eq.~\eqref{eq:ct1channel}).
Denoting arbitrary two-qubit Pauli operators \mbox{$P_2 \in \mathcal{P} \otimes \mathcal{P}$}, we can perform the Pauli twirling analogously to the single-qubit crosstalk by calculating the 16 expressions $P_2 \curlE(P_2 \rho P_2)P_2$. The resulting incoherent channel for one MS crosstalk location (a red gate in Fig.~\ref{fig:circ_ctfaults}) is
\begin{align}
	\tilde{\curlE}(\rho) &= \cos^2 \frac{\varepsilon \theta}{2} \rho \notag \\ 
    &+ \sin^2 \frac{\varepsilon \theta}{2} \left( \cos^2 \varphi_t \cos^2 \varphi_n X_t X_n \rho X_t X_n \right. \notag \\ 
    &~~~~~~~~~~~~+ \left. \sin^2 \varphi_t \sin^2 \varphi_n Y_t Y_n \rho Y_t Y_n \right. \notag \\ 
	&~~~~~~~~~~~~+ \left. \cos^2 \varphi_t \sin^2 \varphi_n X_t Y_n \rho X_t Y_n \right. \notag \\ 
    &~~~~~~~~~~~~+ \left. \sin^2 \varphi_t \cos^2 \varphi_n Y_t X_n \rho Y_t X_n \right)
\end{align}
where we can now define the incoherent noise channel
\begin{align}
	\tilde{\curlE}(\rho) &= (1-p_{\text{c}_2}) \rho + p_{\text{c}_2} \left( r_{xx}X_tX_n\rho X_tX_n \right. \notag \\ 
	&~~~~~~~~~~~~~~~~~~~~~~~~+ \left. r_{xy} X_tY_n \rho X_tY_n \right. \notag \\ 
	&~~~~~~~~~~~~~~~~~~~~~~~~+ \left. r_{yx} Y_tX_n \rho Y_tX_n \right. \notag \\ 
	&~~~~~~~~~~~~~~~~~~~~~~~~+ \left. r_{yy}Y_tY_n \rho Y_tY_n  \right) \label{eq:ctincohallphases}
\end{align}
with phase ratios 
\begin{align}
    r_{xx} &= \cos^2\varphi_t\cos^2\varphi_n \\
    r_{xy} &= \cos^2\varphi_t\sin^2\varphi_n \\
    r_{yx} &= \sin^2\varphi_t\cos^2\varphi_n \\
    r_{yy} &= \sin^2\varphi_t\sin^2\varphi_n.
\end{align}
Averaging over phases of neighbor ions $\varphi_n$, we use Eq.~(\ref{eq:intphi}) to obtain the incoherent noise channel 
\begin{align}
	\langle \tilde{\curlE}\rangle_{\varphi_n}(\rho) &= (1-p_{\text{c}_2})\rho \notag \\ 
    &+ \frac{p_{\text{c}_2}}{2} \left( \cos^2 \varphi_t\,\left( X_t X_n \,\rho\, X_t X_n \right. \right. \notag \\
    &~~~~~~~~~~~~~~~~~+ \left. \left. X_tY_n \, \rho \, X_tY_n  \right) \right. \notag \\
	&~~~~~~+ \left. \sin^2 \varphi_t\,\left( Y_t X_n \,\rho\, Y_t X_n \right. \right. \notag \\
    &~~~~~~~~~~~~~~~~~+ \left. \left. Y_tY_n \, \rho \, Y_tY_n  \right) \right).
\end{align} 

For a simple noise model which -- in the same spirit as depolarizing noise -- does not need to take into account the microscopic nature of the gate, we also average over the target ion phases $\varphi_t$ to obtain the channel
\begin{align}
    \curlE_{\text{c}_2}(\rho) &= (1-p_{\text{c}_2}) \rho + \frac{p_{\text{c}_2}}{4} \left( X_tX_n\rho X_tX_n + X_tY_n\rho X_tY_n \right. \notag  \\
    &~~~~~~~~~~~~~~~~~~~~+ \left. Y_tX_n\rho Y_tX_n + Y_tY_n\rho Y_tY_n\right) \label{eq:pact}
\end{align}
which we use in our numerical simulations.

We now consider the special case where the two target ions share a common neighbor, i.e.~that $n_2(1) = n_1(2)$, which receives laser light from both target ions. If also the phases on target and neighbor ions are the same, the spin operator in Eq.~\eqref{eq:spop_twotermsexpanded} changes to read
\begin{align}
	S_{\vec{\varphi}} &= \frac{1}{2} \left(  \sigma_{\ph,1} + \sigma_{\ph,2} + \varepsilon\left( \sigma_{\ph,n_1(1)} + 2\sigma_{\ph,n_2(1)} + \sigma_{\ph,n_2(2)}\right) \right) \label{eq:spop_twotermsexpanded_commonneighbor}
\end{align}
so we get a coherent rotation of doubled angle $\theta \rightarrow 2\theta$ on the common neighbor ion. This doubling translates to the incoherent model through $\sin \varepsilon \theta = 4 \sin^2 \nicefrac{\varepsilon \theta}{2} \cos^2 \nicefrac{\varepsilon \theta}{2}$ to a shift in probability $p_{\text{c}_2} \rightarrow 4p_{\text{c}_2}$. 

This is, e.g., relevant for the XX-crosstalk discussed in Sec.~\ref{sec:ctr} where for all MS gates $\varphi_1 = \varphi_2 = 0$. On each target-neighbor-pair $t,\,n$ we can expand the unitary evolution operators from the coherent channel 
\begin{align}
	\curlE_\text{cct}(\rho) &= \exp \left( -\ii\frac{\varepsilon}{2}\theta X_tX_n \right) \,\rho\, \exp \left( \ii\frac{\varepsilon}{2}\theta X_tX_n \right) \label{eq:cct}
\end{align}
(cf.~Eq.~\eqref{eq:twoqubitrotation}) to obtain an incoherent noise channel for the MS gate crosstalk after PTA. Every crosstalk location which does not involve a common neighbor ion is then subject to the noise channel
\begin{align}
    \curlE_\text{xct}(\rho) &= (1-p_{\text{c}_2}) \rho + p_{\text{c}_2} X_tX_{n(t)} \rho X_tX_{n(t)}
\end{align}
with \mbox{$p_{\text{c}_2}(\theta) = \sin^2 \varepsilon\theta/2 = 6.2 \times 10^{-5}$ and $\theta = -\pi/2$}. For locations with common neighbor ions the shifts $\theta \rightarrow 2\theta$ and $p_{\text{c}_2} \rightarrow 4p_{\text{c}_2}$ are taken into account in the numerical simulations respectively.

As another special case, let us consider target ions labeled as qubit 2 and 4 so there is a common neighbor 3 and two outer neighbors 1 and 5. We take $\varphi_1 = 0$ and $\varphi_2 = -\pi/4$ and define the operator $F \equiv \frac{X-Y}{\sqrt{2}}$. Under the assumption that the neighbor ion phases were the same as their associated target ion's phase, we now find all operator combinations that contribute to crosstalk from
\begin{align}
	S_{0,-\pi/4} &= \frac{1}{2} \left( X_2 + F_4 + \varepsilon \left( X_1 + X_3 + F_3 + F_5 \right) \right) \\
	S_{0,-\pi/4}^2 &\supset \frac{1}{4} \left( 2 \varepsilon \left( X_1X_2 + X_2X_3 + X_2F_3 + X_2F_5 \right. \right. \notag \\ 
    &~~~~~~~~~~+ \left. \left. X_1F_4 + X_3F_4 + F_3F_4 + F_4F_5 \right) \right) \label{eq:spsq_F}
\end{align}
in the squared spin operator. Note that both terms $X_3F_4$ and $F_3F_4$ occur in Eq.~(\ref{eq:spsq_F}) so there is no angle doubling on the common neighbor qubit 4 since $\varphi_1 \neq \varphi_2$. Adjusting the phases of target ions could also be used in order to cancel the crosstalk on a common neighbor ion completely with the above assumption.

\textbf{Overrotations.}
The above reasoning for deriving noise channels from rotation operators can also be applied for overrotations of a small angle $\xi$ on a rotation about $\theta$ on an MS target qubit pair. This effectively implements a rotation of angle $\theta + \xi$ around an axis parametrized by phases $\varphi_1,\,\varphi_2$ for a two-qubit gate. The incoherent noise channel that we employ for simulations of XX-overrotation in MS gates (Eq.~(\ref{eq:ms0_unitary})) is
\begin{align}
    \curlE^{(2)}_\text{ior}(\rho) &= (1-p_2)\rho + p_2 X_1X_2 \rho X_1X_2 \label{eq:incoh_ovr}
\end{align}
with
\begin{align}
    p_2 &= \sin^2 \frac{\xi}{2}.
\end{align}
The corresponding coherent channel reads
\begin{align}
    \curlE^{(2)}_\text{cor}(\rho) &= \exp\left(-\ii\frac{\xi}{2} X_1X_2\right)\,\rho\, \exp\left(+\ii\frac{\xi}{2}X_1X_2\right).
\end{align}

\section{Simulation methods}\label{sec:sim_methods}
In this Appendix, we provide a detailed description of the theoretical methods employed for numerical simulations of logical failure rates. Depending on the range of physical error rates, we make use of either direct Monte Carlo (MC) simulation or subset sampling (SS) which is an importance sampling technique focusing on just the most important fault-weight-subsets contributing significantly to the logical failure rate.

\textbf{Direct Monte Carlo.}
When using direct Monte Carlo simulations, we model faulty qubit operations by an ideal unitary $U$ which is followed by a fault operator $E$ to form the faulty operation
\begin{align}
    U_\text{faulty} = E \cdot U_\text{ideal}.
\end{align}
The operator $E$ is placed after any ideal unitary gate or qubit initialization (or before a qubit measurement) with probability $\vec{p} = (p_1,\,p_2,\,p_i,\,\dots)$ and then drawn from the set of all possible fault operators according to the noise model. The MC estimator for the logical failure rate $\hat{p}_L$ is given by the number of samples where the stochastic placing of fault operators results in a logical failure divided by the total number of samples
\begin{align}
    \hat{p}_L &= \frac{\# \text{logical failures}}{\# \text{MC samples}}.
\end{align}
The sampling error for MC sampling can be estimated by the Wald interval
\begin{align}
    \varepsilon_\text{MC} = \sqrt{\frac{\hat{p}_L\left(1-\hat{p}_L\right)}{N}} \label{eq:stddev_MC}
\end{align}
so that for a large number of samples $N \rightarrow \infty$ the true logical failure rate $p_L$ is likely to be found in the confidence interval $\left[ \hat{p}_L - \varepsilon_\text{MC},\, \hat{p}_L + \varepsilon_\text{MC}\right]$. It is known that for $\hat{p}_L$ estimations that are close to or equal to zero or one after a finite but potentially small number of samples the Wald interval suffers from irregularities. These can be prevented using the Wilson score interval \cite{wilson1927probable} instead which is bounded by
\begin{align}
    p_\pm &= \frac{1}{1+\frac{z_{\alpha/2}^2}{N}} \left( \hat{p}_L + \frac{z_{\alpha/2}^2}{2N} \pm z_{\alpha/2} \sqrt{ \frac{\hat{p}_L(1-\hat{p}_L)}{N} + \frac{z_{\alpha/2}^2}{4N^2}} \right) \label{eq:wilsonint}
\end{align}
at confidence level $\alpha$ where $z$ is the quantile function of the normal distribution

MC sampling is efficient in a regime of physical failure rates where faults are realized frequently so we only employ it for larger physical failure rates. For low physical failure rates, in MC sampling one would mostly run the fault-free case, e.g.~at $p = 0.1\%$ and a circuit of 100 gates the ideal circuit would be realized $(1-p)^{100} \approx 90\%$ of the time. When realization of fault operations becomes a rare event, we turn towards subset sampling instead.

\textbf{Subset sampling.} The logical failure rate $p_L$ can be written as a sum of so-called subset failure rates $\vec{p}_\text{fail}$ that contribute with different weights $A(\vec{w},\vec{p})$ each, so that

\begin{align}
    p_L &= \sum_{\vec{w}} A(\vec{w},\vec{p}) \vec{p}_\text{fail}(\vec{w})
\end{align}
where we distinguish subsets by the weight $\vec{w} = (w_1, w_2, w_i, ...) $ of the fault operator that is applied to the respective circuit operations. Each subset failure rate $\vec{p}_\text{fail}(\vec{w})$ is obtained by MC sampling fault operations with fixed weight $\vec{w}$. The contribution of each subset is given by the binomial weight
\begin{align}
	A(\vec{w},\vec{p}) = \prod_\mu \binom{N_\mu}{w_\mu} p_\mu^{w_\mu} (1-p_\mu)^{N_\mu-w_\mu} \label{eq:binom_ss}
\end{align}
where $\mu$ iterates over all types of faulty circuit operations since the probability of applying exactly $w_\mu$ fault operators is $p_\mu^{w_\mu} (1-p_\mu)^{N_\mu-w_\mu}$ and there are $\binom{N_\mu}{w_\mu}$ possibilities to arrange these configurations for any type $\mu \in \{1, 2, i,...\}$. The true logical failure rate is bounded by 
\begin{align}
    \hat{p}_L &= \sum_{\vec{w} = \vec{0}}^{\vec{w}_\text{max}} A(\vec{w},\vec{p}) \vec{p}_\text{fail}(\vec{w}) \leq p_L \\
    &\leq \sum_{\vec{w} = \vec{0}}^{\vec{w}_\text{max}} A(\vec{w},\vec{p}) \vec{p}_\text{fail}(\vec{w}) + \sum_{\vec{w}_\text{max}+1}^{\vec{N}} A(\vec{w},\vec{p}) \label{eq:pL_ss_bounds}
\end{align}
where the weight cutoff error 
\begin{align}
    \delta(\vec{p}) &= \sum_{\vec{w}_\text{max}+1}^{\vec{N}} A(\vec{w},\vec{p})
\end{align}
vanishes for low physical failure rates $\delta(\vec{p}) \rightarrow 0$ as $\vec{p} \rightarrow \vec{0}$. However, in the opposite regime $\delta(\vec{p})$ becomes large so one must choose an appropriate weight cutoff $\vec{w}_\text{max}$ to keep the cutoff error below a desired numerical value. For large weight cutoff $|\vec{w}_\text{max}|$ the number of subsets is so large that it becomes advantageous to use MC sampling instead. Subset sampling will be advantageous as long as the fault-free subset $\vec{w} = \vec{0}$ is the largest subset
\begin{align}
    A(\vec{0},\vec{p}) &\geq A(\vec{w},\vec{p})~~~\forall \vec{w}.
\end{align}
The MC sampling errors $\varepsilon_\text{SS}(\vec{w}) \sim \sqrt{\frac{\vec{p}_\text{fail}(\vec{w})\left(1-\vec{p}_\text{fail}(\vec{w})\right)}{N_\text{SS}(\vec{w})}}$ for all subsets accumulate to the sampling error on the logical failure rate
\begin{align}
	\varepsilon_\text{SS} &= \sqrt{\sum_{\vec{w} = \vec{0}}^{\vec{w}_\text{max}} \left[ A(\vec{w},\vec{p}) \varepsilon_\text{SS}(\vec{w}) \right]^2} \label{eq:ss_samplingerror}
\end{align}
so that overall the true logical failure rate $p_L$ will likely be in the interval $\left[ \hat{p}_L - \varepsilon_\text{SS},\, \hat{p}_L + \varepsilon_\text{SS} + \delta \right]$.

\textbf{Practical procedure.}
For the logical failure rates presented in this work we performed the following sampling procedure. First, we fix a scale of interest for the physical failure rates and the crosstalk ratio parametrized by $\lambda$ (see Eq.~\eqref{eq:lambda}). This scale contains the experimental parameters as a reference point at $\lambda = 1$. For the depolarizing noise model we scale the parameters $p_1,\,p_2,\,p_i,\,p_m$ and for the extended noise model we additionally scale the parameters $p_{\text{idle},1},\,p_{\text{idle},2},\,p_{\text{idle},m},\,p_{\text{c}_1}(\pi),\,p_{\text{c}_1}(\pi/2),\,p_{\text{c}_1}(\pi/4),\,p_{\text{c}_2}$. For the XX-type crosstalk model, $p_{\text{c}_2}$ is replaced by $p_{\text{c}_2, \text{com}}$ and $p_{\text{c}_2, \text{non}}$ for common and non-common neighbor ion crosstalk locations.

We then start our numerical simulation by using MC at the largest physical failure rates and sample at decreasing physical failure rate until either the target relative error is reached or the previously specified maximum number of samples has been run. In the latter case or when no logical failure was recorded at all, we repeat the simulation at the present physical failure rates using subset sampling. Here, we now choose the maximum weight such that the cutoff error $\delta$ at the present physical failure rates accounts for at most half of the target relative error. We perform subset sampling uniformly over all relevant subsets until the sampling error $\varepsilon_\text{SS}$ is also at most half of the target relative error or until the maximum number of samples has been reached. The sampling error for all numerical simulations is given as the Wilson score interval \eqref{eq:wilsonint} at a confidence level of 95\% ($z_{0.025} \approx 1.96$) in a symmetric form $[\hat{p}_L - \frac{p_+-p_-}{2}, \hat{p}_L + \frac{p_+-p_-}{2}]$. This prevents us from irregularities of the Wald interval that may occur at subset failure rate estimations that are close to or equal to zero or one after a finite but potentially small number of samples.

In subset sampling we refrain from actually sampling the fault-free subset but fix its subset failure rate and sampling error to be equal to zero. For a non-fault-tolerant circuit we exhaustively place all possible weight-1 faults to obtain the subset failure rates for the subsets with total weight $|\vec{w}|$ equal to one exactly, i.e. without sampling error. We do the same for all crosstalk faults since they do not respect the FT property in general. For faults that do respect FT, we also fix their subset failure rates and sampling error to be equal to zero. 

\section{Deterministic FT magic state preparation} \label{sec:detFTm}

The look up table used for correcting errors during the logical Hadamard measurement as part of the deterministic FT magic state preparation protocol in Fig.~\ref{fig:det_seq} is given in Tab.~\ref{tab:hlut}. The recovery operation $R$ which is applied directly after an $EC$ block depends not only on the measured syndrome but also on the flag pattern $f_0,f_1,f_2,f_3$ measured in the $M_H$ block. Note for example that the syndromes 000 001 may lead to either the recovery operation $R = X_2$ or $R = X_1X_3$ depending on said flag measurements. The full six-bit syndrome is necessary to correct all Hadamard errors despite the symmetry of $X$- and $Z$-stabilizers in the Steane code. To see this, consider the Hadamard error
\begin{align}
    H_1H_3 &= \frac{1}{2} \left( X_1X_3 + Z_1X_3 + X_1Z_3 + Z_1Z_3 \right).
\end{align}
Since $H = \frac{X+Z}{\sqrt{2}}$ the product of two or more Hadamards mixes all possible combinations of $X$- and $Z$-operators which must be distinguished by the syndrome. At the same time, the flag pattern allows us to distinguish weight-2 errors from weight-1 errors that would cause the same syndrome.

\begin{table}[!htbp]
\tiny 
\begin{center}
\begin{tabular}{ c | c | c ||c }
$f_0,f_1,f_2,f_3$ & $K_1^X, K_2^X, K_3^X$ & $K_1^Z, K_2^Z, K_3^Z$ & $R$ \\ \hline 
1100 & 000 & 001 & $X_2$ \\
1110 & 000 & 001 & $X_2$ \\
1010 & 000 & 001 & $X_2$ \\
1011 & 000 & 001 & $X_2$ \\ \hline 
1100 & 001 & 000 & $Z_2$ \\
1110 & 001 & 000 & $Z_2$ \\
1010 & 001 & 000 & $Z_2$ \\
1011 & 001 & 000 & $Z_2$ \\ \hline 
1000 & 000 & 011 & $X_3$ \\
1000 & 011 & 000 & $Z_3$ \\ 1000 & 000 & 111 & $X_7$ \\
1000 & 111 & 000 & $Z_7$ \\ \hline 
1000 & 000 & 001 & $X_1X_3$ \\
1000 & 011 & 010 & $X_1Z_3$ \\
1000 & 010 & 011 & $X_3Z_1$ \\
1000 & 001 & 000 & $Z_1Z_3$ \\ 
1100 & 110 & 100 & $X_4Z_5$ \\
1100 & 100 & 110 & $X_5Z_4$ \\
1000 & 000 & 010 & $X_6X_7$ \\
1100 & 000 & 010 & $X_6X_7$ \\
1000 & 111 & 101 & $X_6Z_7$ \\
1100 & 111 & 101 & $X_6Z_7$ \\
1000 & 101 & 111 & $X_7Z_6$ \\
1100 & 101 & 111 & $X_7Z_6$ \\
1000 & 010 & 000 & $Z_6Z_7$ \\
1100 & 010 & 000 & $Z_6Z_7$ \\ \hline 
1000 & 010 & 001 & $X_1X_3Z_1$ \\
1000 & 001 & 010 & $X_1Z_1Z_3$ \\
1000 & 000 & 101 & $X_1X_3X_4$ \\
1010 & 000 & 101 & $X_1X_3X_4$ \\ 
1000 & 011 & 110 & $X_1X_4Z_3$ \\
1010 & 011 & 110 & $X_1X_4Z_3$ \\
1000 & 010 & 111 & $X_3X_4Z_1$ \\
1010 & 010 & 111 & $X_3X_4Z_1$ \\
1000 & 001 & 100 & $X_4Z_1Z_3$ \\
1010 & 001 & 100 & $X_4Z_1Z_3$ \\
1000 & 100 & 001 & $X_1X_3Z_4$ \\
1010 & 100 & 001 & $X_1X_3Z_4$ \\
1000 & 111 & 010 & $X_1Z_3Z_4$ \\
1010 & 111 & 010 & $X_1Z_3Z_4$ \\
1000 & 110 & 011 & $X_3Z_1Z_4$ \\
1010 & 110 & 011 & $X_3Z_1Z_4$ \\
1000 & 101 & 000 & $Z_1Z_3Z_4$ \\
1010 & 101 & 000 & $Z_1Z_3Z_4$ \\ \hline 
1110 & 000 & 100 & $X_5X_6X_7$ \\
1010 & 000 & 100 & $X_5X_6X_7$ \\
1100 & 000 & 100 & $X_5X_6X_7$ \\
1110 & 101 & 001 & $X_5X_7Z_6$ \\
1010 & 101 & 001 & $X_5X_7Z_6$ \\
1100 & 101 & 001 & $X_5X_7Z_6$ \\
1110 & 110 & 010 & $X_6X_7Z_5$ \\
1010 & 110 & 010 & $X_6X_7Z_5$ \\
1100 & 110 & 010 & $X_6X_7Z_5$ \\
1110 & 011 & 111 & $X_7Z_5Z_6$ \\
1010 & 011 & 111 & $X_7Z_5Z_6$ \\
1100 & 011 & 111 & $X_7Z_5Z_6$ \\
1110 & 111 & 011 & $X_5X_6Z_7$ \\
1010 & 111 & 011 & $X_5X_6Z_7$ \\
1100 & 111 & 011 & $X_5X_6Z_7$ \\
1110 & 010 & 110 & $X_5Z_6Z_7$ \\
1010 & 010 & 110 & $X_5Z_6Z_7$ \\
1100 & 010 & 110 & $X_5Z_6Z_7$ \\
1110 & 001 & 101 & $X_6Z_5Z_7$ \\
1010 & 001 & 101 & $X_6Z_5Z_7$ \\
1100 & 001 & 101 & $X_6Z_5Z_7$ \\
1110 & 100 & 000 & $Z_5Z_6Z_7$ \\
1010 & 100 & 000 & $Z_5Z_6Z_7$ \\
1100 & 100 & 000 & $Z_5Z_6Z_7$ \\
1000 & 101 & 010 & $X_6X_7Z_6$ \\
1000 & 010 & 101 & $X_6Z_6Z_7$ \\ \hline 
1100 & 011 & 001 & $X_5X_7Z_5Z_6$ \\
1100 & 001 & 011 & $X_5X_6Z_5Z_7$ \\
1000 & 100 & 101 & $X_1X_3X_4Z_4$ \\
1000 & 111 & 110 & $X_1X_4Z_3Z_4$ \\
1000 & 110 & 111 & $X_3X_4Z_1Z_4$ \\
1000 & 101 & 100 & $X_4Z_1Z_3Z_4$ 
\end{tabular} 
\end{center}
\caption{Look up table for flag-FT measurement of the logical Hadamard operator in the deterministic scheme given in Fig.~\ref{fig:det_seq}b. $+1$ and $-1$ measurement outcomes of flag and syndrome auxiliary qubits are represented as $0$ and $1$ respectively. The full six bit syndrome needs to be considered in order to choose the appropriate recovery operation $R$ in contrast to the situation where X- and Z-type recoveries are applied independently in standard EC on the Steane code.}
\label{tab:hlut}
\end{table}

\begin{figure}[!htbp]
    \centering
    \includegraphics[width=\linewidth]{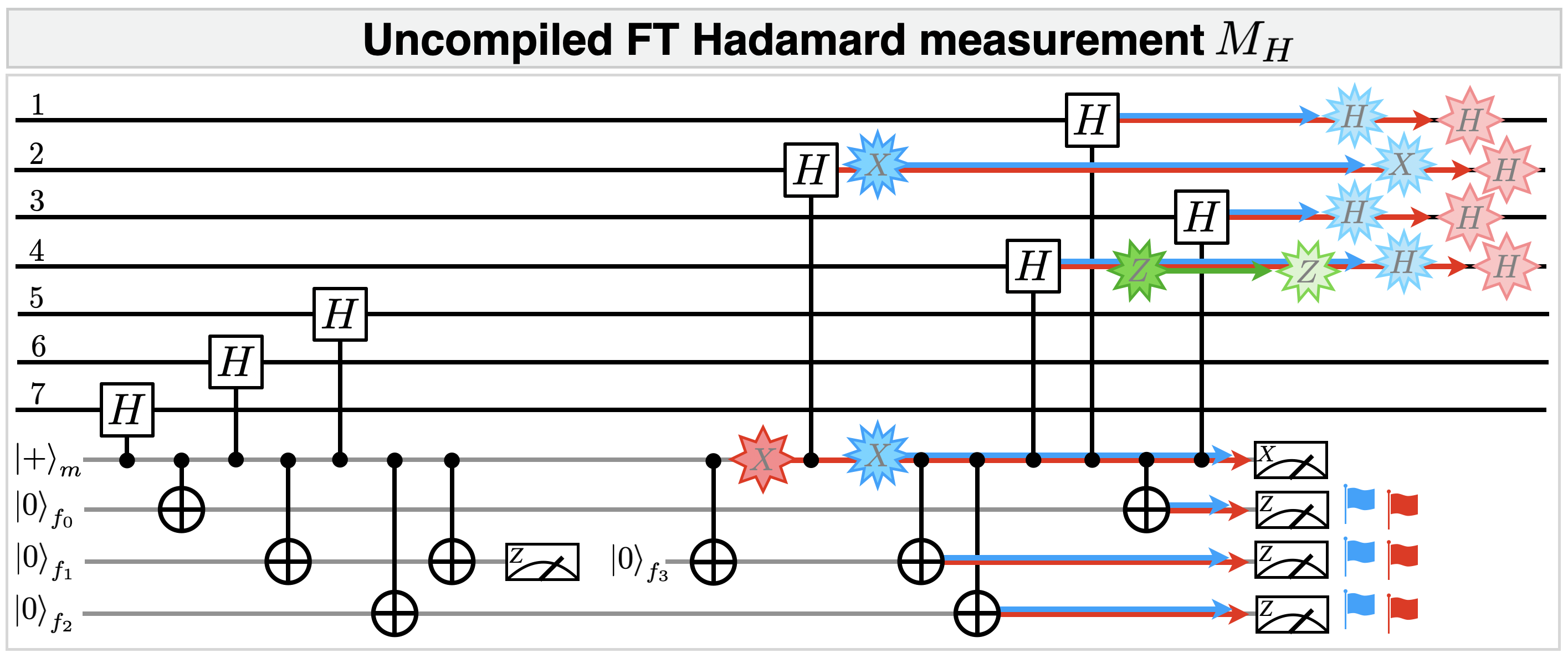}
    \caption{\textbf{Uncompiled logical Hadamard measurement circuit.} Flag FT circuit for measuring the logical Hadamard operator according to Ref.~\cite{chamberland2019fault}. The qubit mapping is not changed because our stabilizers are unchanged compared to Ref.~\cite{chamberland2019fault}. The faults $X_m$ (red, 8-cornered star), $X_2X_m$ (blue, 12-cornered star) and $Z_4$ (green, 10-cornered star) as described in the main text are shown with their respective propagated errors $H_1H_2H_3H_4,\,X_2H_1H_3H_4$ and $Z_4$.}
    \label{fig:mh_ch}
\end{figure}

To see why the correction $F = H_1H_3H_4$ from Fig.~\ref{fig:det_seq}c is necessary, we consider the uncompiled version of the measurement circuit $M_H$ from Ref.~\cite{chamberland2019fault} which is reproduced in Fig.~\ref{fig:mh_ch}. Here, an $X$-fault on the measurement qubit can cause the error $H_1H_2H_3H_4$ at the end of $M_H$ as shown in Fig.~\ref{fig:mh_ch}. This error contains all combinations of X- and Z-type operators on qubits 1 to 4, for instance $X_1X_2X_3Z_4 = X_LZ_4$ and $Z_1Z_2Z_3Z_4 = Z_LZ_4$. Both these constituents of the Hadamard error will lead to the same syndrome measurement in the EC block, namely $-++\,+++$, the one matching $Z_4$, but different logical operators are introduced unnoticed. Applying the $F$-block will transform the error to $F\,H_1H_2H_3H_4 = H_2 \simeq X_2 + Z_2$. By the subsequent EC block, this superposition will collapse so that either the syndrome $+++\,++-$ or $++-\,+++$ will be measured and the respective error can be corrected. If instead we had not applied the $F$ operation, we could confuse the error with another one causing the same syndrome, i.e.~an error that does not contain logical $X$ or logical $Z$ as shown above but a logical identity or logical $Y$ on qubits 1 to 3. Keep in mind that also the flag pattern needs to be identical so that we cannot use it either to distinguish the errors. Consider the $Z_4I_m$ fault on the third to last controlled Hadamard gate. It will cause the error $Z_4$ with syndrome $-++\,+++$ which can be distinguished from the $X_LZ_4$ and $Z_LZ_4$ errors given above because it will not trigger any flag of $M_H$. As an example of two faults that lead to the same flag pattern and syndrome if $F$ were not applied, take the fault $X_2X_m$ on the fourth controlled-Hadamard gate and $X_m$ just before this gate. Both cause the flag pattern $1011$. The former will cause the error $X_2H_1H_3H_4$ and the latter will propagate to $H_1H_2H_3H_4$ which is equivalent when acting on the logical magic state to $H_5H_6H_7$ since the magic state is the eigenstate of the logical Hadamard operator $H_L = H^{\otimes 7}$. These two errors can, e.g., be collapsed to $X_1X_2X_3X_4$ and $X_5X_6X_7$ by the EC block and the syndrome $+++\,-++$ will be measured. Since they differ by a logical $X$ they cannot be distinguished by the Hadamard look up table. Applying $F$ will transform the errors according to
\begin{align}
    F X_2H_1H_3H_4 &= X_2\\
    F H_5H_6H_7 &= H_1H_3H_4H_5H_6H_7 \notag \\ \Leftrightarrow F H_1H_2H_3H_4 &= H_2
\end{align}
which can both be corrected. As another example, these two faults could also collapse to $X_1X_2X_3Z_4$ and $Z_1Z_2Z_3Z_4$ respectively by the EC block yielding syndrome $-++\,+++$. Confusing one for the other we would in total apply a logical $Y$ operation to the logical magic state which is prevented by the $F$ flip.

\begin{figure}[!htbp]
    \centering
    \includegraphics[width=\linewidth]{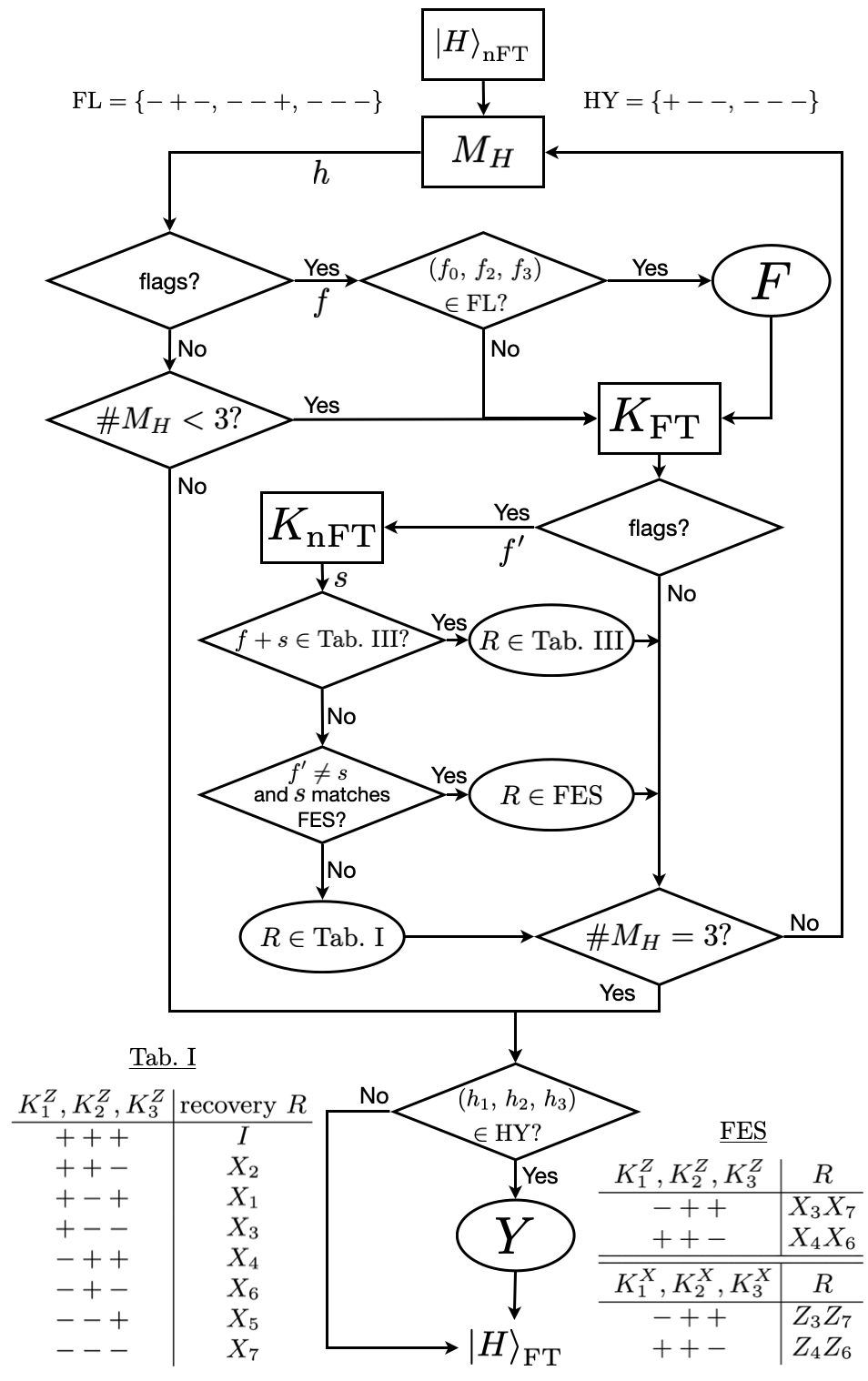}
    \caption{\textbf{Flowchart for the deterministic FT magic state preparation procedure.} In-sequence measurements determine the circuits of the noisy protocol chosen dynamically during runtime (rectangles). The measurement outcome of an individual $M_H$ circuit is labeled $h$. Measurements of flag circuits are shown as outputs $f$ and $f'$. Denoted by $s$ is the syndrome output by $K_\text{nFT}$. Additional corrections need to be applied (ovals) depending on the intermediate measurement results: $R$ is drawn from the Steane look up table \ref{tab:steane_lut}, the Hadamard error set (Tab.~\ref{tab:hlut}) or the flag error set (FES) and $F$ and $Y$ are given in Fig.~\ref{fig:det_seq}. The sets of measurement results that cause application of $F$ and $Y$ are labeled FL and HY respectively.}
    \label{fig:flowchart}
\end{figure}

\begin{figure*}[!htbp]
    \centering
    \includegraphics[width=0.7\linewidth]{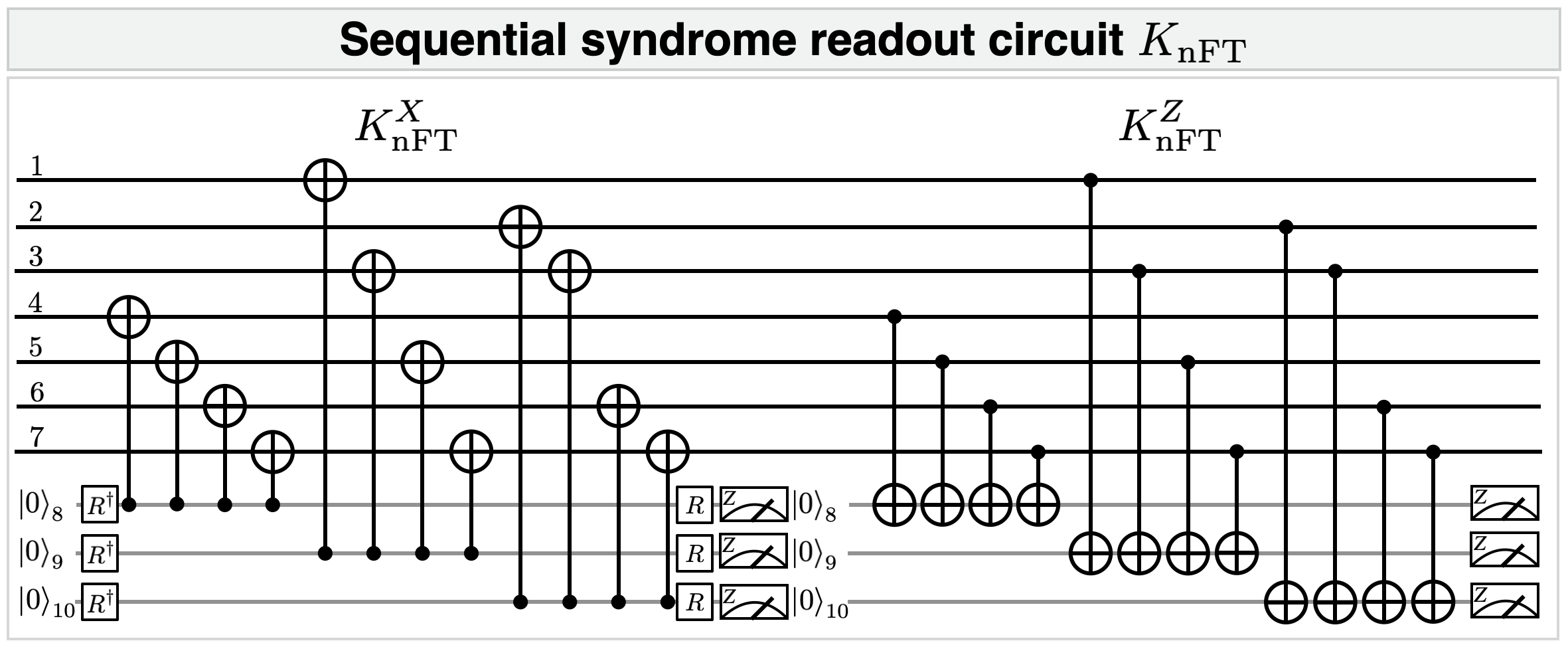}
    \caption{\textbf{Sequential stabilizer readout.} The circuit is used for non-FT stabilizer readout of the six bit syndrome as part of the deterministic FT magic state preparation protocol shown in Fig.~\ref{fig:det_seq}. Each CNOT gate is directly compiled into the sequence of MS gates and local rotations as given in Fig.~\ref{fig:architecture}c.}
    \label{fig:circ_knft}
\end{figure*}

The steps involved in the deterministic FT magic state preparation are depicted as a flowchart in Fig.~\ref{fig:flowchart}. The overall correction strategy works as follows: If the Hadamard measurement flags and there exists an entry in the Hadamard look up table \ref{tab:hlut} for the measured flag pattern and syndrome, apply the corresponding recovery operation. Else, if the error correction flags, run the non-FT syndrome readout (Fig.~\ref{fig:circ_knft}) and apply the correction according to the flag error set if the flags and syndrome disagree. Otherwise, apply the standard Steane look up table recovery operation. For the EC block, X- and Z-type syndromes can be read out independently from each other.

\begin{figure*}[!htbp]
    \centering
    \includegraphics[width=0.7\linewidth]{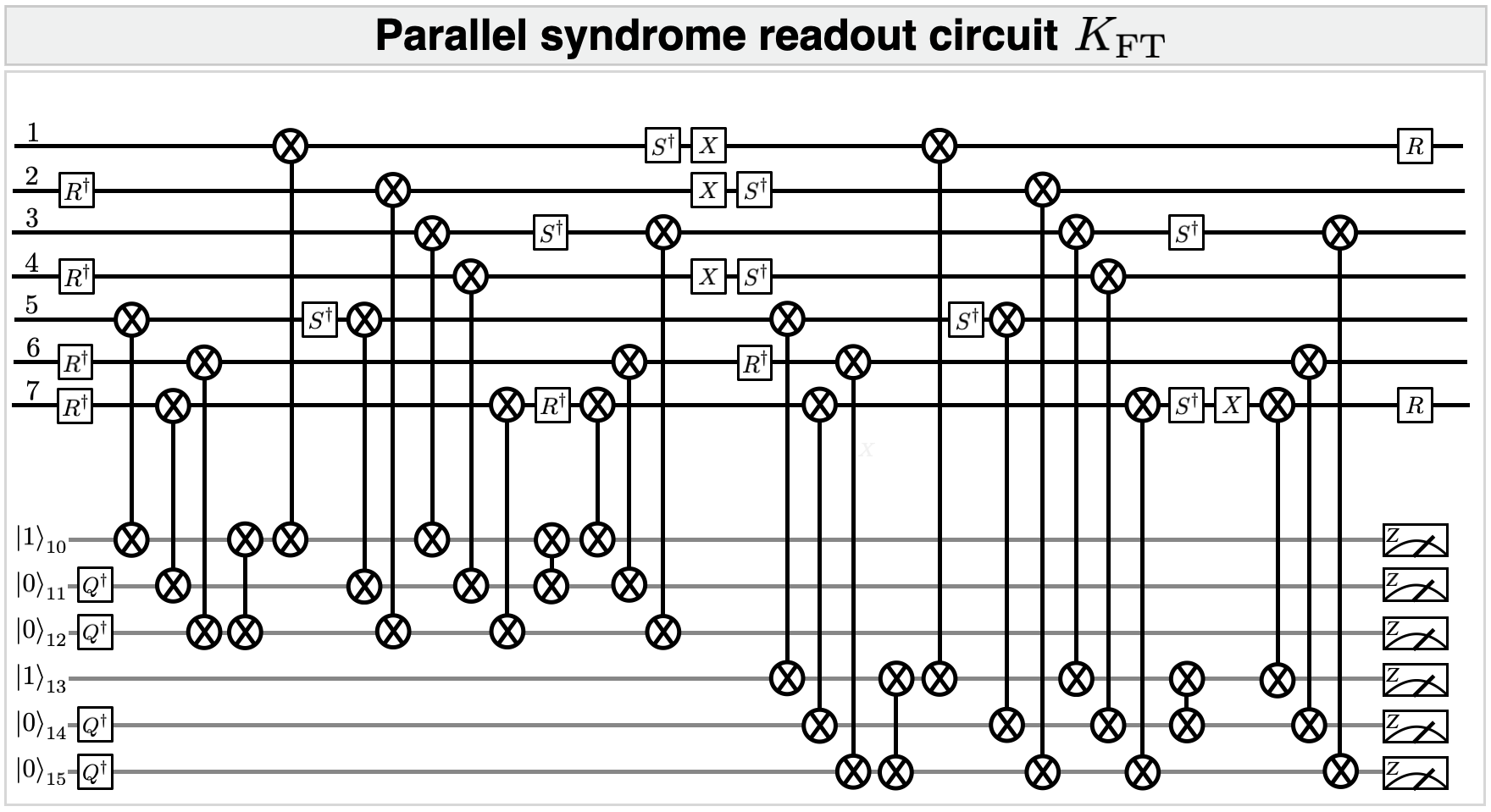}
    \caption{\textbf{Parallel stabilizer readout.} FT circuit for interleaved measurement of all six stabilizers. Auxiliary qubits simultaneously act as measurement and flag qubits for the deterministic FT magic state preparation protocol (see Fig.~\ref{fig:det_seq}).}
    \label{fig:circ_kft}
\end{figure*}

\section{Quantum state fidelity}\label{sec:qsf}

The full quantum state fidelity of the data qubit state is defined as

\begin{align}
    \mathcal{F}(\rho_t, \rho) &= \tr (\rho_t \rho) = \langle \rho_t \rangle \label{eq:statefid}
\end{align}
the expectation value of our logical target state $\rho_t = \ket{t}\bra{t}$. Eq.~(\ref{eq:statefid}) is in contrast to the logical fidelity which is the overlap of the output state with the desired logical Bloch vector. 
The quantum state fidelity is the standard quantity that characterizes a quantum state independently of any QEC framework. It contains information about the full state including local properties which the logical fidelity fails to provide since it is merely understood as the overlap of the logical Bloch vector with the desired logical target state, i.e. the projection onto this state. Although the logical fidelity reflects the probability to successfully recover the state after a noisy circuit, it is defined only in the code space but not the full $n$-qubit Hilbert space \cite{nigg2014quantum}. Since the stabilizers subdivide the Hilbert space to form the code space in the first place, it is important to quantify how well the code space itself is prepared, i.e.~how close to unity are the expectation values of the stabilizer generators. 

We expand the general $n$-qubit target state $\rho_t$ in the operator basis formed by all possible $n$-qubit Pauli operators $W_k$ where $k = 1,\,...,\,4^n$. The quantum state fidelity then reads

\begin{align}
    \mathcal{F}(\rho_t,\rho) &= \frac{1}{4^n} \tr \left( \sum_{k=1}^{4^n} \left[ \tr \left( W_k \rho_t \right) W_k \right] \rho \right).
\end{align}

For stabilizer states $\rho = \ket{\psi}\bra{\psi}$ with $W_k \ket{\psi} = \pm \ket{\psi}$ being the elements of the stabilizer group, only the $2^n$ coefficients corresponding to the set of all stabilizer elements $W_k$ are non-zero $\tr(W_k\rho) = \pm 1$. The fidelity can then be expressed as
\begin{align}
    \mathcal{F}(\rho_t,\rho) = \frac{1}{2^n} \sum_{k=1}^{2^n} \langle W_k \rangle \label{eq:stabfidelity}
\end{align}
where the $W_k$ are all possible products of combinations of stabilizer elements of the logical state, i.e.~combinations of code stabilizer generators with the respective logical operators or the identity: 
\begin{align}
    \sum_{k=1}^{2^n} W_k &= \prod_{i=1}^n \frac{I + S_i}{2} \\
    S_i &\in \{ K_1^X, K_1^Z, K_2^X, K_2^Z, K_3^X, K_3^Z, O_t \}
\end{align}

For stabilizer states we only need to evaluate Eq.~(\ref{eq:stabfidelity}) to obtain the quantum state fidelity. For a single logical qubit in an $n=7$-qubit register
$\rho_t$ may be factorized by projectors onto the code space and the logical subspace
\begin{align}
    \rho_t &= P_{\pm O_t} P_{\text{CS}} \\
    P_\text{CS} &= \prod_{i=1}^6 \frac{I + K_i}{2} \\
    P_{\pm O_t} &= \frac{I \pm O_t}{2} \text{.}
\end{align}
The density operator for the logical zero state $\ket{0}_L$ reads
\begin{align}
    \rho_{\ket{0}_L} &= \ket{0}\bra{0}_L = \frac{I + Z_L }{2} P_\text{CS}
\end{align}
and the state fidelity for each of these cases reduces to
\begin{align}
    \mathcal{F}(\rho_t,\rho) = \frac{1}{128} \sum_{k=1}^{128} \langle W_k \rangle 
\end{align}
with the respective \mbox{$W_k = I, \dots, Z_LK_1^XK_1^ZK_2^XK_2^ZK_3^XK_3^Z$}. The code space population $p_\text{CS}$ and the fidelity within the code space $\mathcal{F}_\text{CS}$ is obtained via
\begin{align}
    p_\text{CS} &= \tr\left(P_\text{CS}\rho_t\right) = \frac{1}{64} \sum_{k=1}^{64} \langle W_k \rangle \\
    \mathcal{F}_\text{CS}(\rho_t) &= \frac{\tr\left(\rho_t \rho \right)}{p_\text{CS}}
\end{align}
where the 64 terms for the code space population are the Pauli operators $W_k$ which do not contain the logical operator \mbox{$W_k = I, \dots, K_1^XK_1^ZK_2^XK_2^ZK_3^XK_3^Z$}.

\section{Single-qubit randomized benchmarking}\label{sec:rb}

\begin{figure*}[!htbp]
    \centering
    \includegraphics[width=0.9\linewidth]{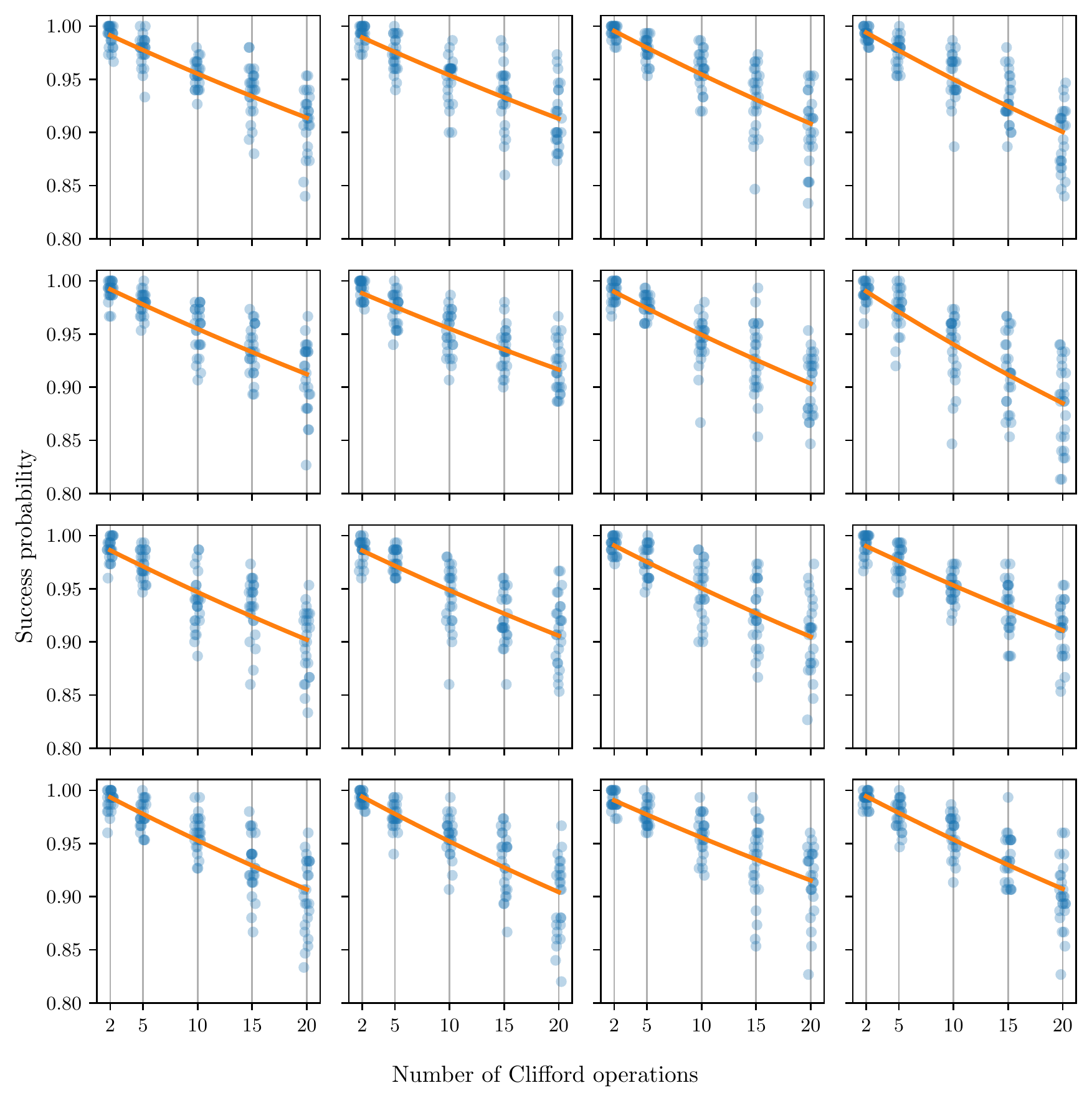}
    \caption{\textbf{Single-qubit gate benchmarking.} Success probability decay of randomized benchmarking sequences in a 16-qubit register (qubit 1 in the top-left corner, qubit 16 in the bottom-right corner). The scatter on the horizontal axis around the sequence lengths 2, 5, 10, 15 and 20 is introduced for better visibility of the success probability of the individual random sequences. The discretization on the vertical axis is given by averaging over 150 executions per random sequence.}
    \label{fig:rb_individual}
\end{figure*}

The fidelity of single-qubit operations is extracted from randomized benchmarking experiments as described in Ref.~\cite{chen2018metrology}, where a single Clifford operation is decomposed into 2.167 laser pulses on average. In Fig.~\ref{fig:rb} we combined data for all 16 qubits to a single dataset, whereas in Fig.~\ref{fig:rb_individual} we show the underlying datasets for all qubits individually. The numerical values for single-qubit gate fidelities are given in Table~\ref{tab:rb}. As there is no pattern of single-qubit gate fidelity with respect to the position in the ion chain apparent, all error models discussed in this work feature only a single fidelity for all single-qubit gates.

\begin{table}[!htbp]
\begin{center}
\begin{tabular}{ c | c}
Qubit number & Single-qubit gate fidelity \\ \hline 
1 & $0.9978(3)$ \\
2 & $0.9978(3)$ \\
3 & $0.9975(3)$ \\
4 & $0.9973(3)$ \\
5 & $0.9977(3)$ \\
6 & $0.9980(3)$ \\
7 & $0.9975(3)$ \\
8 & $0.9969(4)$ \\
9 & $0.9976(3)$ \\
10 & $0.9977(3)$ \\
11 & $0.9975(3)$ \\
12 & $0.9977(3)$ \\
13 & $0.9975(3)$ \\
14 & $0.9974(3)$ \\
15 & $0.9979(3)$ \\
16 & $0.9975(3)$ \\
\end{tabular} 
\end{center}
\caption{Single-qubit gate fidelities estimated from randomized benchmarking in a 16-qubit register. The number of Clifford operations used in the generation of the benchmarking sequences ranges from 2 to 20. The given errors are 95\% confidence intervals.}
\label{tab:rb}
\end{table}

\section{Estimation of entangling gate fidelity}\label{sec:ghz}

\begin{figure}[!htbp]
    \centering
    \includegraphics[width=\linewidth]{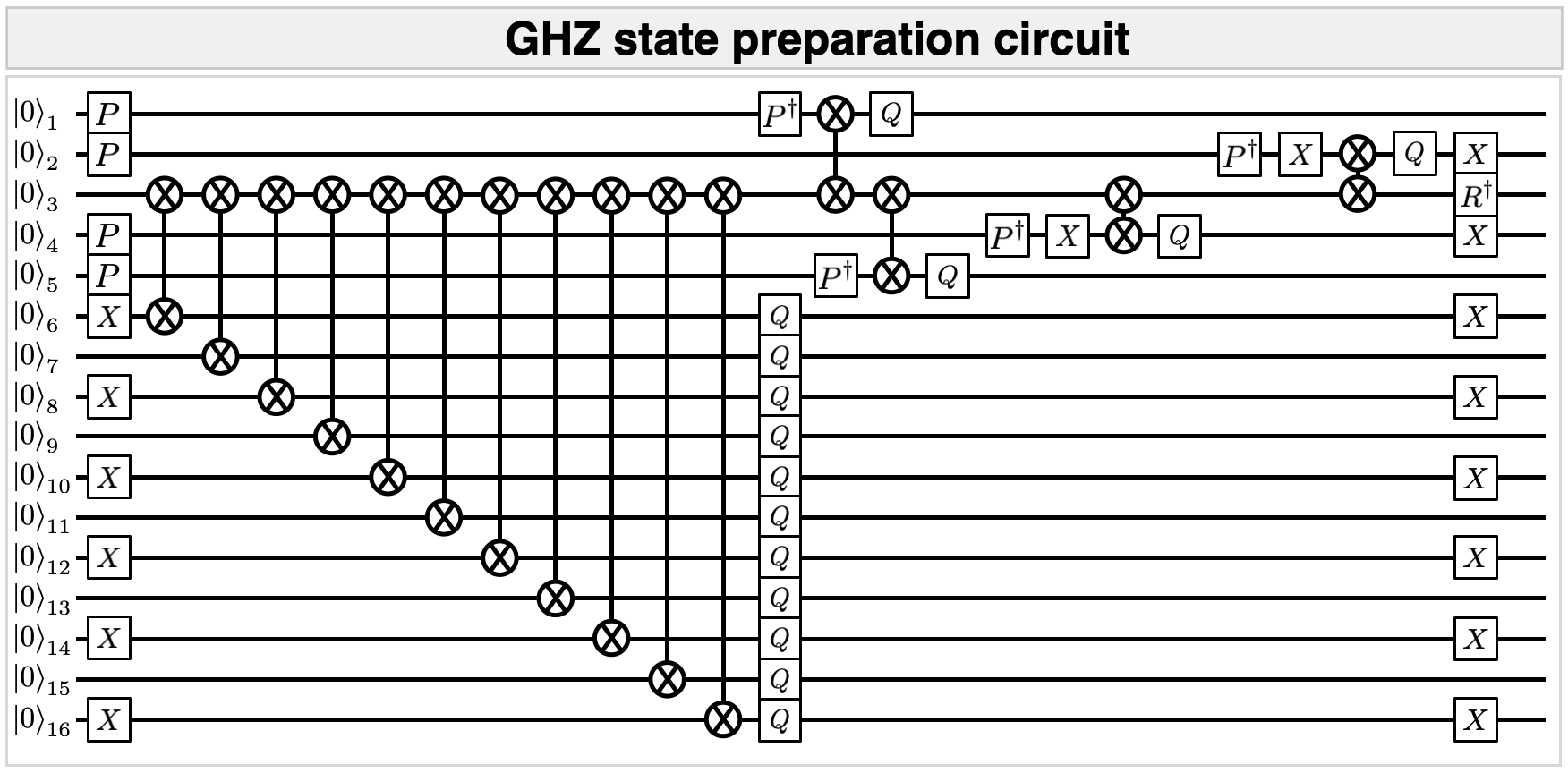}
    \caption{\textbf{16-qubit GHZ state preparation circuit.} The circuit is used to estimate the fidelity of a single entangling gate. The operations $P^{(\dagger)}$
    are resonant pulses with a rotation angle of $\pi$ (and opposite rotation direction) on the transition 4$S_{\nicefrac{1}{2}, m_j = -\nicefrac{1}{2}}$ to 3$D_{\nicefrac{5}{2}, m_j = -\nicefrac{3}{2}}$ used for (un)hiding of qubits. This shelving procedure reduces noise due to crosstalk from multiple entangling gates acting on qubit 3.}
    \label{fig:ghz}
\end{figure}

To estimate the mean fidelity of entangling operations without using time-consuming benchmarking techniques the following approach is used: We prepare the 16-qubit GHZ state $\ket{\psi_{\mathrm{GHZ}}} = (\ket{0}^{\otimes 16} - i \ket{1}^{\otimes 16})/\sqrt{2}$ across the entire register by using 15 two-qubit MS gates and 40 single-qubit resonant operations. The corresponding circuit is depicted in Fig.~\ref{fig:ghz}. For the analysis of the fidelity of the prepared GHZ state we perform two measurements: The probabilities to project to the basis states $\ket{0}^{\otimes 16}$ and $\ket{1}^{\otimes 16}$ are determined by a direct projective measurement in the $Z$-basis. The off-diagonal elements of the density matrix of the GHZ state instead are measured by applying single-qubit gates \mbox{$R_{\varphi}^{(i)}(\pi / 2)$} to all qubits after preparing the GHZ state. For different phases $\varphi$ the parity of the prepared state is measured via a projective measurement and a sinusoidal model is fitted to the observed parity oscillations~\cite{monz2011entanglement}. The mean of the sum of the populations in $\ket{0}^{\otimes 16}$ and $\ket{1}^{\otimes 16}$ and the contrast of the parity oscillations of the coherence measurement gives the fidelity of the GHZ state. The fidelity of a single two-qubit gate is estimated as~\cite{baldwin2022reexamining}
\begin{equation}
    \mathcal{F}_\mathrm{tq} = \left( \frac{\mathcal{F}_\mathrm{GHZ}}{\mathcal{F}_\mathrm{sq}^{40}} \right) ^{\frac{1}{15}},
\end{equation}
where $\mathcal{F}_\mathrm{GHZ} = 0.62(3)$ and $\mathcal{F}_\mathrm{sq} = 0.99760(8)$ are fidelity of the GHZ state and mean single-qubit gate fidelity estimated from randomized benchmarking respectively. The estimated two-qubit gate fidelity in the 16-qubit register is $\mathcal{F}_\mathrm{tq} = 0.975(3)$.

\bibliographystyle{bibstyle}\bibliography{references}

\end{document}